\documentclass[traditabstract]{aa} 

\usepackage{graphicx}
\usepackage{rotating}
\usepackage{epsfig}
\usepackage{float} 
\usepackage{subfig}

\usepackage{natbib}
\bibpunct{(}{)}{;}{a}{}{,} % to follow the A&A style
\usepackage{dcolumn}
\newcolumntype{.}{D{.}{.}{-1}}
\newcolumntype{;}{D{;}{.}{7}}

\begin{document}

\authorrunning{Shahzamanian et al.} 
\titlerunning{NIR polarization of Sgr A*}
\title{Polarized light from Sgr~A* in the near-infrared $K_{s}$-band \thanks{Based on 
NACO observations collected between 2004 and 2012 at the Very Large Telescope (VLT) of the 
European Organization for Astronomical Research in the Southern Hemisphere (ESO), Chile}}
\subtitle{}

\author{B. Shahzamanian\inst{1,2} 
          \and
A. Eckart\inst{1,2}
         \and
M. Valencia-S.$^{1}$
          \and
G. Witzel\inst{3}
          \and
M. Zamaninasab\inst{2}
          \and
N. Sabha\inst{1,2}
          \and
M. Garc\'{i}a-Mar\'{i}n$^{1}$
          \and
V. Karas$^{4}$
          \and
G. D. Karssen$^{1}$
          \and                    
A. Borkar$^{2,1}$
          \and
M. Dov\u{c}iak$^{4}$
          \and
D. Kunneriath$^{4}$          
          \and
M. Bursa$^{4}$
          \and
R. Buchholz$^{1}$
          \and
J. Moultaka$^{5}$ 
          \and
C. Straubmeier$^{1}$
}

%\offprints{B. Shahzamanian\inst{1,2} (shahzaman@ph1.uni-koeln.de)}

   \institute{ I.Physikalisches Institut, Universit\"at zu K\"oln,
              Z\"ulpicher Str.77, 50937 K\"oln, Germany\\
              \email{shahzaman@ph1.uni-koeln.de}
         \and
             Max-Planck-Institut f\"ur Radioastronomie, 
             Auf dem H\"ugel 69, 53121 Bonn, Germany
         \and
            Department of Physics and Astronomy,
            University of California, Los Angeles, CA 90095, USA
	\and
          Astronomical Institute, Academy of Sciences, 
          Bo\u{c}n\'{i} II 1401, CZ-14131 Prague, Czech Republic     
	\and
Institut de Recherche en Astrophysique et Planetologie (IRAP)
Université de Toulouse, CNRS
Observatoire Midi-Pyrénées (OMP)
14, avenue Edouard Belin
F-31400 Toulouse, France
             }

\date{Received:?? / Accepted: ??  }

%0000000000000000000000000000000000000000000000000000000000000000000000000000000000000000000000000000000000000000

\abstract {
We present a statistical analysis of polarized near-infrared (NIR) light from Sgr A*, the radio source 
associated with the supermassive black hole at the center of the Milky Way. 
The observations have been carried out using the adaptive optics
instrument NACO at the VLT UT4 in the infrared $K_\mathrm{s}$-band from 2004 to 2012. 
Several polarized flux excursions were observed during these years.
Linear polarization at 2.2 $\mu m$, its statistics and time variation, can be used 
constrain the physical conditions of the accretion process onto this supermassive black hole.
With an exponent of about 4 for the number density histogram of fluxes above 5~mJy, the 
distribution of polarized flux density is closely linked to the single state power-law distribution of the total 
$K_\mathrm{s}$-band flux densities reported earlier.
We find typical polarization degrees of the order of 20\%$\pm$10\% and 
a preferred polarization angle of 13$^o$$\pm$15$^o$. Simulations show the uncertainties under a total flux density of $\sim 2\,{\rm mJy}$ are probably dominated by observational effects. At higher flux densities there are intrinsic variations of polarization degree and angle within rather well constrained ranges.
Since the emission is most likely due to optically 
thin synchrotron radiation, this preferred polarization angle we find is very likely coupled to the intrinsic orientation of the
Sgr A* system i.e. a disk or jet/wind scenario associated with the super massive black hole.
If they are indeed linked to structural features of the 
source the data imply a rather stable geometry and accretion process for the Sgr A* system.
}


\maketitle

%0000000000000000000000000000000000000000000000000000000000000000000000000000000000000000000000000000000000000000
\section{Introduction}
\label{section:Introduction}

Sagittarius A* (Sgr~A*) is a bright and compact radio source associated with a supermassive black hole 
(M$_{BH}$ $\sim$ $4 \times$ $10^{6}$ \textit{$M_{\odot}$}) located at the center of our galaxy 
\citep{Eckart&Genzel1996, Eckart&Genzel1997, Eckart2002, Schoedel2002, Eisenhauer2003, Ghez1998, 
Ghez2000, Ghez2005b, Ghez2008, Gillessen2009stars} which is the best example of a low-luminosity galactic nucleus accessible to observations. 
Sgr~A* shows time variability in high spatial resolution observations in the Near-infrared (NIR) 
and X-ray regime compared to a lower degree of variability in the radio to sub-mm domain. 
The NIR counterpart to Sgr~A* shows short bursts of increased radiation which can occur four 
to six times per day and last about 100 minutes.

Analyzing the polarization of the electromagnetic radiation can help us to reveal 
the nature of emission processes detected from Sgr~A*. Therefore, this source has been observed since 2004 
in the polarimetric imaging mode with NACO using its Wollaston prism  
\citep{Eckart2006a, Meyer2006a, Meyer2006b, Eckart2008a, Zamaninasab2010, Witzel2012}.
Multi-wavelength observations have been conducted by different research groups 
to study the spectral energy distribution (SED) and the variable emission process 
of Sgr~A* from the radio to the X-ray domain
\citep{Baganoff2001, Porquet2003, Genzel2003flares, Eckart2004, Eckart2006a, Eckart2006b, Eckart2006c, Eckart2008a, Eckart2008b, Eckart2008c, Meyer2006a, Meyer2006b, Meyer2007, yusefzadeh2006a, yusefzadeh2006b, yusefzadeh2007, yusefzadeh2008, Dodds-Eden2009, Sabha2010, Eckart2012}.  
Observations at $1.6~\mu m$ and $1.7~\mu m$ wavelengths using the Hubble Space Telescope (HST) 
\citep{yusefzadeh2009} indicate that the activity of Sgr A* is above the noise level 
more than 40\% of the time. 
The highly polarized NIR flux density excursions have usually X-ray counterparts 
which suggests a synchrotron-self-Compton (SSC) or inverse Compton emission as 
the responsible radiation mechanism \citep{Eckart2004, Eckart2006a, Eckart2006c, Yuan2004, Liu2006, Eckart2012}. 
Several relativistic models that successfully describe the observations 
presume the variability to be related to the emission from single or multiple spots 
close to the last stable orbit of the black hole \citep{Meyer2006a, Meyer2006b, Meyer2007, Zamani2008}.

Based on relativistic models, \cite{Zamaninasab2010} predict and 
explore a correlation between the modulations of the observed flux 
density light curves and changes in polarimetric data. This information 
should in principle allow us to constrain the spin of the black hole 
(assuming that the gravitational field is indeed described by the Kerr 
metric). However, the question of whether timescales comparable to the 
orbital period near the inner edge of the accretion flow (in 
particular, near the radius of the innermost stable orbit) play a role 
in the variability, was (and still remains) impossible to decide on the 
basis of available data. Although the geometrical effects of strong 
gravitational fields act on photons independently of their energy, the 
intrinsic emissivity of accretion discs and the influence of magnetic 
field are energy dependent. Therefore, the variability amplitudes of 
both the polarization degree and the polarization angle are expected to be 
energy dependent as well. These dependencies suffer from some degeneracy. These degeneracies and the interdependencies of the observables require both time-resolved observations \citep[e.g.][]{Zamaninasab2010} and a statistical analysis as presented here.

\cite{Witzel2012} show that the time variable NIR emission from Sgr A* can be understood as
a consequence of a single continuous power-law process with a break time scale between 
500 and 700 minutes. This continuous process shows extreme flux density excursions 
that typically last for about 100 minutes. In the following we will refer to these excursions as 
flares and to the fact that they occur as flaring activity. 
On the base of multi-wavelength observations in 2009, \cite{Eckart2012} show that the 
flaring activity can be modeled as a signal from a synchrotron/synchrotron-self-Compton component. 

Several authors have studied the statistical properties of flaring activity of Sgr A* instead of concentrating on investigating the individual flares. \cite{do2009flares} do not find quasiperiodic oscillations (QSOs), which can be related to the orbital time of the matter in the inner part of an accretion disk, against the pure red noise while probing 7 total intensity NIR light curves taken with Keck telescope. The authors also conclude that Sgr A* is continuously variable. The red power-law distribution of the variable emission at NIR can be described by fluctuations in the accretion disk \citep{chan2009}. However, the correlation between flux density modulations and changes in the degree of polarization, the delayed sub-mm emission, and the spectral energy distribution show that the emission is coming from a compact flaring region with a size close to the Schwarzschild radius. This compact region can be a jet with blobs of ejected material \citep{markoff2001} or a radiating hot spot(s) falling into the black hole \citep[see e.g.][]{ Genzel2003flares, Dovciak2004, Dovciak2008, Eckart2006b, gillessen2006, Meyer2006a, Zamaninasab2010}. 

The statistics of NIR $K_\mathrm{s}$-band total intensity variability of Sgr~A* observed from 
2004 to 2009 with the VLT, has been investigated by \cite{Dodds-eden2011}. 
The authors interpret the time variability of Sgr~A* as a two state process, a quiescent 
state for low fluxes (below 5~mJy) which has a log-normal distribution and a flaring state 
for high fluxes (above 5~mJy) that has a power-law distribution. 
From their analysis 
they claim that the physical processes responsible for the low and high flux densities 
from Sgr~A* are different. 
However, their conclusions for the low flux densities are based on data at or below the detection limit, and therefore is biased by the measurements 
uncertainties and source crowding.
On the other hand, \cite{Witzel2012} show, for their slightly larger dataset taken 
between 2003 and early 2010, that the variability of Sgr A* is well described by a 
single power-law distribution, and conclude that there is no evidence 
for a second intrinsic state based on the distribution of flux densities.  
\cite{meyer2014} come to a similar result modeling the data by a rigorous 
two state regime switching time series that additionally included the information 
on the timing properties of Sgr A*. These results unambiguously show that in the 
range of reliably measurable fluxes the variability process can be described as a continuous, single 
state red-noise process with a characteristic timescale of several hours, without any characteristic flux density.

The analysis of the intrinsic polarization degree and polarization angle of the emission 
from Sgr A* and their changes during the 
flaring activity is another important aspect of the time variability. 
In this paper we analyze the most comprehensive sample of NIR polarimetric
light curves of Sgr A*.
In Sect. 2 we provide details about the observations and data reduction. 
In Sect. 3, we present the statistical analysis of polarized flux densities, 
a comparison with total flux densities and their distribution as provided by 
\cite{Witzel2012}.
In Sect. 4 we summarize the results and discuss their implications.

%xxxxxxxxxxxxxxxx
%#############################################################

\section{Observations and data reduction}
\label{Observations}
All observations for this paper have been carried out with the adaptive optics (AO) module
NAOS and NIR camera CONICA \citep[together NACO;][]{Lenzen2003, Rousset2003}
at the UT4 (Yepun) at the Very Large Telescope (VLT) 
of the European Southern Observatory (ESO) on Paranal, Chile.
We collected all $K_\mathrm{s}$-band (2.2~$\mu$m) observational data of the central cluster 
of the Galactic Center (GC) in 13 mas pixel scale polarimetry with the camera 
S13 from mid-2004 to mid-2012 that have flare events. 
In all the selected observations the infrared wavefront sensor of NAOS was used 
for locking the AO loop on the NIR supergiant IRS7 with $K_\mathrm{s}$ $\sim 6.5-7.0$~mag, located $\sim 5.5''$ north from Sgr~A*.
NACO is equipped with a Wollaston prism combined with a 
half-wave retarder plate that provide simultaneous measurements of two orthogonal 
directions of the electric field vector and a rapid change between different angles 
of the electric field vector. 

In the following we present a short summary of the reduction steps.
For 2004 to 2009  we used the reduced datasets as presented in \cite{Witzel2012}.
The 27 May 2011 and 17 May 2012 data have not been published before and
we applied an observational strategy and data reduction steps similar to \cite{Witzel2012} 
to these datasets. 
The AO correction for the 27 May 2011 and 17 May 2012 nights, was most of the time stable and in 
good seeing condition. We had Sgr~A* and a sufficient number of flux secondary density calibrators 
in the innermost arcsecond. 
The observing dates, integration times, sampling rate and mean flux densities of the data sets 
used for our analysis are presented in Table \ref{table:nonlin1}.
%======================================================================
\begin{table*}[]% * makes wide table in 2 columns
\caption{Observations Log.}% title of Table
\centering% used for centering table
\begin{tabular}{c c c c c c c c}% centered columns (4 columns)
\hline
\hline %inserts double horizontal lines
Date & Start & Stop & Length & Number of& Maximum flux& Average& Integration\\[0.5ex]% inserts table
%heading
&&&&frames&density&sampling rate&time\\
&(UT time)&(UT time)&(min)&&(mJy)&(min)&(sec)\\
\hline% inserts single horizontal line
\\
13.06.2004 & 07:54:22.95 & 09:15:08.79 & 80.76  & 70  & 3.17 & 1.17 & 20\\% inserting body of the table% max(f) of data set(dat file) * 0.75 (because of extinction) = max. flux density
30.07.2005 & 02:07:36.13 & 06:21:40:41 & 254.07 & 187 & 8.94 & 1.36 &30\\
01.06.2006 & 06:39:49.16 & 10:44:27.63 & 378.41 & 244 & 14.5 & 1.55 &30 \\
15.05.2007 & 05:29:55.42 & 08:31:48.45 & 181.88 & 116 & 16.7 & 1.58 & 40\\
17.05.2007& 04:24:14.84 & 09:34:40.15 & 292.42 & 192 & 9.78  & 1.53 & 40\\
25.05.2008 & 06:05:20.32 & 10:35:38.65 & 270.31 & 250 & 10.25& 1.085& 40\\
27.05.2008 & 04:52:04.92 & 08:29:38.07 & 217.55 & 184 & 4.32 & 1.18 & 40\\
30.05.2008 & 08:24:33.51 & 09:45:25.69 & 80.87 & 80 & 12.39 & 1.023 & 40\\
01.06.2008 & 06:04:51.56 & 10:10:26.78 & 245.59 & 240 & 7.08 & 1.027 & 40\\
03.06.2008 & 08:37:23.56 & 09:58:58.85 & 81.59 & 80 & 10.02& 1.032 & 40\\
18.05.2009 & 04:37:55.08 & 10:19:54.10 & 341.98 & 286 & 12.53 & 1.19 & 40\\
27.05.2011 & 04:49:39.82 & 10:27:25.65 & 337.77 & 334 & 7.55 & 1.2 & 45\\
17.05.2012 & 04:49:20.72 & 09:52:57.08 & 303.62 & 256 & 6.64 & 1.2 & 45\\[1ex]% [1ex] adds vertical space

\hline%inserts single line
\end{tabular}
\label{table:nonlin1}
\end{table*}
%======================================================================

All the exposures were sky subtracted, flat fielded and bad pixel corrected. 
We used lamp flat fields, instead of sky flat fields to avoid polarimetric effects produced 
by the sky. Since the exposures were dithered, all the polarization channels 
(0\degr, 45\degr, 90\degr, 135\degr) of the individual data set were aligned with using a cross-correlation method with sub-pixel accuracy \citep{Devillard1999}. 
The Point Spread Functions (PSFs) were extracted from the images with the IDL 
routine Starfinder \citep{Diolaiti2000} using isolated stars close to Sgr~A*. 
We used the Lucy-Richardson algorithm to deconvolve the images. 

Image restoration was done by convolving the deconvolved images with a Gaussian beam 
of a FWHM of about 60~mas corresponding to the diffraction limit at $2.2 \mu m$.

%======================================================================
\begin{figure*}[!ht]
     \begin{center}

        \subfloat{%
            \includegraphics[width=0.45\textwidth]{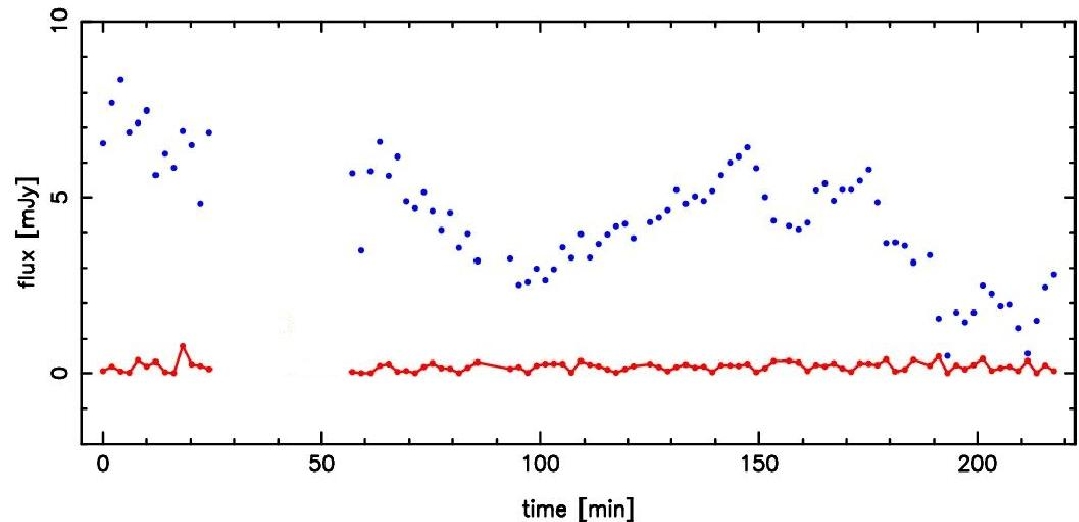}
        }
        \subfloat{%
           \includegraphics[width=0.45\textwidth]{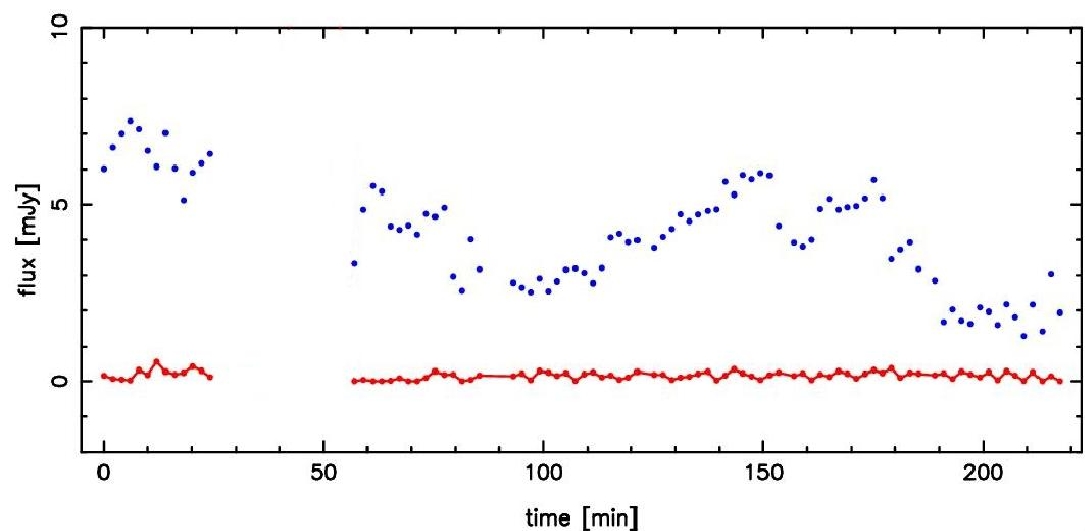}
        }\\ %  ------- End of the first row ----------------------%
        \subfloat{%
            \includegraphics[width=0.45\textwidth]{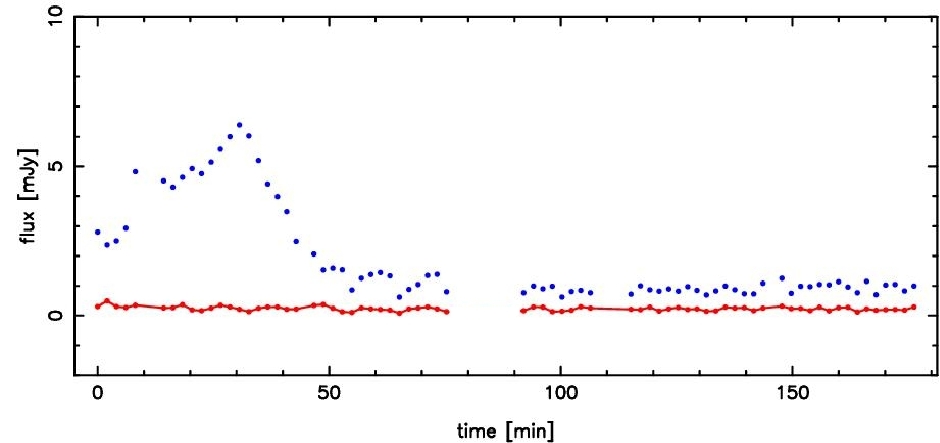}
        }
        \subfloat{%
           \includegraphics[width=0.45\textwidth]{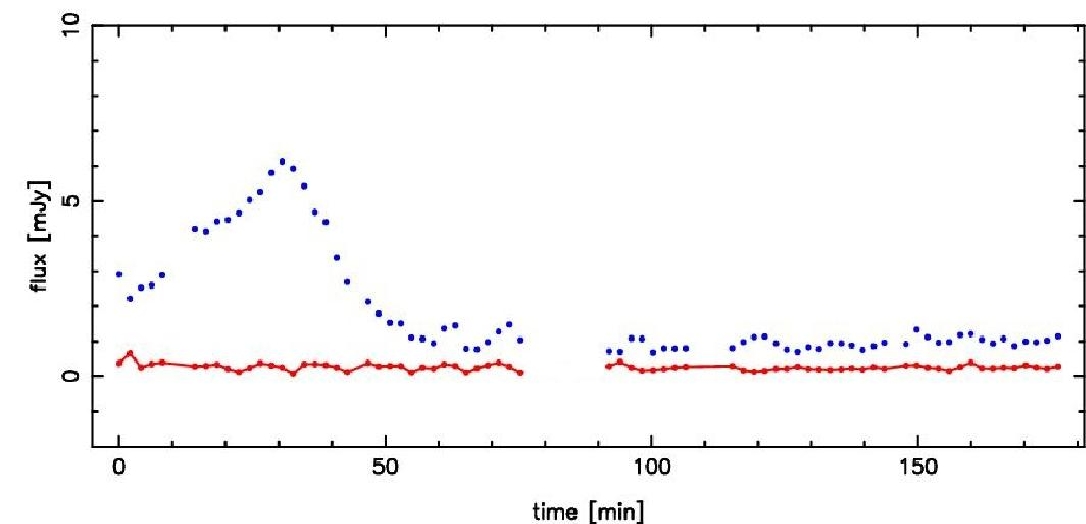}
        }\\
        
 \end{center}
    \caption{NIR Ks-band (2.2 $\mu$m) light curves of Sgr A* observed in polarimetry mode 
on 27 May 2011 (top) and 17 May 2012 (bottom) produced by combining pairs of orthogonal polarization channels; 
left: $0^\circ$,90$^\circ$ and right: 45$^\circ$,135$^\circ$. 
The blue dots show Sgr A* flux density measured in mJy; 
While the red connected dots show the background flux densities.
     }
   \label{fig:lightcurve}
\end{figure*}
%======================================================================

%%%%%%%%%%%%%%%%%%%%

\subsection{Flux density calibration}
\label{section:calibration}

We measured the flux densities of Sgr~A* and other compact sources in the 
field by aperture photometry using circular apertures of 40~mas radius. 
The flux density calibration was carried out using the known $K_\mathrm{s}$-band 
flux densities of 13 S-stars \citep {schoedel2010}. 
Furthermore, 6 comparison stars and 8 background apertures placed at positions 
where no individual sources are detected.
For more details about the positions of the apertures and the list of calibrators see 
Figure~2 of this paper and Table~1 in \citet{Witzel2012}. To get the total flux densities we added up the photon counts in each aperture 
and then added the resultant values of two orthogonal polarimetry channels. 
These values were corrected for the background contribution. 
We calculated the flux densities of the calibrators close to Sgr~A* and also at the 
position of Sgr~A* and then corrected them for extinction using 
$A_{K_\mathrm{s}} = 2.46$, derived for the inner arcsecond by \cite{schoedel2010}.
Applying aperture photometry on all frames, results in the light curves obtained for Sgr~A* in 
two orthogonal channels for 27 May 2011 and 17 May 2012 data, as shown in Fig.\ref{fig:lightcurve}.

The gaps in the measurements are due to AO reconfiguration or sky measurements. 
For 27 May 2012 data the flux density of Sgr A* varied between about 5 and 10~mJy over the entire observing run.
For 17 May 2012 data, over the first 50 minutes, the flux density of Sgr A* increased to about 
7~mJy and then decreased again.

\begin{figure}

    \includegraphics[width=\columnwidth]{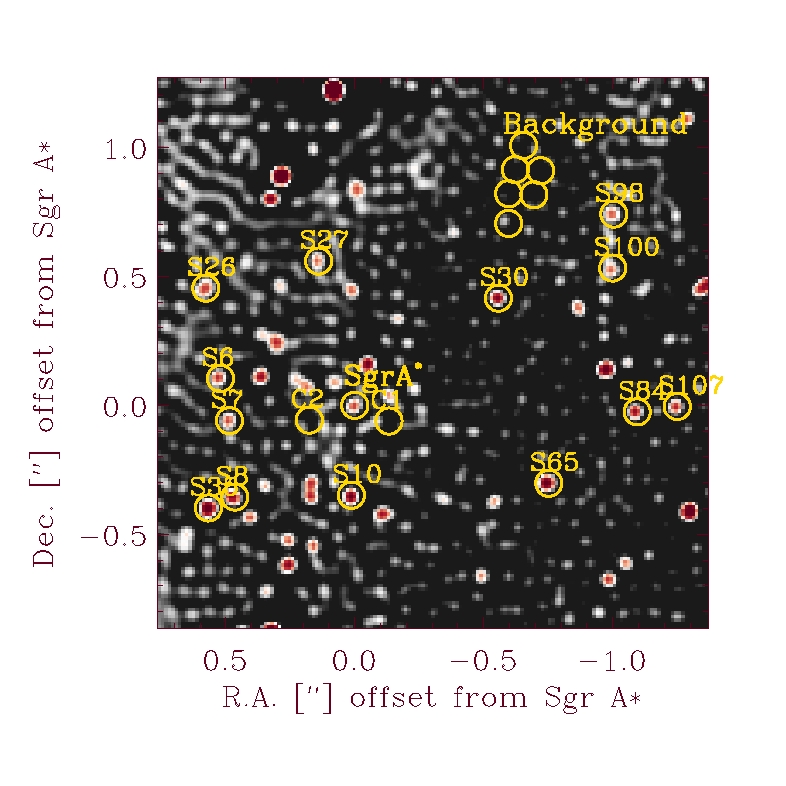}

 \caption{$K_\mathrm{s}$-band deconvolved image of the Galactic Center on 17 May 2012
showing the positions of Sgr~A*, calibration stars and comparison apertures for background 
estimates marked by yellow circles.}
 \label{fig:apertures}
\end{figure}

Fig.\ref{fig:apertures} shows a
$K_\mathrm{s}$-band deconvolved image of the Galactic Center on 17 May 2012. 
The image is taken with the ordinary beam of the Wollaston prism. 
The positions of Sgr~A*, calibration stars and comparison apertures for background 
estimates are shown.
For comparison see also Fig.1 from 30 September 2004 by \cite{Witzel2012}.
Source identification has been done using the nomenclature by \cite{Gillessen2009stars}.

\begin{figure}
    
        \subfloat{%
            \includegraphics[width=\columnwidth]{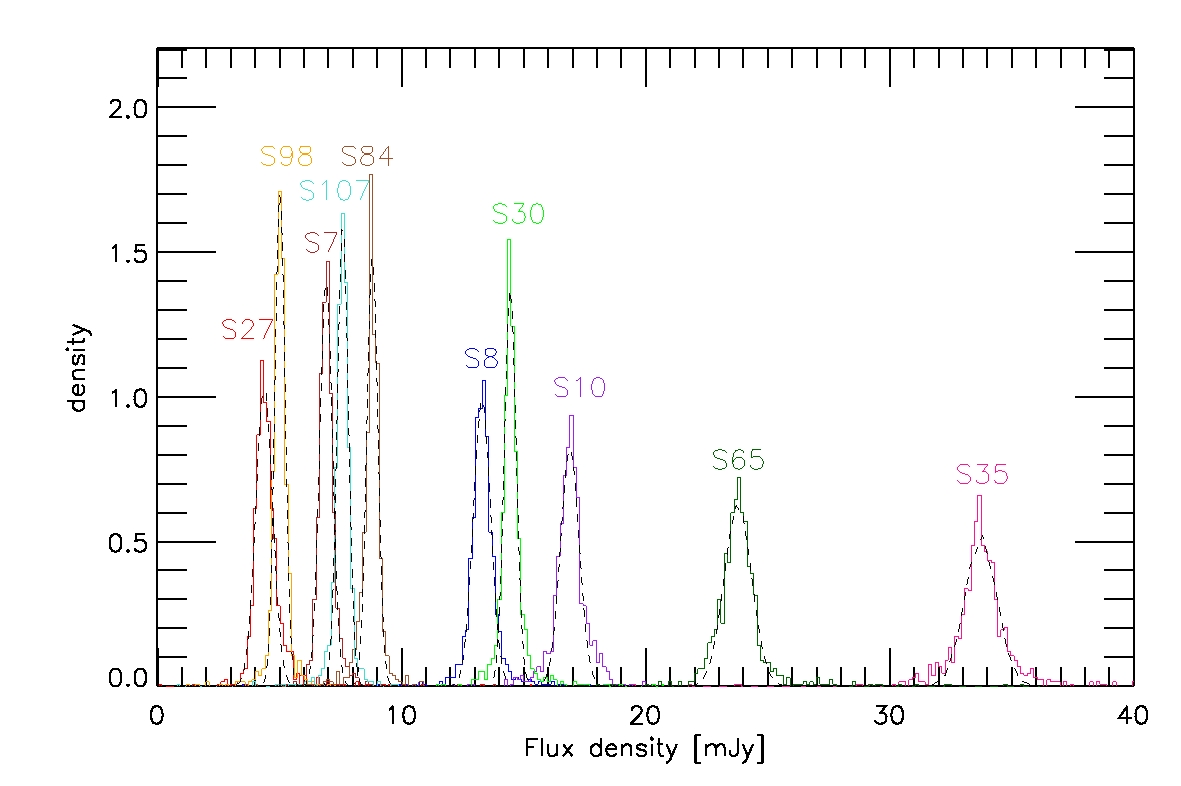}
        }
        \label{fig:hist_calibrators}
        \subfloat{%
            \includegraphics[width=\columnwidth]{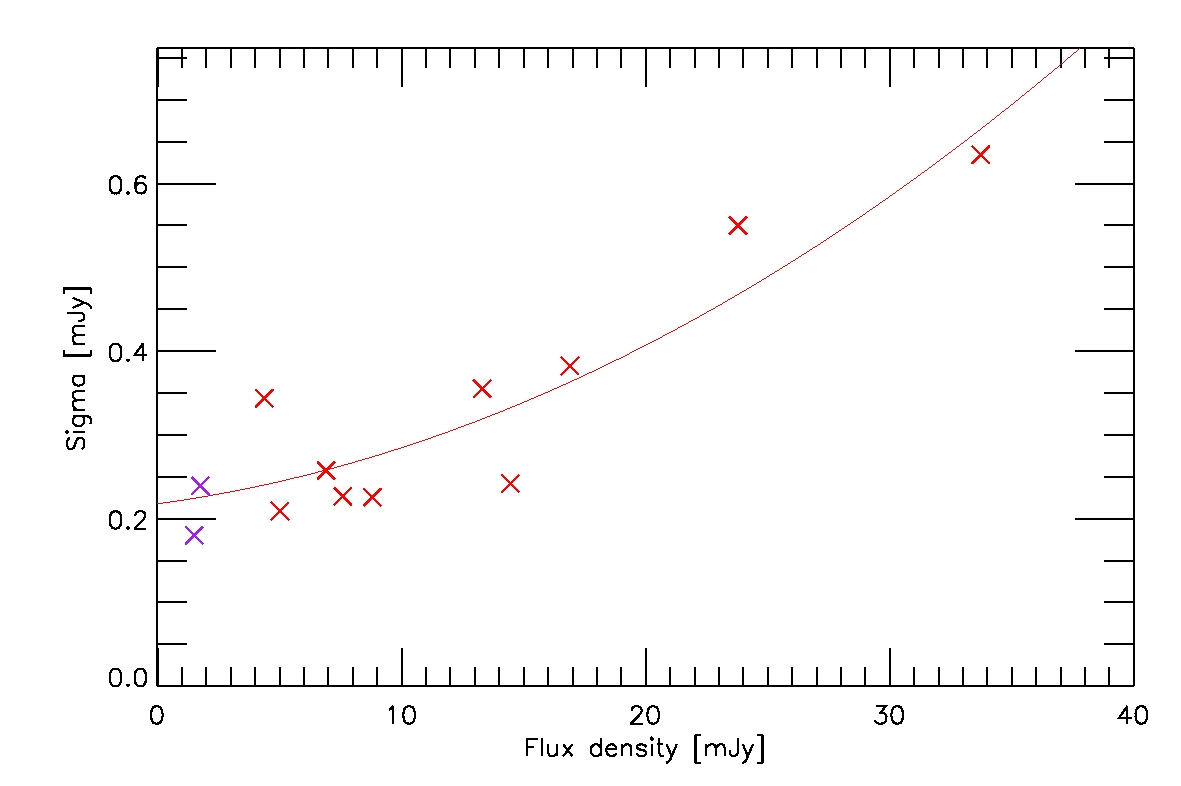}
        \label{fig:hist_calibrators_2} 
        }
        
  \caption{Top: Normalized flux density distributions of 10 flux calibrators of Sgr~A*. 
The dashed lines are Gaussian fits to the distributions. 
Bottom: Standard deviation of flux densities of calibration stars versus flux densities of them. 
The red line is the polynomial fit to the measured $\sigma$ values of the calibrators shown in the 
upper panel (red crosses). The two purple $x$ symbols present the measured error values 
(obtained by \citealt{Witzel2012}) at the position of the comparison apertures for the background 
emission close to the position of Sgr~A* (see Fig.\ref{fig:apertures}). 
     }
\label{fig:hist_calibrators_total}
\end{figure}

The flux density of Sgr A* was calculated from the flux densities measured in the four different
polarization channels and corrected for possible background contributions.
The top panel of Fig.\ref{fig:hist_calibrators_total} shows the measured 
flux density distributions of different 
calibration stars close to Sgr~A* fitted by Gaussian functions. The scatter of the flux densities 
around the mean value originates from the observational uncertainties and can be estimated from these 
fits. 
The standard deviations $\sigma$ of the Gaussians fitted to the distributions are presented 
as a function of the mean flux density in the bottom panel of Fig.\ref{fig:hist_calibrators_total}.
The function that best describes the dependency of $\sigma$ values with the flux densities up to 
33~mJy is a second degree polynomial that tends to be flatter at small flux density values. 
The measured flux densities and scatter of the two background apertures (C1 and C2 in Fig.\ref{fig:apertures})
have been added to the plot and included the fit.
The uncertainties up to 10~mJy total flux are $\sim$0.25~mJy and are mostly introduced by 
a combination of a variation in the AO performance, imperfectly subtracted PSF 
seeing halos of surrounding, brighter stars and differential tilt jitter.
Within the uncertainties and for a total flux density value below 10~mJy this relation is in good agreement with 
that found by \cite{Witzel2012} using a larger sample of total flux density measurements 
shown in their Fig.7. 
However, our data shows 20\% to 25\% narrower flux density distributions at
higher flux values around 30~mJy. This is due to the fact that polarization data tend to be 
observationally biased towards higher Strehl values compared to AO imaging in 
standard 'observer mode', i.e. without selection of preferred atmospheric conditions.
The result shown in Fig.\ref{fig:hist_calibrators_total} also implies that statements on the 
source intrinsic total or polarized flux density of Sgr A* can only be made with certainty if the total flux density is
significantly larger than the limit of $\sim$0.25~mJy.

%##############################################################
%=======================================================================

%=======================================================================

%%%%%%%%%%%%%%%%%%%%
\subsection{Polarimetry}
\label{section:polarimetry}
Using non-normalized analog-to-digital converter (ADC) values from the detector (see details in \citealt{Witzel2011}) 
to obtain normalized stokes parameters we can derive the polarization degree and angle:\\

\noindent
\begin{equation}
F = f_{0} + f_{90} = f_{45} + f_{135}\\
\end{equation}
\begin{equation}
Q =\frac{f_{0}-f_{90}}{f_{0}+f_{90}}
\end{equation}
\begin{equation}
U =\frac{f_{45}-f_{135}}{f_{45}+f_{135}}
\end{equation}

\noindent
\begin{equation}\label{eq:deg}
p =\sqrt{Q^{2} + U^{2}}
\end{equation}
\begin{equation}
\phi =\frac{1}{2}~\arctan\left(\frac{U}{Q} \right) 
\end{equation}
\noindent
Where $f_{0}$, $f_{45}$, $f_{90}$ and $f_{135}$ are the four polarimetric channels flux densities with $f_{0}$, $f_{90}$ and $f_{45}$, $f_{135}$ 
being pairs of orthogonally polarized channels.
The variable $F$ represents the total flux density and $Q$ and $U$ are the Stokes parameters. 
No information on the circular polarization in Stokes $V$ is available with NACO, hence the circular polarization is assigned to zero \citep[see][]{Witzel2011} for a detailed discussion).
The quantity $p$ is the degree of polarization and $\phi$ is the polarization angle
which is measured from the North to the East and samples a range between 0$^\circ$ and 180$^\circ$. 
We compute the polarized flux density as the product of the degree of polarization and the 
total flux density.
Uncertainties for F, Q, U and the obtained p and $\phi$ were determined from the flux density 
uncertainties.
Since NACO is a Nasmyth focus camera-system, instrumental effects can influence our 
results and a careful calibration is needed. \cite{Witzel2011} used the Stokes/Mueller 
formalism to describe the instrumental polarization. Their analytical model applies 
Mueller matrices to the measured Stokes parameters to get the intrinsic Stokes parameters. 
We used their model to diminish the systematical uncertainties of polarization angles 
and degrees caused by instrumental polarization to about $\sim1\%$ and $\sim5^\circ$ respectively. 
Foreground polarization has been obtained for the stars in the innermost arcsecond 
to Sgr A* \citep[see e.g.][]{buchholz2013}, but its value is of course not known for the exact line of sight towards Sgr A* itself. 
With the current instrumentation it is not possible to clearly disentangle line of sight effects from
the foreground polarization produced by the stars close to Sgr A*.

Since the Galactic center region is very crowded \citep{sabha2012}, confusion 
with stellar sources can happen in determining the flux density of Sgr~A*. 
To eliminate this confusion in order to be able to compare the polarized 
flux densities of different years without offsets, we subtract the minimum flux density of the four 
polarization channels from the flux densities of the corresponding channels 
for each data set and then obtain the polarization degrees and polarized flux densities of our data sample. 
Here we assume that the polarized flux density contributions of confusing stars 
are on the same level as the foreground polarization.
Therefore, subtracting the minimum of all four channels in each epoch is conservative. 
Moreover, subtraction of the faint stellar contribution was needed to 
compare the polarized flux density distribution with the total flux density 
distribution in \citet{Witzel2011}.  
The mentioned change in the flux densities did not significantly affect the value of polarization angle.

%#############################################################

\section{Data analysis}
\label{section:analysis}

In Fig.\ref{fig:allepochs}
we show the light curves that represent all flaring activities observed in
NIR $K_\mathrm{s}$-band by NACO in polarization mode. 
Typical strong flux density excursions last for about 100 minutes. 
The changes in total flux density, polarization degree and polarization angle are more 
apparent when bright flare states occur. Although the flare events are different in terms 
of the maximum flux density, it is interesting to investigate if there are preferred values or
ranges of values for the polarization degree and angle that are independent of the flare flux density.

The flux density uncertainties for our statistical analysis are determined via the relation shown
in Fig.\ref{fig:hist_calibrators_total} and we use 
the results of our analysis presented in the following for determining the uncertainties of the polarization degree and angle.

%=================================================================================
    
\begin{figure*}[h!tb]
     \begin{center}

        \subfloat{%
            \includegraphics[width=0.45\textwidth]{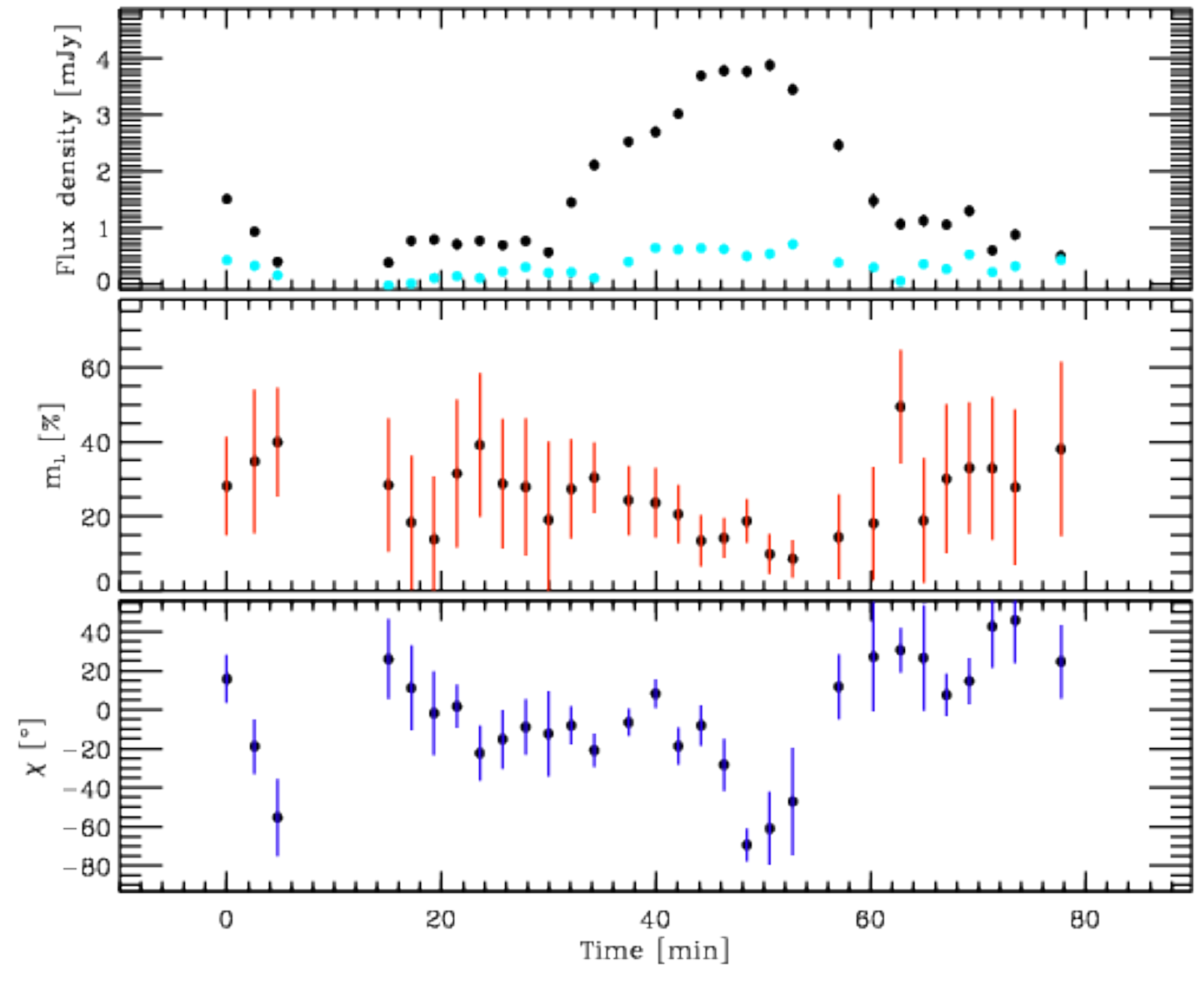}
        }
        \subfloat{%
           \includegraphics[width=0.45\textwidth]{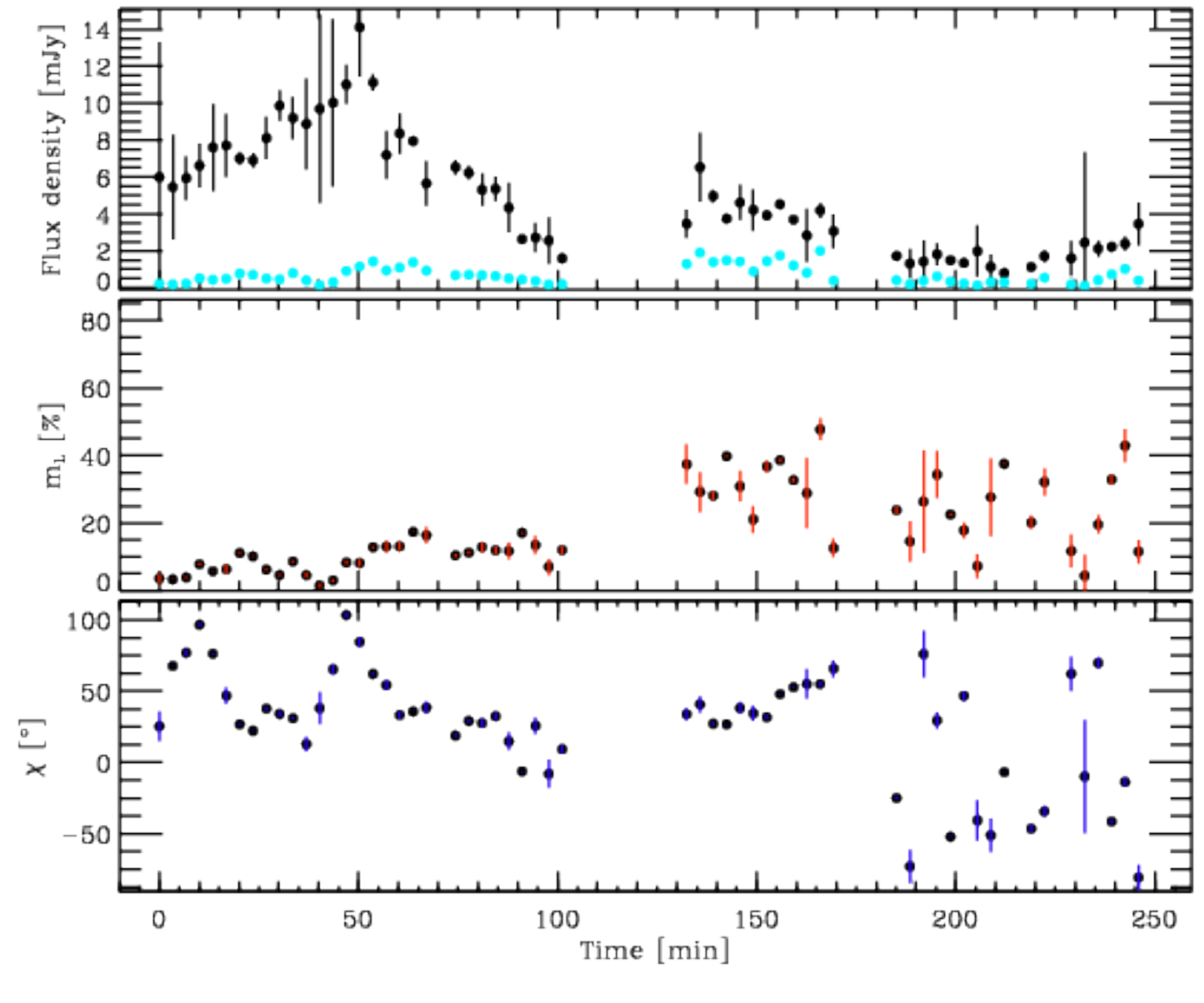}
        }\\ %  ------- End of the first row ----------------------%
        \subfloat{%
            \includegraphics[width=0.45\textwidth]{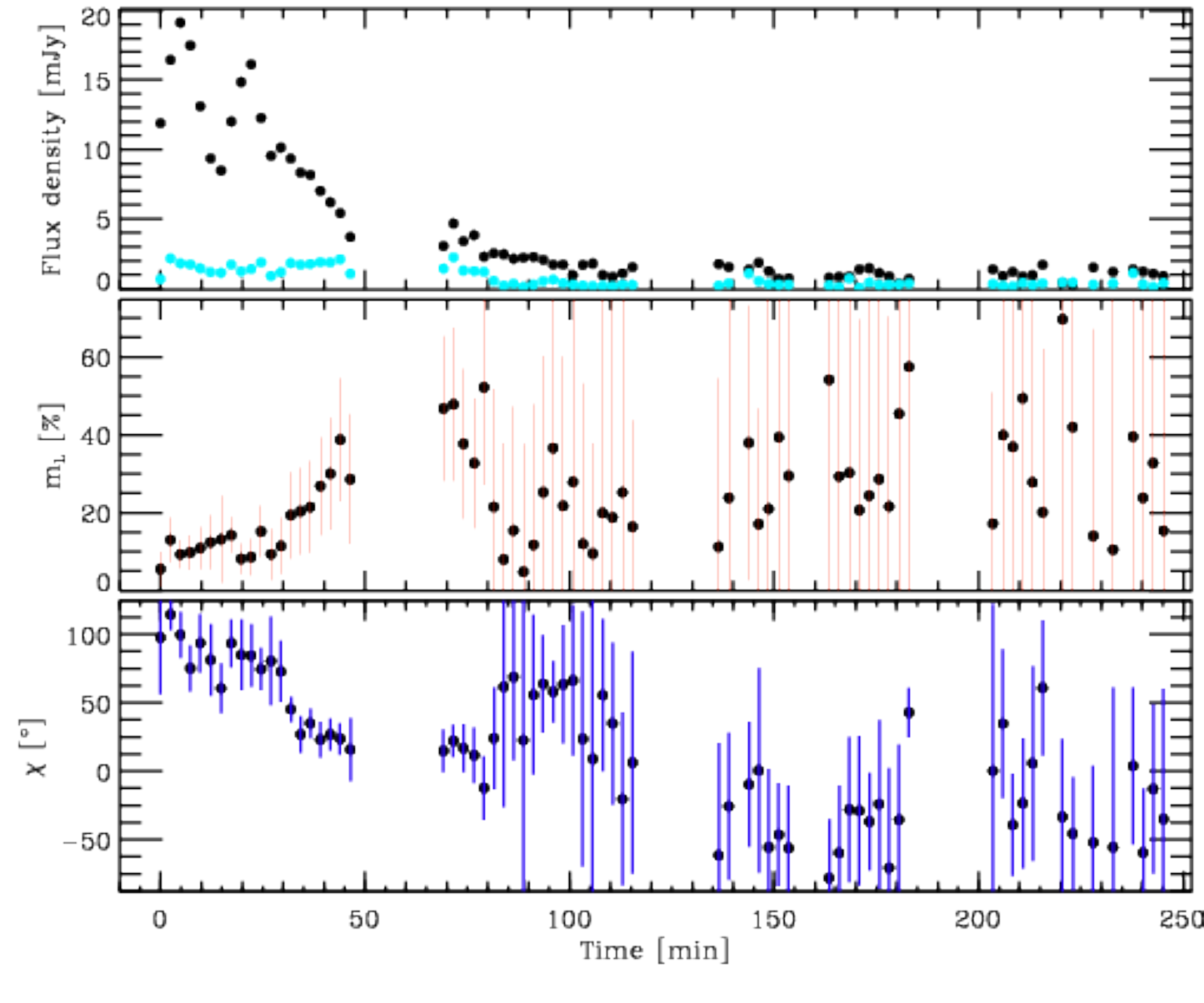}
        }
        \subfloat{%
            \includegraphics[width=0.45\textwidth]{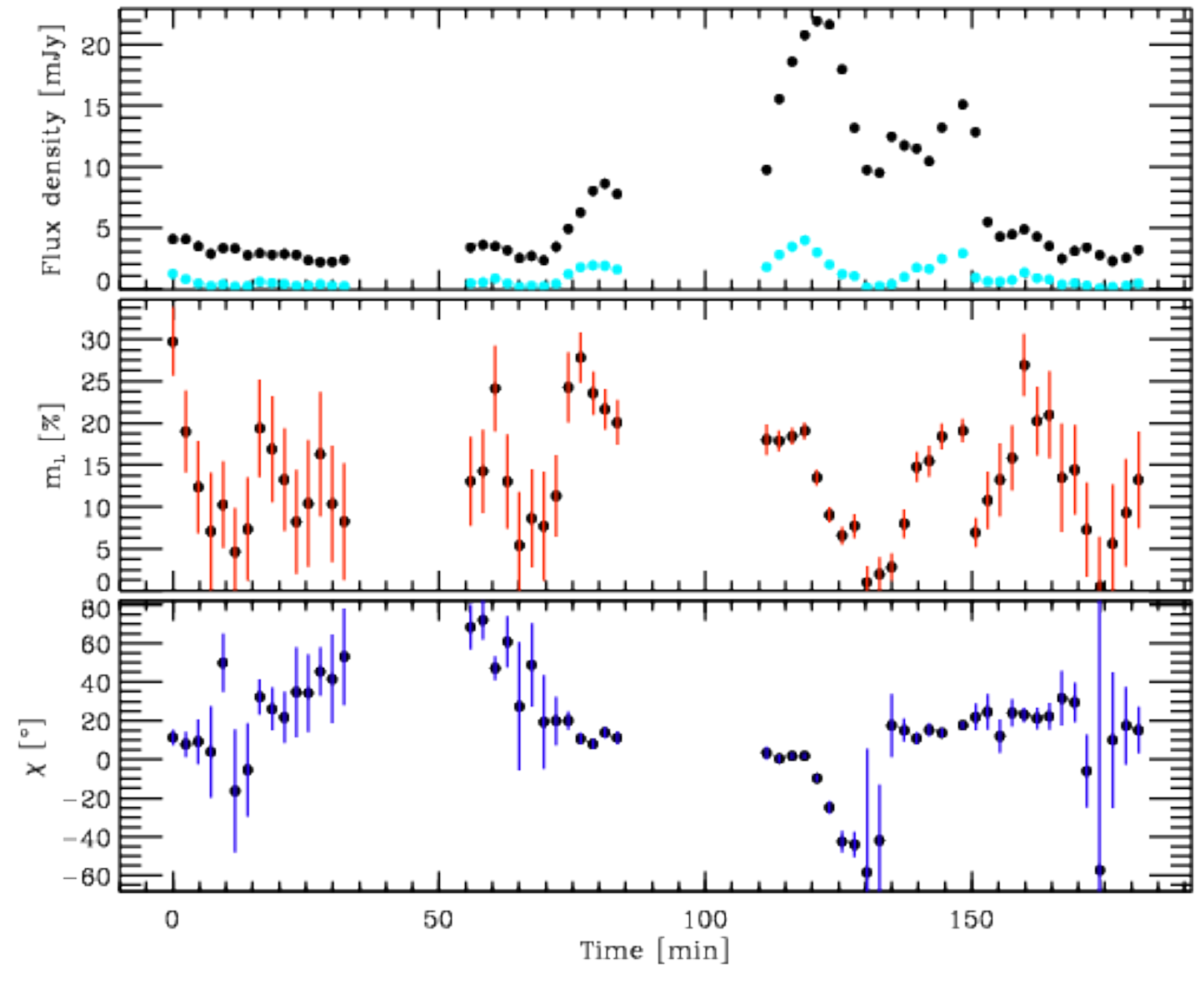}
        }\\
         \subfloat{%
            \includegraphics[width=0.45\textwidth]{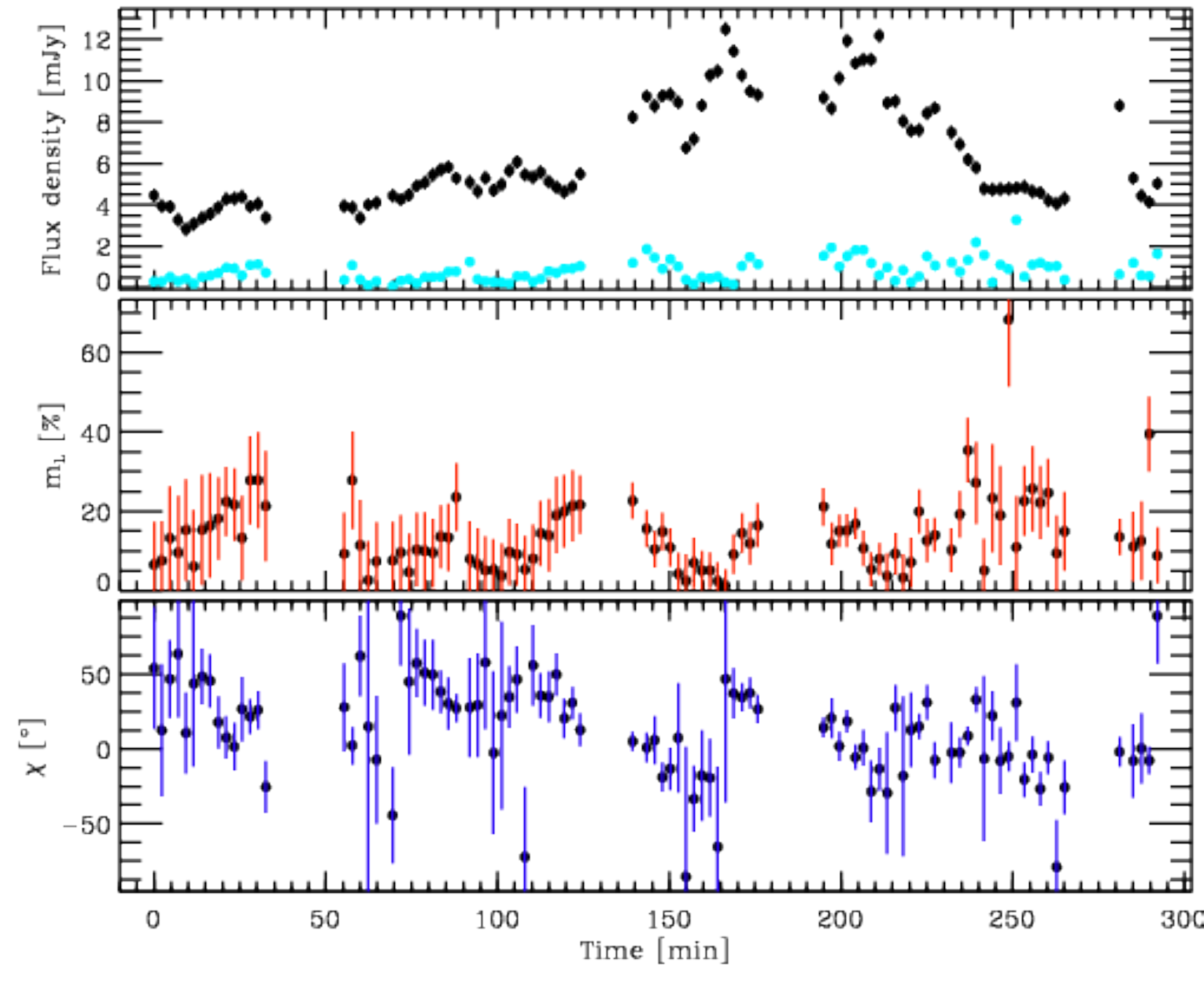}   
        }
        \subfloat{%
            \includegraphics[width=0.45\textwidth]{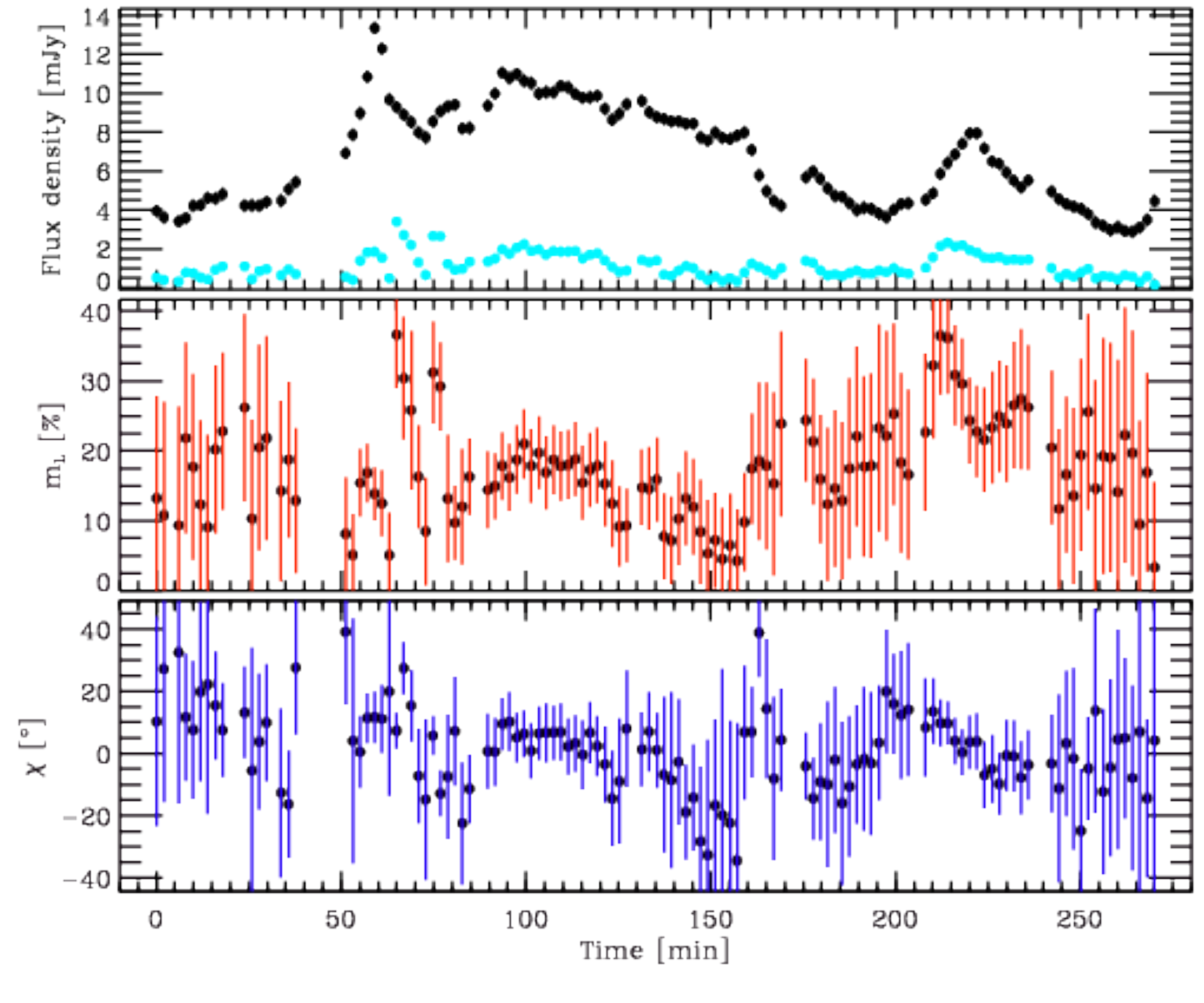}
        }\\

    \end{center}
    \caption{ Flux density excesses (flares) observed in NIR $K_\mathrm{s}$-band polarimetry mode of Sgr~A*. These events were observed on 2004 June 13, 2005 July 30, 2006 June 1, 2007 May 15, 2007 May 17, 2008 May 25, 2008 May 27, 2008 May 30, 2008 June 1, 2008 June 3, 2009 May 18, 2011 May 27, 2012 May 17 (The order of the images starts from top left to bottom right).
In each panel: Top: total flux density (black) and polarized flux density (cyan; polarization degree times total flux density) measured in mJy;
Middle: degree of linear polarization (red);
Bottom: polarization angle (blue).
     }
\label{fig:allepochs}
\end{figure*}

\begin{figure*}[h!tb]
     \begin{center}
     
        \ContinuedFloat

         \subfloat{%
            \includegraphics[width=0.45\textwidth]{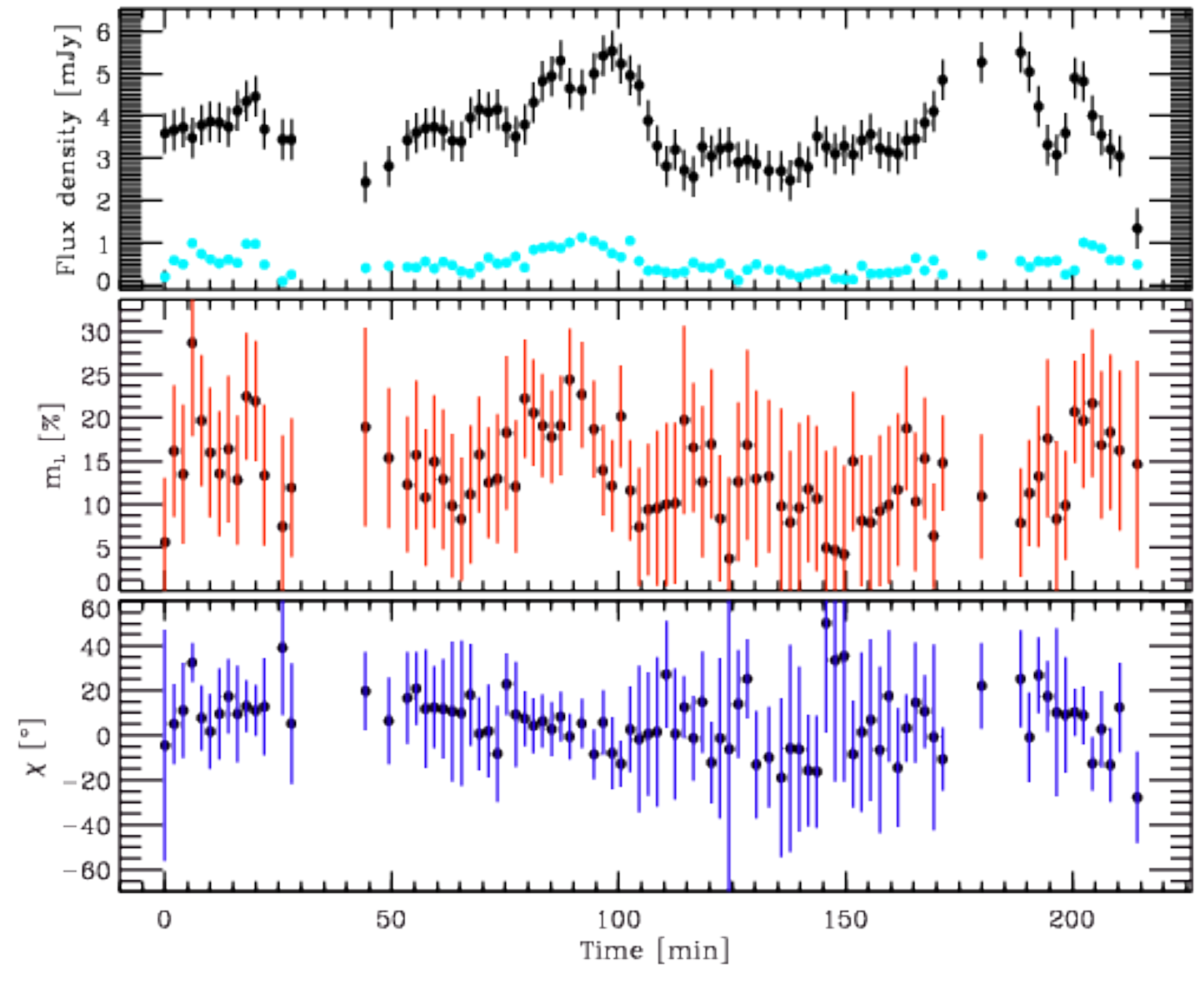}
        }
         \subfloat{%
            \includegraphics[width=0.45\textwidth]{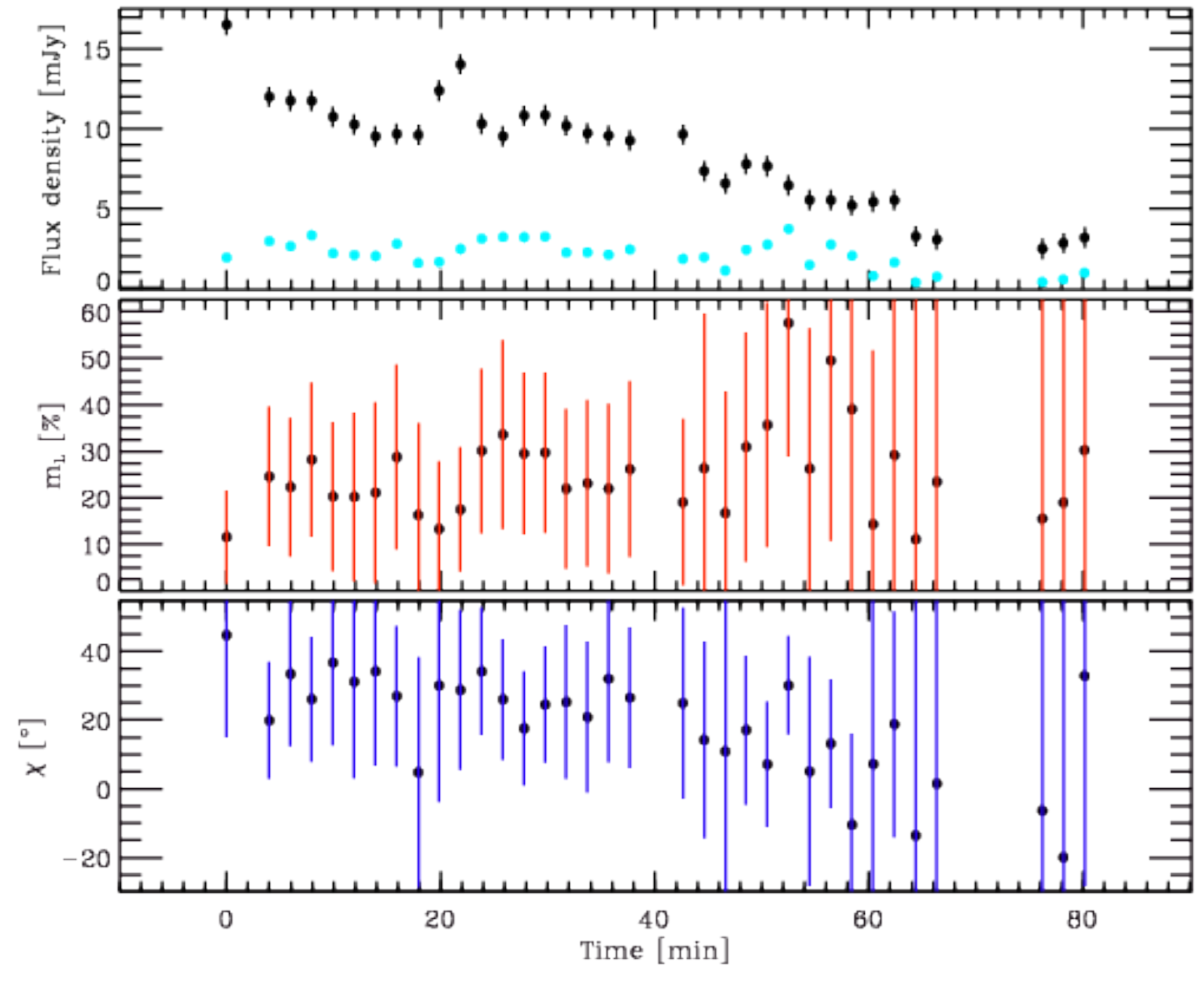}
        }\\
         \subfloat{%
            \includegraphics[width=0.45\textwidth]{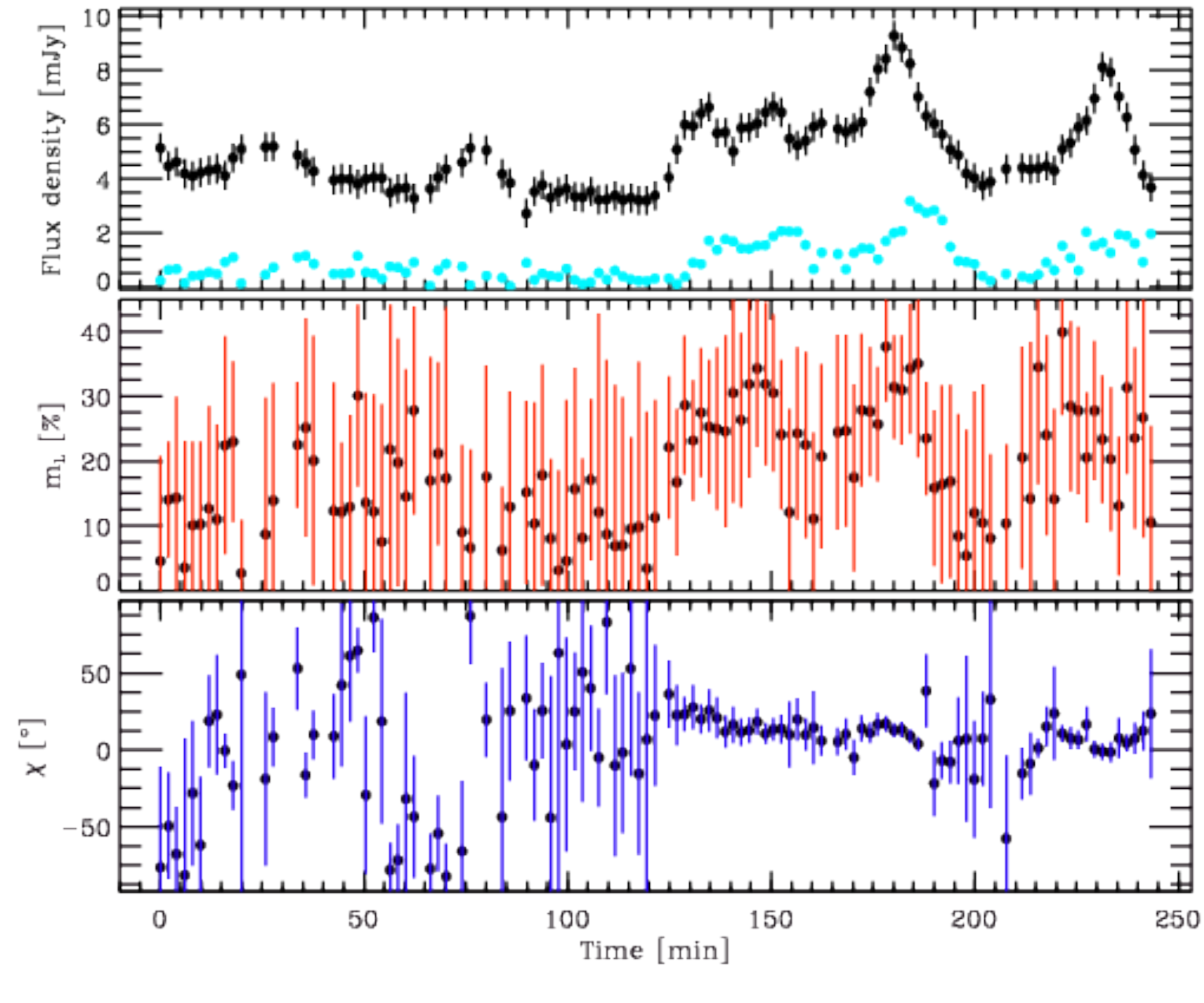}
        }       
         \subfloat{%
            \includegraphics[width=0.45\textwidth]{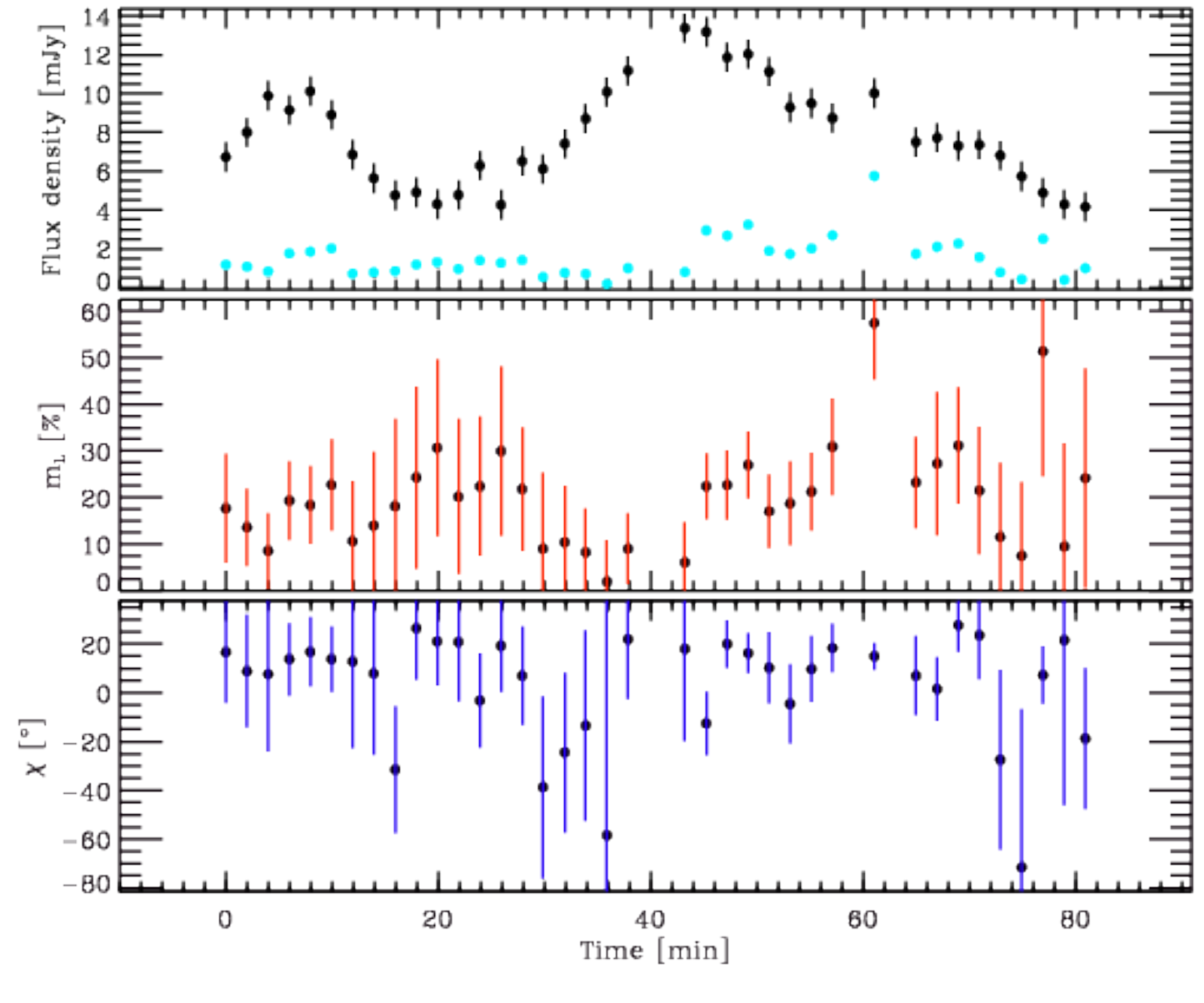}
        }\\
        \subfloat{%
            \includegraphics[width=0.45\textwidth]{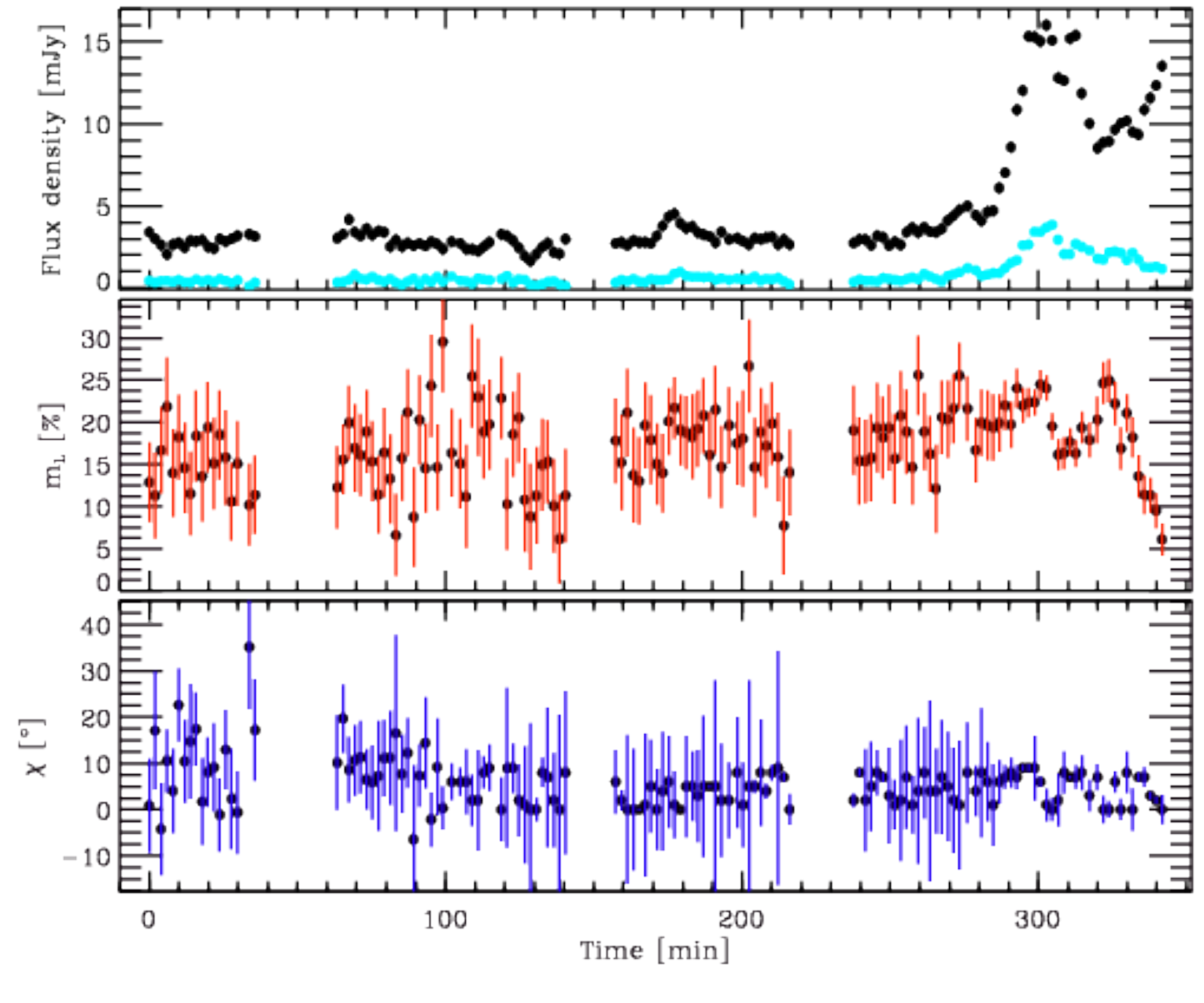}
        }       
         \subfloat{%
            \includegraphics[width=0.45\textwidth]{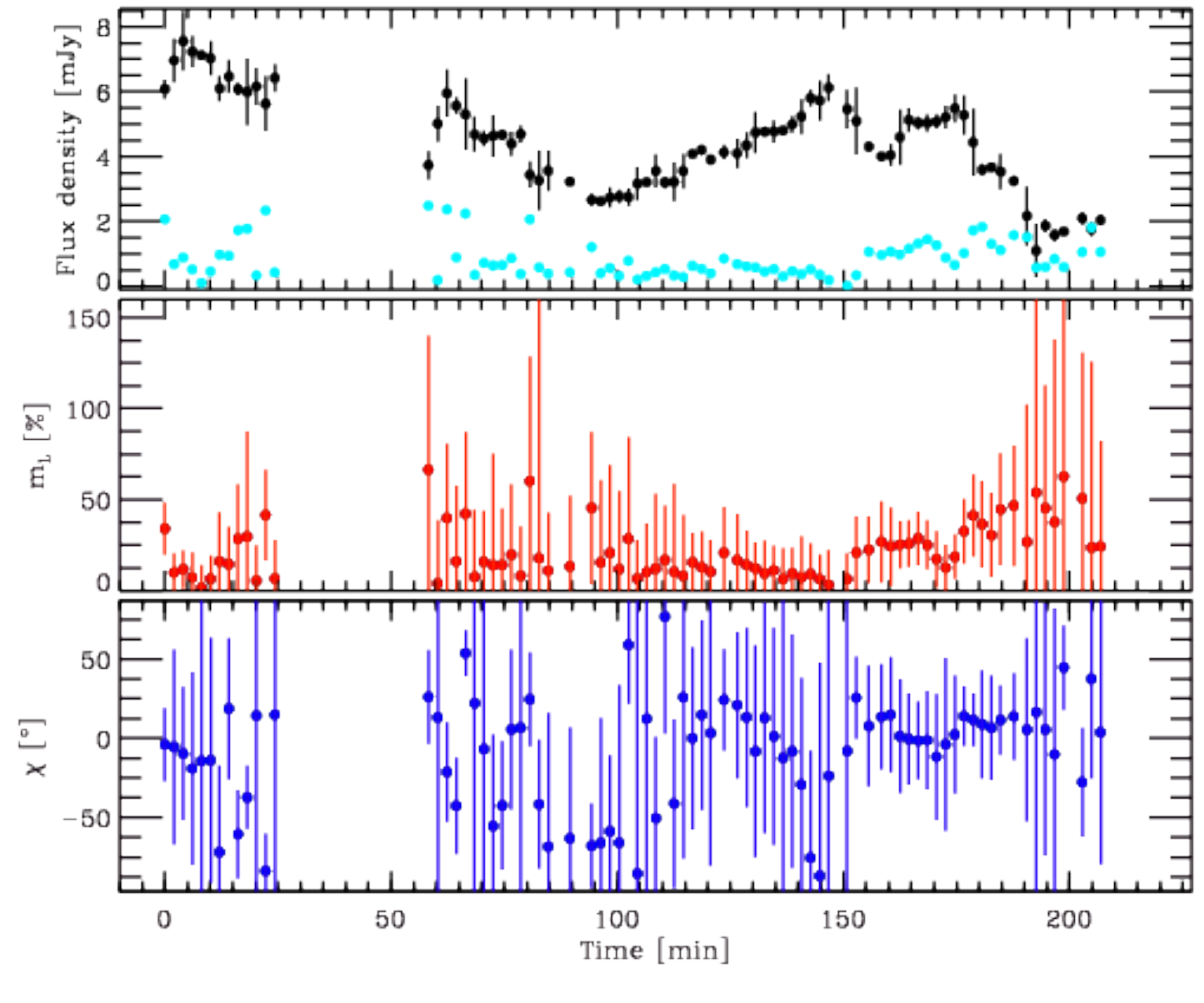}
        }\\

    \end{center}
    \caption{Continued.
     }
\label{fig:allepochs}
\end{figure*}

\begin{figure*}[h!tb]
     \begin{center}
     
        \ContinuedFloat

\subfloat{%
            \includegraphics[width=0.45\textwidth]{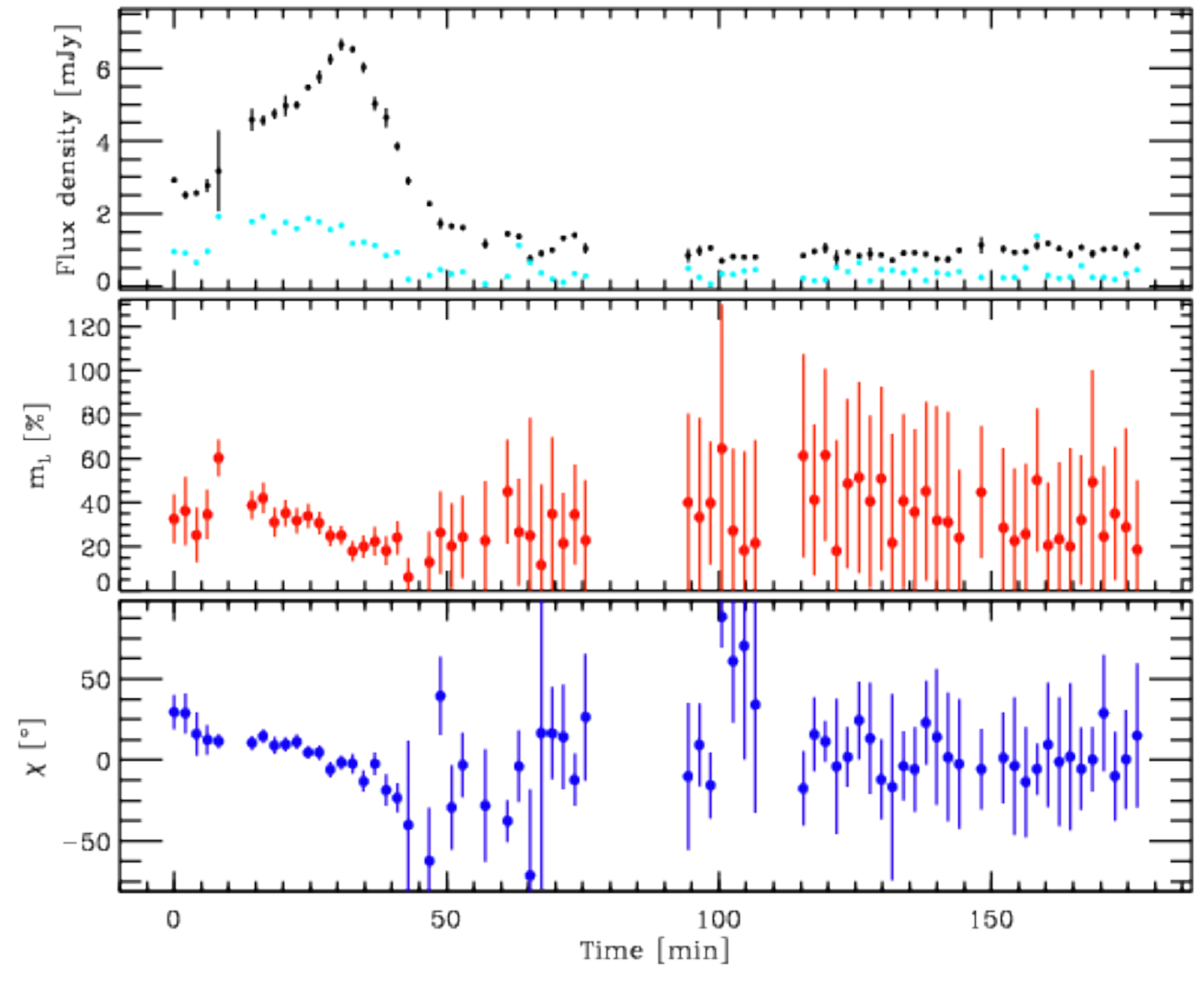}
        }  \\ 

    \end{center}
    \caption{Continued.
     }
\label{fig:allepochs}
\end{figure*}

%===================================================================================================

\subsection{Expected statistical behaviour of polarization measurements}
\label{section:Statistical}

In order to adequately present and interpret the data we need to know what the expected statistical behavior of polarization measurements is.
The polarization statistics has been studied by several authors, 
e.g. \cite{serkowski1958, serkowski1962, vinokur1965, simmons1985, naghizadeh1993, Clarke2010}.
From Eq. \ref{eq:deg} it is clear that the polarization degree $p$ is a positive quantity, that takes values between zero and one (or equivalently $0\% - 100\%$). 
The uncertainties in $Q$ and $U$, which in our case are the result of observational noise in the polarization channels, biases the value of $p$. This leads to an overestimation
of $p$ at low signal-to-noise ($S/N$) measurements. 
In general, polarimetric observations require a higher $S/N$ compared to photometric measurements. 
To first order the $S/N$ of the total intensity is related to that of the polarized intensity like $(S/N)_{polarized~intensity} \approx p\times(S/N)_{total~intensity}$, 
where the degree of polarization is usually smaller than one. 
As a result, weak polarization signal can be detected in case of having high $(S/N)_{total~intensity}$ \citep{trippe2014}. 

\begin{figure*}
  \centering
  \includegraphics[width = 6in]{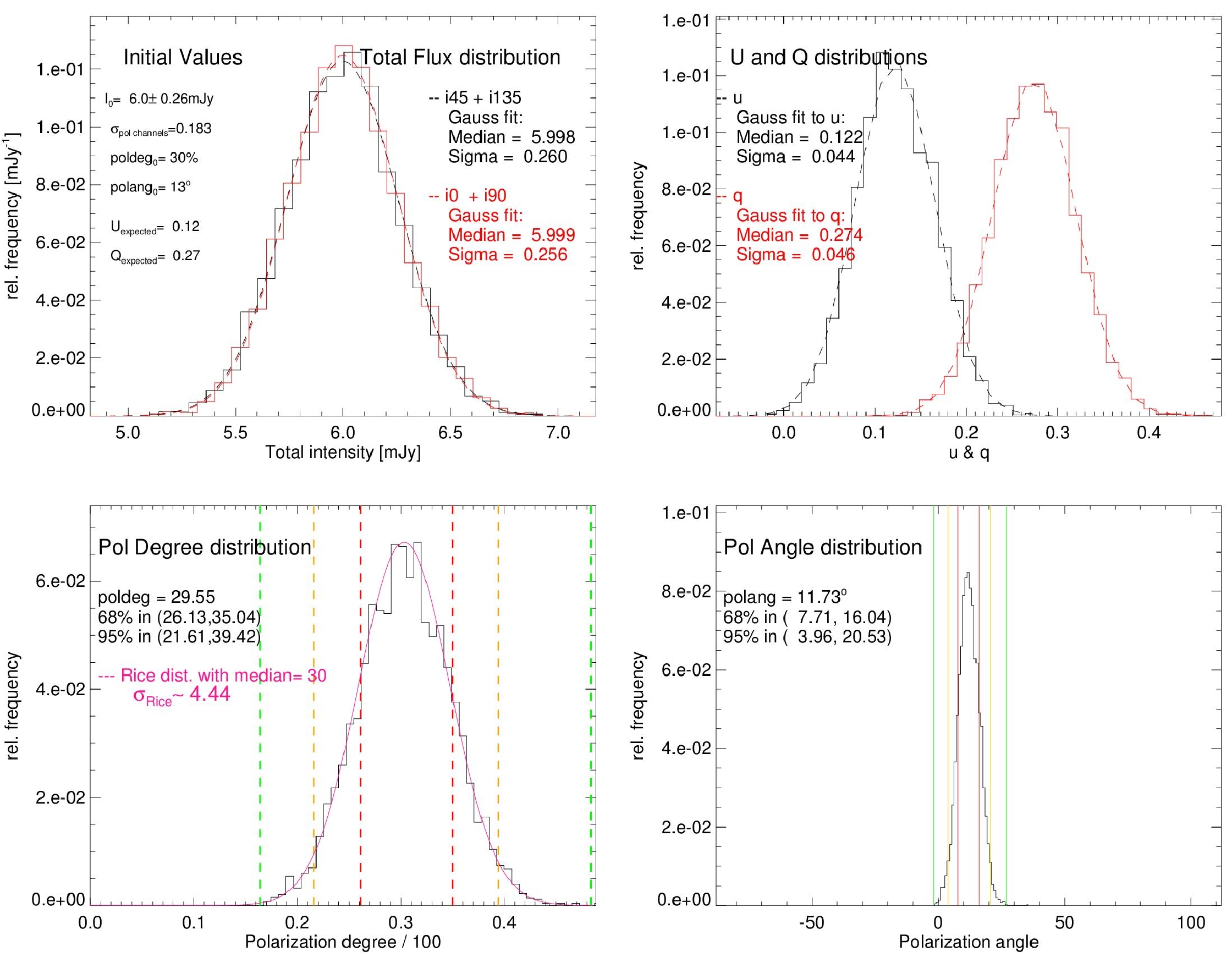}
  \caption{Initial values for simulation: total flux= 6~mJy, polarization degree= 30\% and polarization angle= 13$^o$
  }
  \label{fig:simulation-1}    
\end{figure*}

\begin{figure*}
  \centering
  \includegraphics[width = 6in]{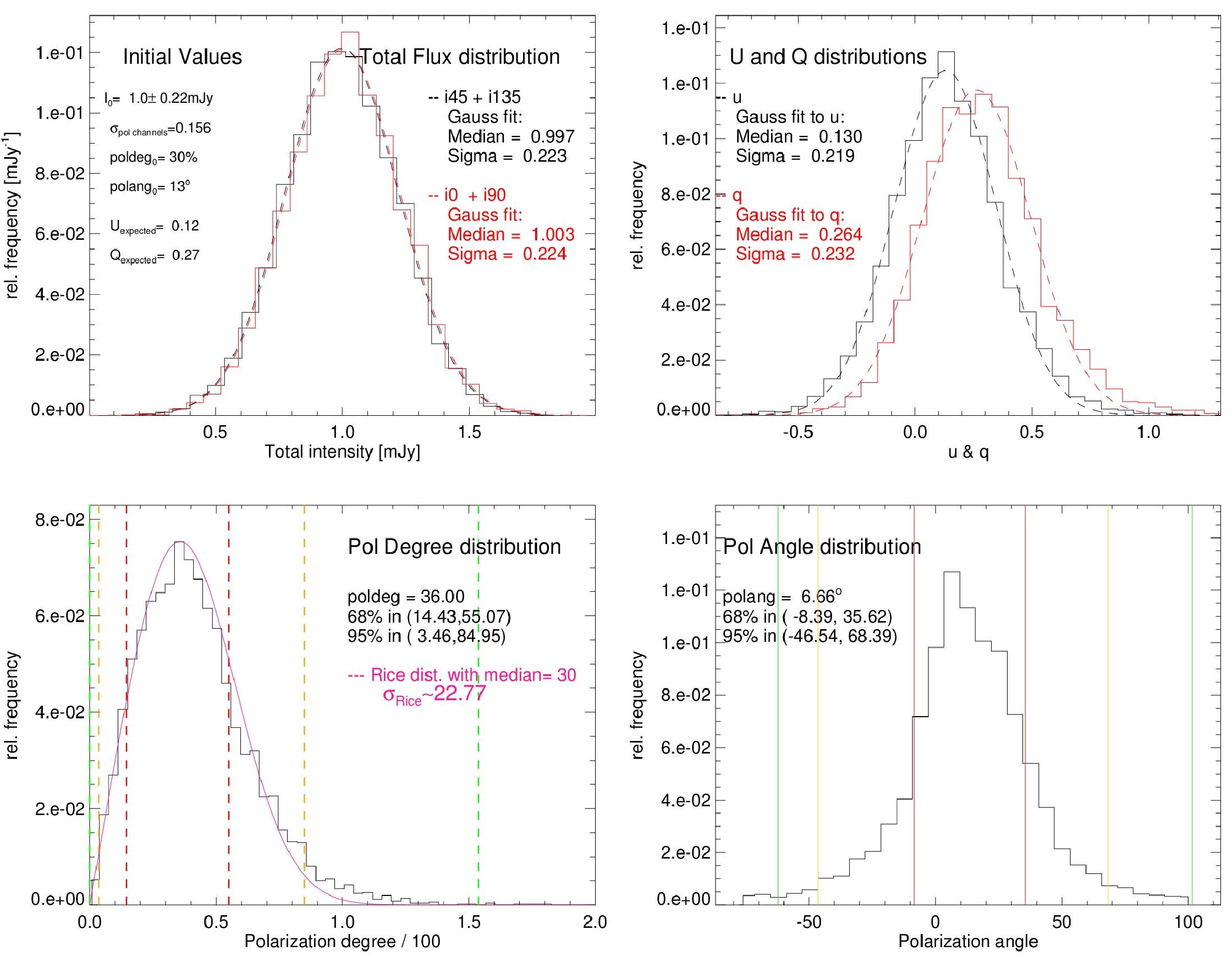}
  \caption{Initial values for simulation: total flux= 1~mJy, polarization degree= 30\% and polarization angle= 13$^o$. 
  }
  \label{fig:simulation-2}    
\end{figure*}

The polarization degree distribution does not follow a Gaussian distribution in the presence of random noise except for large $S/N$. In Sect.\,2.1, we showed that total intensity measurements of stars around Sgr A* are, in very good approximation, Gaussian distributed; this means that the noise in the polarization channels must have the same distribution. Hence, $U$ and $Q$ follow Cauchy distributions that, at medium $S/N$, can already  be approximated by normal distributions around the intrinsic values $U_0$ and $Q_0$. 
In this case, and assuming that $U$ and $Q$ are independent variables with associated variances equal to $\sigma_0^2$ ,  the polarization-degree distribution $F(p;p_0,\sigma_0)$  for a particular value of the  intrinsic polarization degree $p_0=(U_0^2 + Q_0^2)^{1/2}$ can be described by a Rice distribution

\begin{equation}F(p;p_0,\sigma_0) = \frac{p}{\sigma_0^{2}} \, J_{0}\left(i\frac{p\, p_{0}}{\sigma_0^{2}}\right)\,  \exp\left( -\frac{p^{2} + p_{0}^2}{2\sigma_0^{2}} \right) ,\end{equation}

where  $J_{0}(ix)$ is the zero-order Bessel function of the imaginary argument \citep{serkowski1958, vinokur1965} . \citet{ simmons1985} studied the bias on the observed value of $p$ for different $(S/N)_{\rm polarization~degree}$ defined as $P_0=p_0/\sigma_0$ (see their Fig.\,1). The most probable observed value of $p$ is the peak of the $F(p;p_0,\sigma_0)$distribution, which is always larger than $p_{0}$ for low $S/N$, and approaches the intrinsic value $p_0$ with increasing $S/N$.
At low $S/N$, the polarization-angle  distribution is multimodal with a spread that covers the whole range of possible values of $\phi$. At medium $S/N$ and under the same assumptions as previously, the probability distribution $F(\phi; \phi_0,P_0)$ for a particular intrinsic polarization angle $\phi_0$   is symmetric around its most probable value, and depends on the $(S/N)_{\rm polarization~degree}$. It can be expressed as

\begin{equation}F(\phi; \phi_0, P_0) =  \left\{  \frac{1}{\pi} + \frac{\eta_0 }{\sqrt{\pi}} \, {\rm e^{\eta_0^2}} \left[  1+ {\rm erf}(\eta_0) \right]  \right\} \, \exp{\left( -\frac{P_0^2}{2}  \right)} ,\end{equation}

with $\eta_0=(P_0/\sqrt{2}) \cos(\phi-\phi_0)$, and ``${\rm erf}$'' the Gaussian error function \citep{vinokur1965,  naghizadeh1993}. When $S/N$ is high, the distribution tends toward a Gaussian distribution with standard deviation  $\sigma _{\phi} = 28.^{\circ}65(\sigma_p/p) =  \sigma_p/(2p)$(radians), where the dispersion in the polarization degree is $\sigma_p=\sigma_0$ (\citealp{serkowski1958}, \citeyear{serkowski1962}). \citet{simmons1985} and \citet{stewart1991} have proposed several methods to remove the bias in the observed $p$ measurements for a source with a constant polarization state. However, whether this condition is fulfilled in the case of Sgr A* is unknown. Furthermore, there are scenarios that predict variability of its intrinsic polarization degree and angle.  Therefore, without any apriori assumptions on the polarization properties of Sgr A*, it is necessary to follow the propagation of the uncertainties from the observables, measured quantities i.e. flux densities in the polarization channels, to the calculated polarization properties $p$ and $\phi$.
We performed Monte Carlo simulations of the measured quantities and the observational noise, and used them to statistically analyze the calculated polarization degree and angle distributions and their uncertainties. The initial parameters of the simulation are the  total  flux density $F_0$ and its uncertainty $\sigma_F$, and the intrinsic polarization degree $p_0$ and angle $\phi_0$.  For each set of these parameters, the corresponding polarization-channel fluxes $f_0, f_{90}, f_{45}$, and $f_{135}$ can be  calculated from Eqs.\,(1)-(5). In order to associate an uncertainty with the flux measured in each polarization channel $f_X$, we considered the relation between the total flux densities  and their uncertainties presented in Sect.\,2.1.  Given that the distributions of $F$ are  in a very good approximation Gaussian, and that the noise in each par of orthogonal polarization channels is about the same, then the standard deviation of the Gaussian function that describes the distribution of each $f_X$ can be expressed as 
$\sigma_{f_X}=\sigma_F/\sqrt{2}$, i.e. 
$\sigma_{f_0}= \sigma_{f_{90}}=\sigma_{f_{45}}= \sigma_{f_{135}}=\sigma_F/\sqrt{2}$. For each set $(F_0, p_0, \phi_0)$,  we draw from the $f_X$ distributions, $10^4$ tetrads of polarization-channel fluxes and use them to calculate $U$, $Q$, $p$ and $\phi$, as it is done with real data.  From the simulations it is possible to establish the most probable observed values of $p$ and $\phi$, as well as the ranges in which certain percentages of the values are contained. We consider intrinsic total flux densities ranging from $0.8\,{\rm mJy}$, that is the photometry detection limit \citep{Witzel2012}, to $15.0\,{\rm mJy}$, that is approximately the maximum value in our data set. The initial values for polarization degree (the amplitude of the intrinsic polarization, $p_{0}$) are set to a range from 5\% to 70\%, while $\phi_{0}$ is fixed to a preferred polarization angle of $13^{\circ}$, based on the following data analysis.   Figures \ref{fig:simulation-1} and \ref{fig:simulation-2} show the resulting distributions of $U$, $Q$, $p$ and $\phi$ for two different initial total flux values, one with medium $S/N$ and the other with high $S/N$ ratio, as an example. The distributions in the plots correspond to the values that an observer would measure for a source whose intrinsic total flux, polarization degree and polarization angle are the initial values given at beginning of the simulation. We use the limits of the intervals containing 68\%, 95 \% and 99\% of all values as our effective 1$\sigma$, 2$\sigma$ and 3$\sigma$ error intervals. This gives us asymmetric errors for $p$ in the form $p_{-\sigma_p1} ^ {+\sigma_p2}$. From these two examples it is clear that the most probable value of $p$ is larger than its true value $p_0$, but $p$ becomes closer to $p_0$ for higher $S/N$ ratios. 

Moreover, we have fitted a Rice function to the $p$ distribution and obtained the $\sigma$ value of 
Rice distribution. 
Table~\ref{table:nonlin}
presents the simulated polarization degree values for different 
initial flux and polarization degree values. In the case of polarization angle, we have considered the confidence intervals and obtained 
$\sigma$ values only if the (S/N) ratio is larger than 4.5, since for the lower (S/N) ratios, 
the $\phi$ distribution has a non-Gaussian shape. Table~\ref{table:nonlin}
presents the simulated polarization degree values for different 
initial flux and polarization degree values.
 
\begin{figure*}[]
     \begin{center}
        \subfloat{%
           \includegraphics[width=0.45\textwidth]{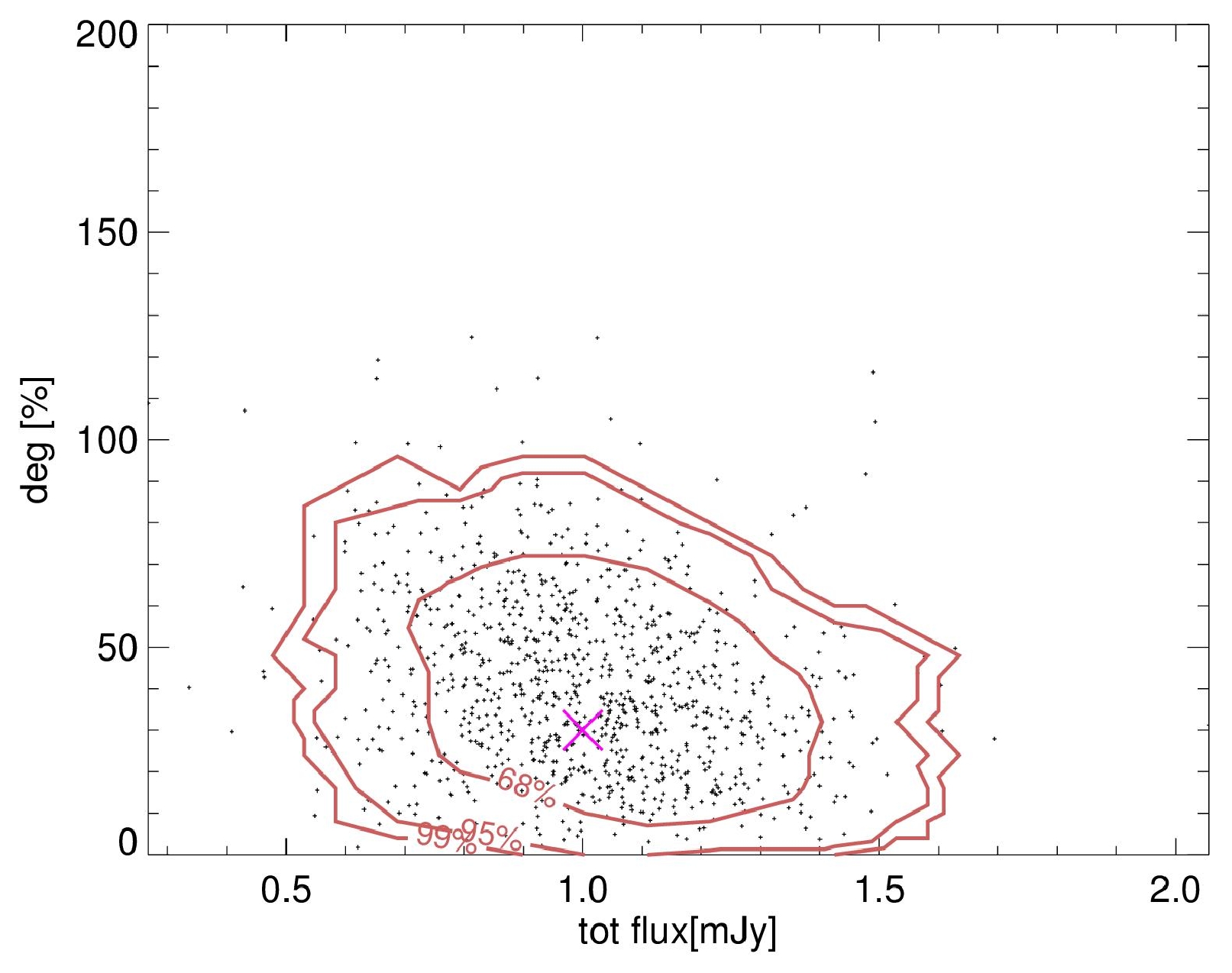}
        }
        \subfloat{%
            \includegraphics[width=0.45\textwidth]{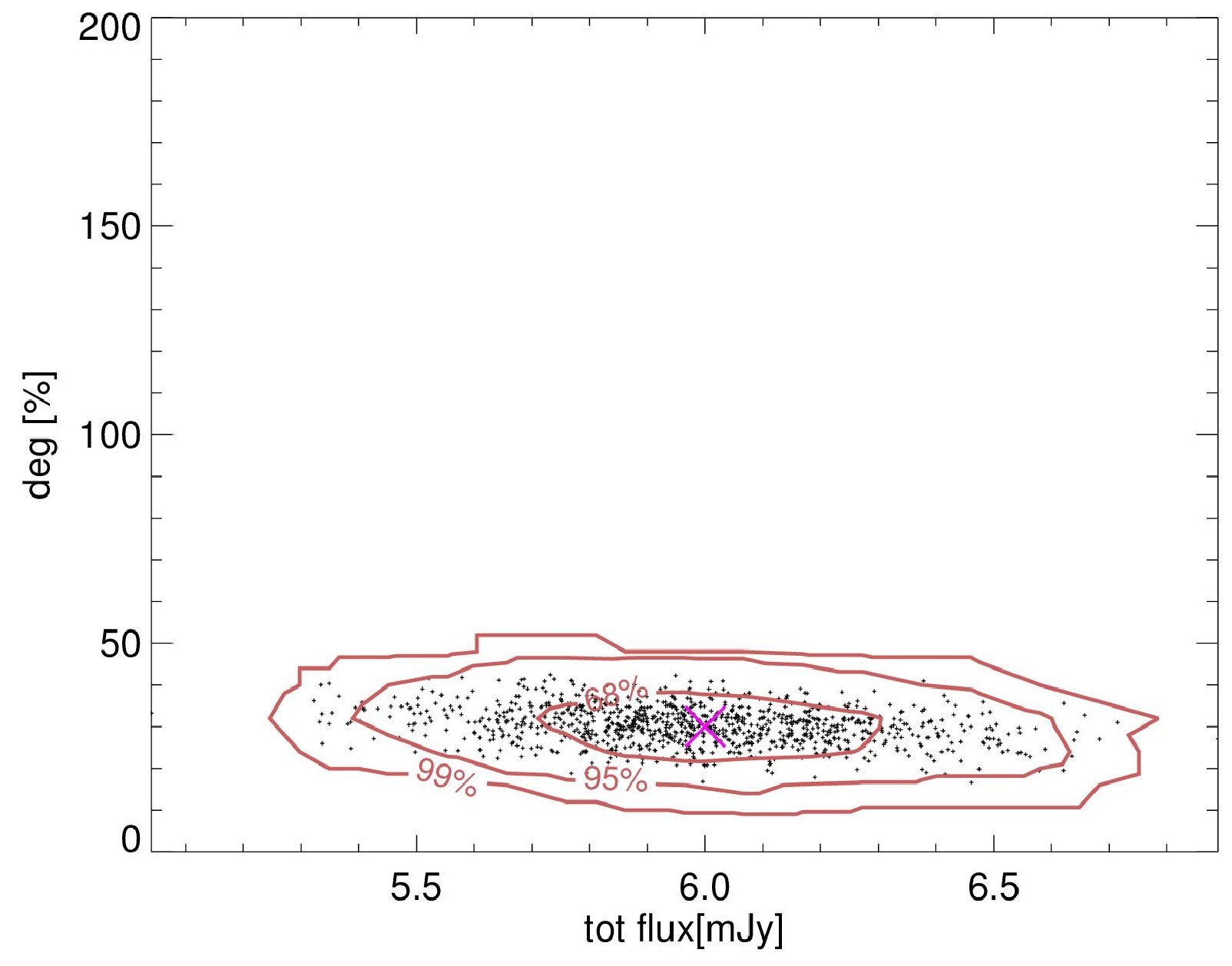}
        }\\ %  ------- End of the first row ----------------------%
 \end{center}
    \caption{The relation between simulated values of total flux density and polarization degree for initial values of : 
Left: total flux = 1~mJy, polarization degree = 30$\%$, polarization angle = 13$^o$, 
Right: total flux = 6~mJy, polarization degree = 30$\%$, polarization angle = 13$^o$. 
     }
   \label{fig:relations_simulation}
\end{figure*}

Fig.\ref{fig:relations_simulation}
represents the relation between the simulated 
values of the total flux and the polarization degree. 
The contours shown in the figure 
correspond to 1$\sigma$, 2$\sigma$ and 3$\sigma$ values 
that enclose 68\%, 95\%, and 99\% of points in the distribution, i.e. if the observer 
could measure these quantities more than 1000 times then, the central contour would enclose 
the 68\% of measured values that are closest to the intrinsic value.

The following results of the expected statistical properties of our polarization data
will be used for interpretation in the upcoming section:

From Fig.\ref{fig:simulation-1} and Table \ref{table:nonlin} one can see 
that for our NIR data for strong flare fluxes the recovered degree of polarization 
is Gaussian distributed around a very well central value close to the intrinsic degree with a $\sim$5\% uncertainty. 
Hence, if the intrinsic polarization degree is centered around a fixed expectation value the 
statistical properties of bright flare samples will be very similar to those of
the total flux density measurements as presented by \cite{Witzel2012}. 

One can see from Fig.\ref{fig:simulation-1} and Table \ref{table:nonlin} 
that for weak intrinsic fluxes the recovered degree of polarization 
is not any longer Gaussian distributed and especially for moderate or weaker intrinsic polarization 
degrees the intrinsic value is not well recovered and the uncertainties are very large, such that unrealistic polarization degrees of above 100\% can be obtained.
Only for intrinsically strongly polarized weak flares the intrinsic polarization
degrees are statistically recovered but the un-symmetric uncertainties remain very large.
Therefore, the total statistical behavior of observed polarization data can be thought of
as being composed of the properties of subsamples of different polarization degree 
and flare fluxes.

%======================================================================
\begin{table*}% * makes wide table in 2 columns
%\caption{}% title of Table
\centering% used for centering table
\begin{tabular}{cccccccccc}% centered columns (4 columns)

\hline %inserts double horizontal lines
\hline

&&&&&F'[mJy]&&&&\\[1ex]% inserts table
p'[\%] & 0.8& 1 & 1.3 & 1.5 & 2 & 3 & 4 & 6  & 10   \\

%heading

\hline% inserts single horizontal line
\\
%\hline% inserts single horizontal line
 5&$    27_{-17}^{+22}$&$    24_{-15}^{+16}$&$    19_{-11}^{+12}$&$    15_{ -8}^{+12}$&$    12_{ -6}^{ +9}$
  &$     9_{ -5}^{ +6}$&$     8_{ -5}^{ +4}$&$     6_{ -4}^{ +3}$&$\bf  5_{ -2}^{ +3}$\\[1.5ex] 

10&$    28_{-18}^{+23}$&$    22_{-13}^{+18}$&$    18_{-10}^{+14}$&$    16_{ -9}^{+12}$&$    14_{ -8}^{ +9}$
  &$    12_{ -6}^{ +7}$&$\bf 12_{ -5}^{ +5}$&$\bf 10_{ -3}^{ +4}$&$\bf 10_{ -3}^{ +3}$\\ [1.5ex]

20&$    30_{-18}^{+26}$&$    22_{-11}^{+24}$&$    24_{-13}^{+15}$&$    24_{-13}^{+13}$&$\bf 21_{ -9}^{+10}$
  &$\bf 22_{ -8}^{ +6}$&$\bf 21_{ -6}^{ +6}$&$    21_{ -5}^{ +4}$&$\bf 20_{ -3}^{ +3}$\\[1.5ex]

30&$    33_{-19}^{+30}$&$    34_{-18}^{+22}$&$\bf 34_{-16}^{+17}$&$\bf 32_{-13}^{+15}$&$\bf 31_{-10}^{+12}$
  &$\bf 31_{ -8}^{ +7}$&$\bf 31_{ -6}^{ +6}$&$\bf 30_{ -4}^{ +4}$&$\bf 30_{ -3}^{ +3}$\\ [1.5ex]% [1ex] 

40&$    40_{-20}^{+33}$&$    40_{-19}^{+26}$&$\bf 42_{-17}^{+18}$&$\bf 44_{-16}^{+14}$&$\bf 41_{-11}^{+12}$
  &$\bf 41_{ -8}^{ +7}$&$\bf 40_{ -6}^{ +6}$&$\bf 40_{ -4}^{ +4}$&$\bf 41_{ -3}^{ +3}$\\ [1.5ex]

50&$    56_{-25}^{+32}$&$    49_{-29}^{+29}$&$\bf 49_{-16}^{+21}$&$\bf 52_{-16}^{+21}$&$\bf 50_{-15}^{+13}$
  &$\bf 49_{ -9}^{+10}$&$\bf 51_{ -6}^{ +9}$&$\bf 50_{ -6}^{ +4}$&$\bf 50_{ -3}^{ +3}$\\ [1.5ex]

60&$    61_{-28}^{+34}$&$\bf 60_{-22}^{+27}$&$\bf 61_{-16}^{+20}$&$\bf 60_{-15}^{+17}$&$\bf 61_{-12}^{+13}$
  &$\bf 60_{ -7}^{ +9}$&$\bf 60_{ -6}^{ +6}$&$\bf 60_{ -4}^{ +4}$&$\bf 60_{ -3}^{ +3}$\\ [1.5ex]

70&$\bf 73_{-30}^{+33}$&$\bf 63_{-26}^{+31}$&$\bf 71_{-19}^{+20}$&$\bf 69_{-18}^{+23}$&$\bf 70_{-15}^{+16}$
  &$\bf 70_{-10}^{+11}$&$\bf 70_{ -8}^{ +8}$&$\bf 69_{ -5}^{ +6}$&$\bf 70_{ -4}^{ +4}$\\ [1.5ex]

\hline

\end{tabular}
\caption {Average values of recovered polarization degrees obtained by simulation for the combination of 
different sets of intrinsic total flux $F'$ and polarization degree $p'$ as initial values.
Table entries for which both the upper and lower uncertainty of the recovered polarization degree are smaller
or equal than half of the actual recovered value are printed in boldface.
}
\label{table:nonlin}
\end{table*}

%======================================================================

\begin{table*}% * makes wide table in 2 columns
%\caption{}% title of Table
\centering% used for centering table
\begin{tabular}{cccccccccc}% centered columns (4 columns)

\hline %inserts double horizontal lines
%\backslashbox{Int. p[\%]}{Int. flux}
\hline

&&&&&F'[mJy]&&&&\\[1ex]% inserts table
p'[\%]& 0.8& 1 & 1.3& 1.5  & 2  & 3 & 4 & 6  & 10   \\

%heading

\hline% inserts single horizontal line
\\
%\hline% inserts single horizontal line
 5&$ 4_{-47}^{+61}$&$ 5_{-42}^{+63}$&$ 9_{-48}^{+60}$&$ 7_{-46}^{+53}$&$ 8_{-38}^{+49}$&$ 8_{-34}^{+42}$&$13_{-35}^{+32}$&$ 8_{-26}^{+24}$&$13_{-18}^{+16}$\\[1.5ex] % inserting body of the table%

10&$ 0_{-38}^{+58}$&$ 5_{-40}^{+53}$&$ 9_{-39}^{+43}$&$ 9_{-36}^{+40}$&$10_{-30}^{+33}$&$13_{-24}^{+21}$&$14_{-19}^{+15}$&$10_{-12}^{+13}$&$13_{ -9}^{ +8}$\\ [1.5ex]

20&$ 5_{-26}^{+46}$&$10_{-27}^{+35}$&$11_{-23}^{+27}$&$11_{-21}^{+24}$&$12_{-17}^{+17}$&$12_{-11}^{+11}$&$12_{ -9}^{ +8}$&$12_{ -6}^{ +6}$&$12_{ -4}^{ +4}$\\[1.5ex]

30&$ 8_{-22}^{+31}$&$11_{-20}^{+24}$&$10_{-15}^{+20}$&$12_{-16}^{+14}$&$12_{-11}^{+11}$&$11_{ -6}^{ +8}$&$12_{ -6}^{ +6}$&$12_{ -3}^{ +5}$&$12_{ -3}^{ +3}$\\ [1.5ex]% [1ex] adds vertical space

40&$10_{-17}^{+27}$&$10_{-15}^{+19}$&$11_{-13}^{+13}$&$11_{-11}^{+11}$&$11_{ -8}^{ +9}$&$11_{ -5}^{ +7}$&$12_{ -4}^{ +5}$&$11_{ -3}^{ +3}$&$12_{ -2}^{ +2}$\\ [1.5ex]

50&$ 6_{ -9}^{+18}$&$10_{-10}^{+13}$&$11_{-10}^{+12}$&$11_{ -9}^{+12}$&$12_{ -7}^{ +9}$&$11_{ -5}^{ +6}$&$12_{ -4}^{ +4}$&$12_{ -3}^{ +3}$&$12_{ -2}^{ +2}$\\ [1.5ex]

60&$ 9_{-12}^{+17}$&$ 9_{-10}^{+13}$&$12_{ -9}^{ +9}$&$12_{ -8}^{ +8}$&$12_{ -6}^{ +6}$&$12_{ -4}^{ +4}$&$12_{ -3}^{ +3}$&$12_{ -2}^{ +2}$&$12_{ -1}^{ +1}$\\ [1.5ex]

70&$ 6_{ -8}^{+16}$&$ 9_{ -8}^{ +9}$&$11_{ -8}^{ +9}$&$11_{ -7}^{ +9}$&$13_{ -7}^{ +6}$&$11_{ -4}^{ +5}$&$12_{ -3}^{ +3}$&$13_{ -2}^{ +2}$&$12_{ -2}^{ +2}$\\ [1.5ex]

\hline%inserts single line

\end{tabular}
\caption {Average values of recovered polarization angles obtained by simulation for the combination of 
different sets of intrinsic total flux $F'$ and polarization degree $p'$ as initial values.}
\label{table:nonlin2}
\end{table*}

%======================================================================

%##############################################################

%=======================================================================

%=======================================================================

\begin{figure*}[]
     \begin{center}
        \subfloat{%
           \includegraphics[width=0.45\textwidth]{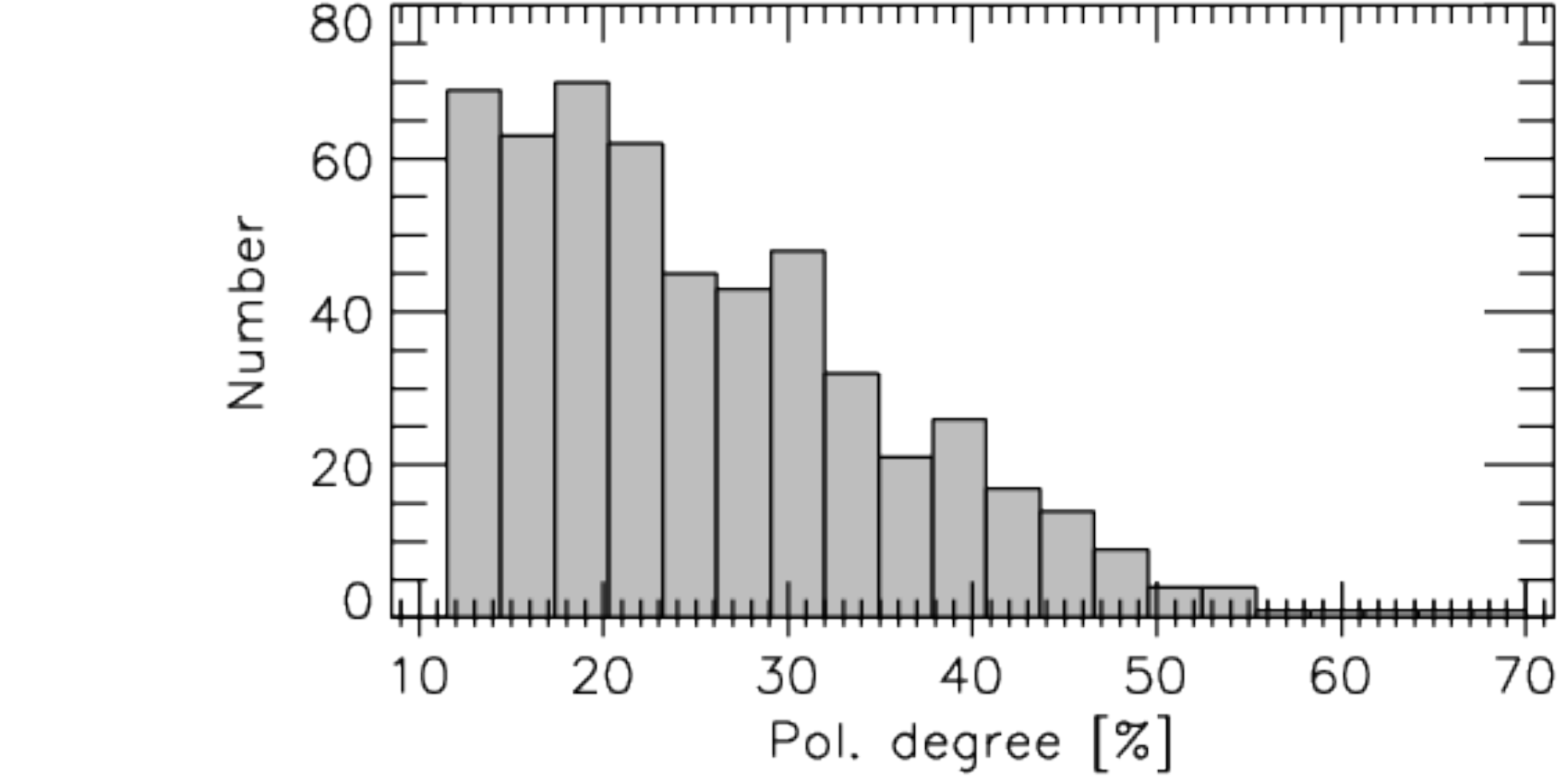}
        }
        \subfloat{%
           \includegraphics[width=0.45\textwidth]{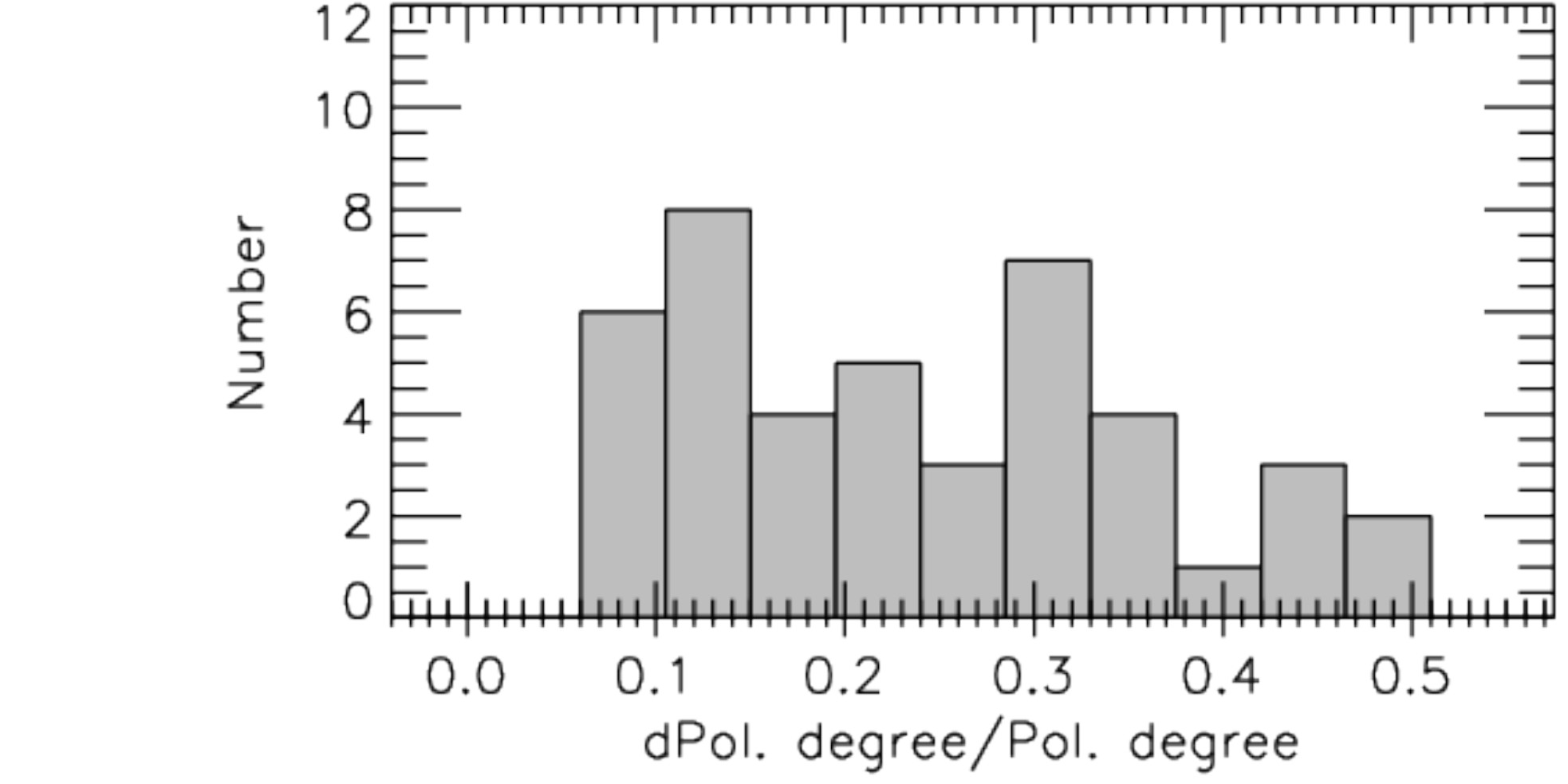}
           
        }\\ %  ------- End of the first row ----------------------%
 \end{center}
    \caption{
Left: Distribution of $K_\mathrm{s}$-band polarization degrees of Sgr~A* for our data set considering the 
significant data points (based on Table \ref{table:nonlin}).   
Right: Distribution of relative uncertainties of the polarization degrees.
     }
   \label{fig:degrees}
\end{figure*}
%======================================
\begin{figure*}[]
     \begin{center}
        \subfloat{%
            \includegraphics[width=0.45\textwidth]{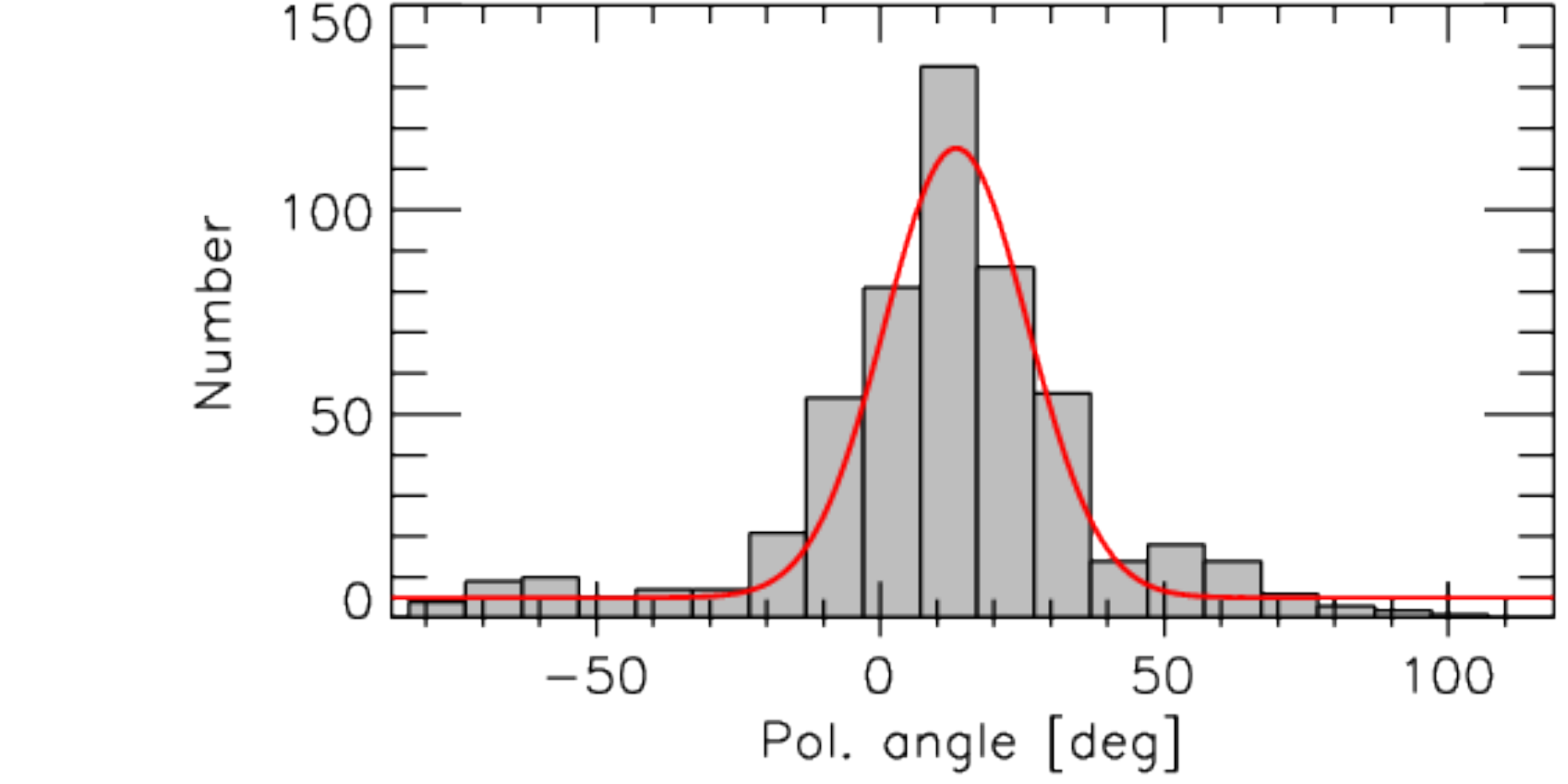}
        }
        \subfloat{%
           \includegraphics[width=0.45\textwidth]{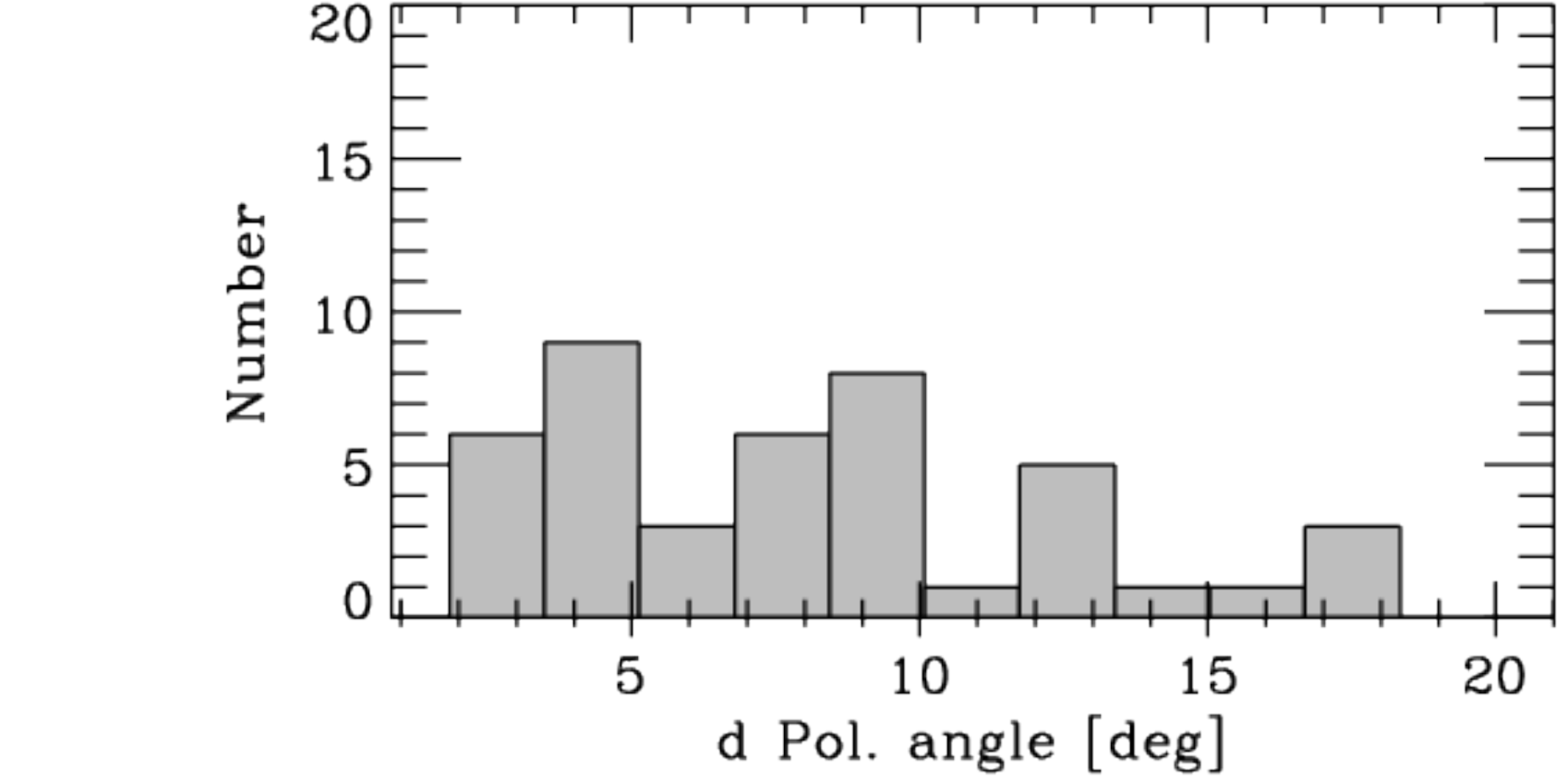}
        }\\ %  ------- End of the first row ----------------------%
 \end{center}
    \caption{
Left: Distribution of significant $K_\mathrm{s}$-band polarization angles of Sgr A*. The red line shows the fit with a Gaussian distribution. 
Right: Distribution of absolute errors of the polarization angles.
     }
   \label{fig:angles}
\end{figure*}
%======================================
\begin{figure*}[]
     \begin{center}
        \subfloat{%
            \includegraphics[width=0.45\textwidth]{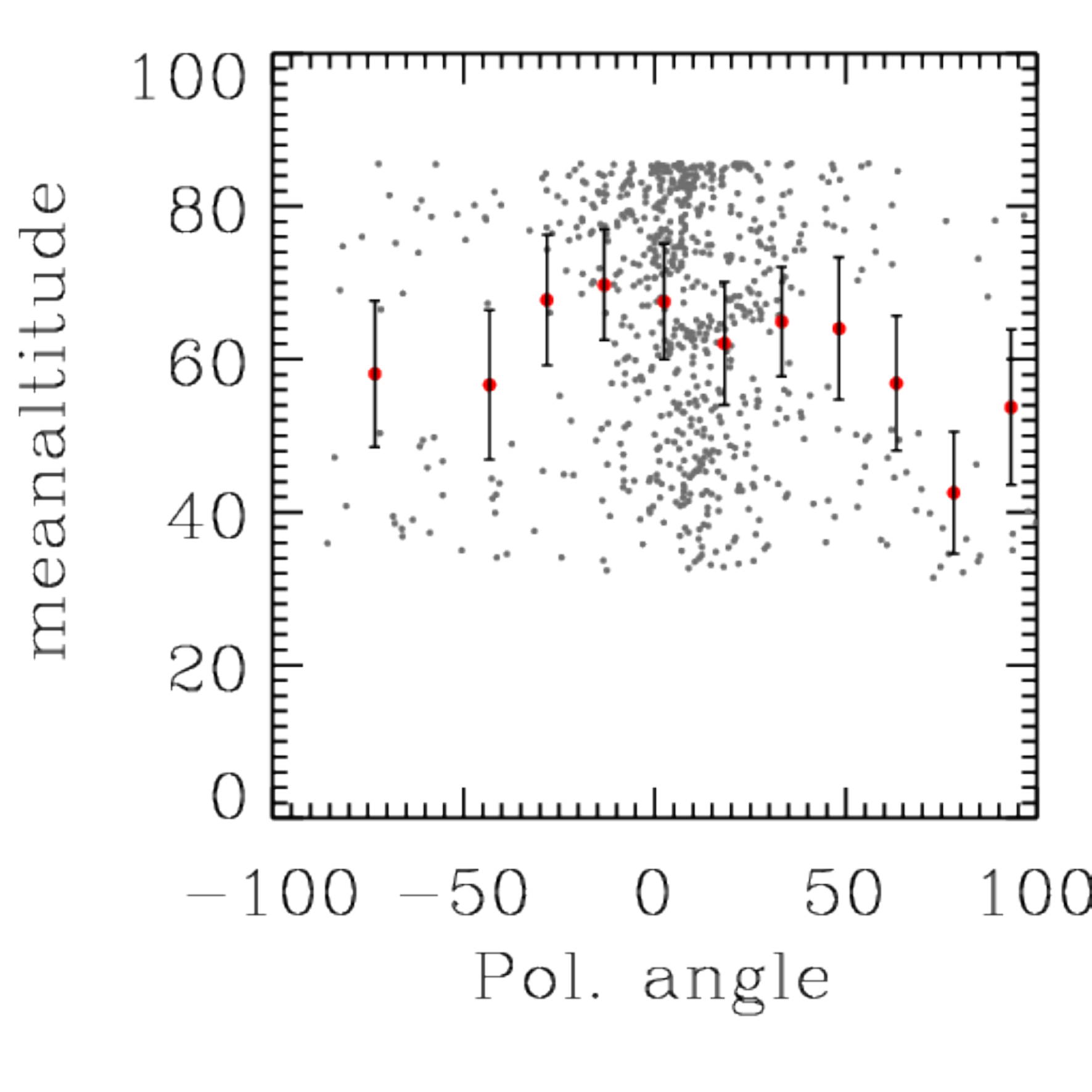}
        }
        \subfloat{%
           \includegraphics[width=0.45\textwidth]{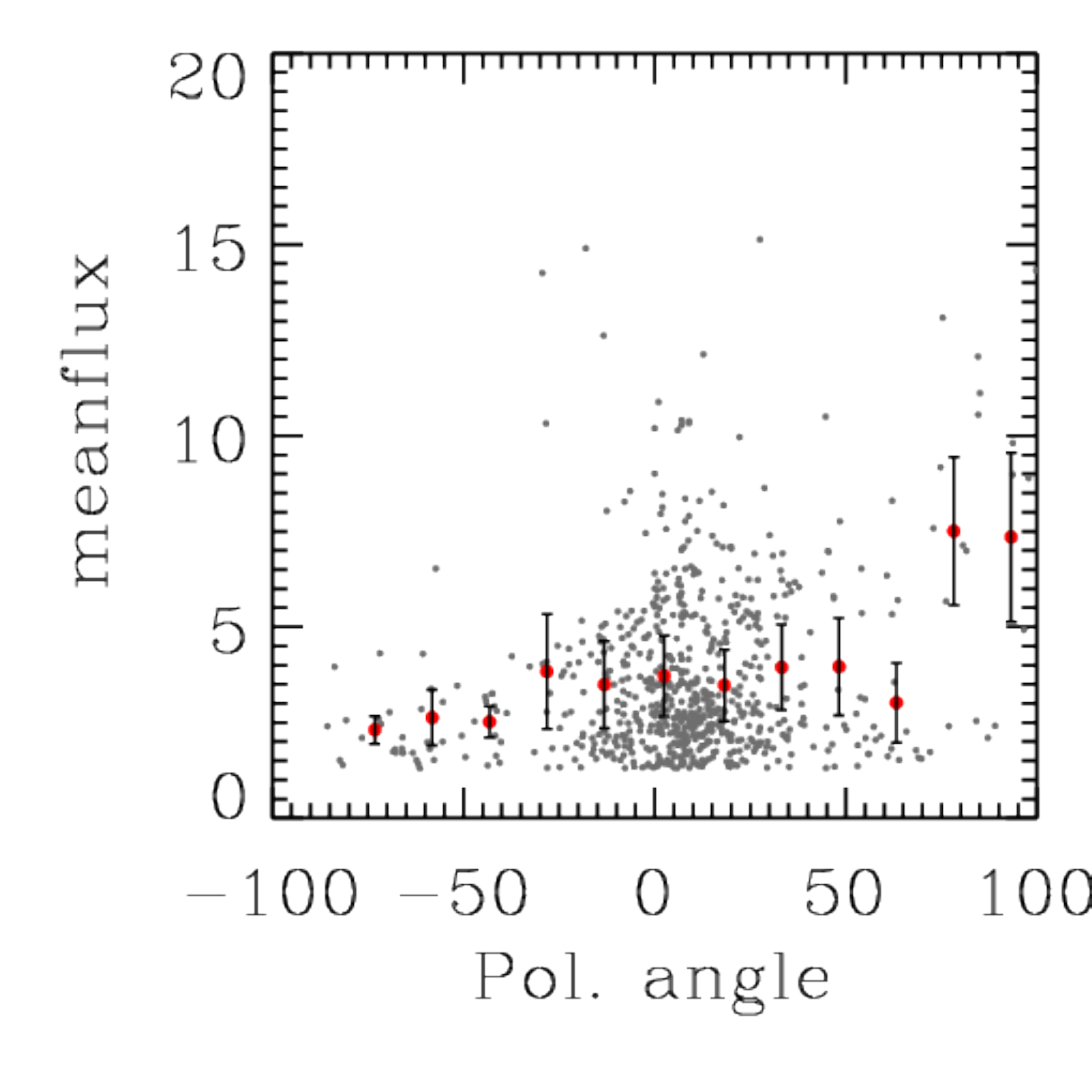}
        }\\ %  ------- End of the first row ----------------------%
 \end{center}
    \caption{Left: Altitude (elevation) of Sgr~A* as a function of polarization angle; 
    Right: Total flux density as a function of polarization angle; 
    The bin width in polarization angle is 15$^o$.
    The values for individual measurements are shown as black dots. The mean values per bin
are shown as red dots with error bars indicating the standard deviation, if the number of data points
per bin is larger than 2.
     }
   \label{fig:altflux}
\end{figure*}
%======================================

\subsection{Polarization degree and polarization angle}
\label{section:polfluxangle}

In Fig.\ref{fig:degrees} we show the distribution of $K_\mathrm{s}$-band polarization degrees of Sgr A* (left) 
as well as the distribution of their uncertainties (right) 
obtained following the results of our statistical analysis presented 
in Section \ref{section:Statistical}.
In these figures we only plot the data points for which - based on our simulations - both the upper and lower 
uncertainty of the recovered polarization degree are smaller than half of the actual recovered value.
In the following we will refer to these values as being significant measurements i.e. successful retrievals
of the intrinsic polarization degree (and angle; see below).
The distribution of polarization degrees has a peak close to $\sim$ 20$\%$. 
It does not have the shape of a Gaussian and is strongly influenced by systematic effects 
with uncertainties ranging from 10\% to 50$\%$ (see section \ref{section:Statistical}).

Fig.\ref{fig:angles} shows the distribution of significant $K_\mathrm{s}$-band polarization angles of Sgr A* (left) and their uncertainties (right) as determined for the corresponding flare fluxes following
the statistical analysis presented in Section \ref{section:Statistical}.
For table entries with significant polarization degrees the corresponding uncertainties in the 
recovered polarization angle are below $\pm$20$^o$ i.e. below about 1/3 of a radian.
The distribution of polarization angles shows a peak at 13$^o$.
The overall width of the distribution is of the order of 30$^o$, hence, the preferred polarization angle 
that we can derive from the distribution is 13$^o$ $\pm$15$^o$.
In Fig.\ref{fig:relations}~(right) we show for all data with significant 
polarization degrees the angle as a function of total flux density.
The same plot for the entire data set is shown in Fig.\ref{fig:App4}.
The distribution of polarization angle and degree for the entire data set are shown in Fig.\ref{fig:App2}.

The data shown on the left side of Figs.\ref{fig:degrees} and \ref{fig:angles} 
are consistent in the sense that the uncertainty of $\Delta$$\phi$=15$^o$ 
in polarization angle is reflected in the width of the distribution of 
uncertainties derived from our simulations Fig.\ref{fig:degrees}~(right).
This implies that the uncertainty in angle is dominated by the measurement error 
and the physical variabilities of the angle is probably much smaller. The measurement error is the combined 
uncertainty of recording the data and retrieving the polarization information out of it. 
An upper limit for the uncertainties in polarization angles (Fig.\ref{fig:angles}, right) 
is $\Delta$$\phi$=20$^o$.
This value $\Delta$$\phi$ also implies a corresponding expected relative 
uncertainty of the polarization degree of about 
$\frac{\Delta p}{p}$$\sim$tan(15$^o$)=0.36 (i.e. 36\%, for explanation 
see sketch in Fig.\ref{fig:App5}).
This is in good agreement with the approximate center value of 0.3 (i.e. 30\%) 
found for the slightly skewed distribution of relative uncertainties of the 
polarization degree (Fig.\ref{fig:degrees}, right) for the entire set of 
significant data as judged from the simulations.
Since the uncertainty in angle only accounts for the lower portion of the 
distribution of relative uncertainties in polarization degree,
this implies that the polarization degree is indeed dominated by intrinsic fluctuations of that quantity.
Under the assumption that the measurement and intrinsic uncertainty add quadratically, the intrinsic
variability of the relative 2 $\mu$m NIR polarization degree for SgrA* is of the order of about 30\%.

In order to investigate if the distribution of polarization angles of Sgr~A* is affected 
by the strength of the flare and the position of this source in the sky, we plotted for all 
data with significant polarization degrees the average of flux densities versus the 
polarization angles binned in 15$^o$ intervals (Fig.\ref{fig:altflux}, right) and the average elevation 
of Sgr~A* in the sky for each polarization measurement (Fig.\ref{fig:altflux}, left). 
The corresponding plots for all data are shown in Fig.\ref{fig:App1}.
The region in which a significant correction due to instrumental polarization needs to be applied is located
at about $\pm$0.5~hours with respect to the meridian \citep[see Fig.9 in][]{Witzel2011}. 
This corresponds to an elevation of higher than about 80$^o$. 
The distribution of data in Fig.\ref{fig:altflux}~(left) indicates that most 
of the measurements were done at elevations below 75$^o$, hence the correction 
for instrumental polarization is very small. 
The distribution of data in Fig.\ref{fig:altflux}~(right) indicates for polarization angles close to the preferred angle of 13$^o$ the corresponding mean flux density is within about 1$\sigma$ from 
the mean flux density values of all 15$^o$ intervals. Hence, the flux density values that correspond to polarization angles
around the preferred value are not exceptionally high or low.
In summary we can exclude that 
the preferred polarization angle of about 13$^o$ is related to flux 
density excursions of particular brightness or to a particular location in the 
sky and instrumental orientation.
Therefore, we conclude that the preferred polarization angle is a source intrinsic property.

\subsection{Polarized flux density distribution}
\label{section:polfluxdistrib}

We produce the histogram of polarized flux density distribution of Sgr A*
in its linear form
(Fig.\ref{fig:App3}, left) 
for the entire data 
and double logarithmic representation 
for the entire data 
(see Fig.\ref{fig:App3}, right) 
as well as for the fraction of the data which is significant, based on our simulation,
showed in Fig.\ref{histban3}.
The distribution is normalized by the total number of points and bin size. 
We use the logarithmic histogram in order to better display 
the distribution of values (Fig.\ref{histban3}).
In the following we present two different approaches to formally describe and physically explain the measured polarized flux density distribution shown in Fig. \ref{histban3}.

\subsection{The polarized flux density distribution for bright flare fluxes}
\label{section:fit}

\begin{figure}
  \centering
  \includegraphics[width=0.45\textwidth]{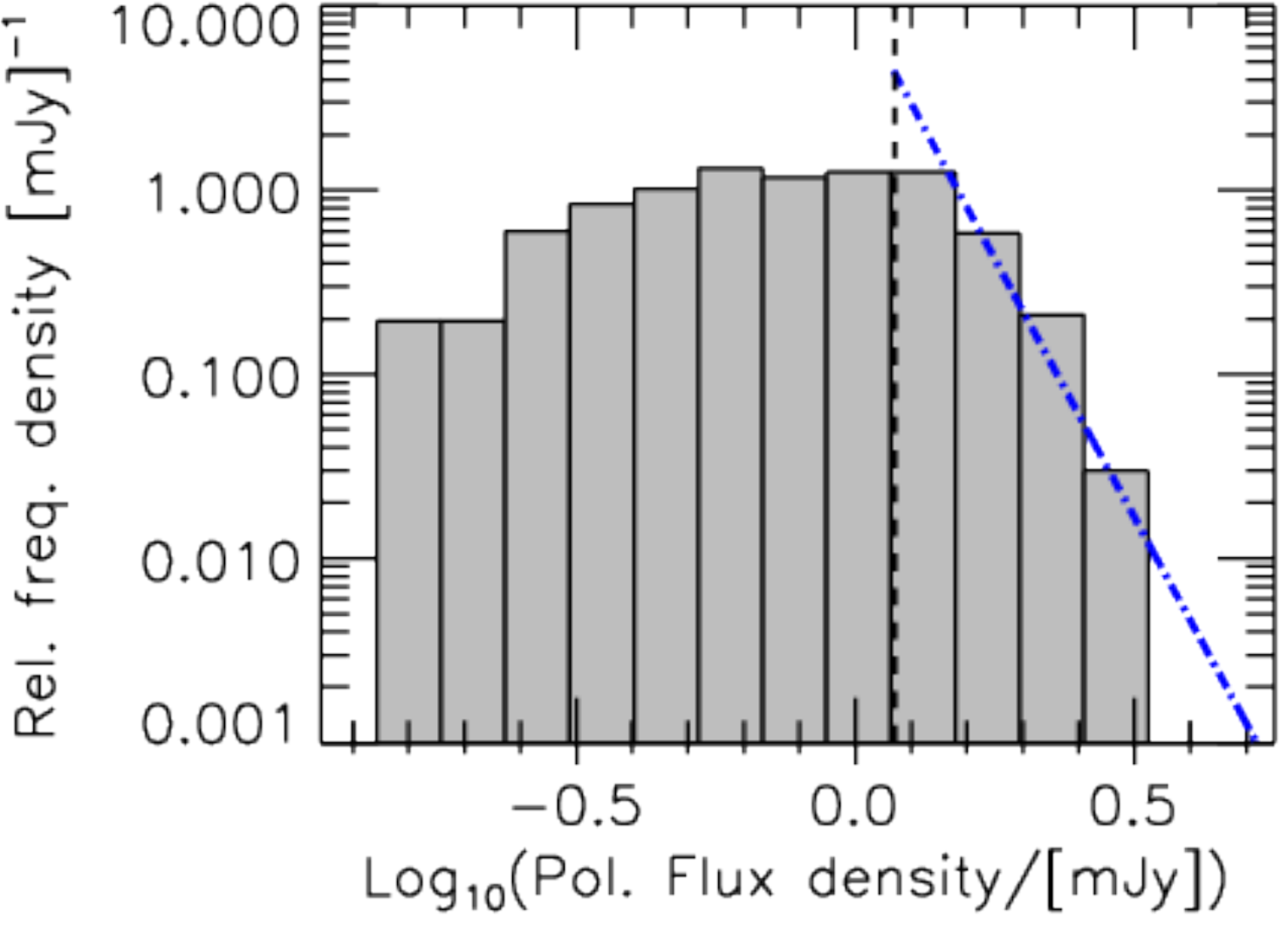}
  \caption{\small Histogram of polarized flux density 
(total flux density times polarization degree) for all significant data in logarithmic scale after correction for stellar contamination. The thick black dashed line shows the position where the powerlaw starts to fit to the histogram.
  }
  \label{histban3}    
\end{figure}

In Fig.\ref{histban3} we show the histogram of polarized flux density. Our simulations have shown that 
only for bright flare fluxes the polarization degree can be recovered with a small uncertainty.
Therefore, the properties of the polarized flux density distribution (i.e. the product
of the polarization degree and the total flux density) can be investigated best for high
polarized flare fluxes. A powerlaw fit to the data at high flux densities is shown as a dot-dashed blue line.
For high flare fluxes the slope $\alpha$ is fitted to a value of 4.00$\pm$0.15 which is very close to the value of
4.21$\pm$0.05 obtained for the total flux densities by \cite{Witzel2012}.
The behavior was  predicted by the simulations (see Fig.\ref{fig:simulation-1} and Table \ref{table:nonlin}).
Recovering this exponent for the polarized flare flux density distribution indicates that 
the intrinsic polarization degree is centered around a fixed 
expectation value (see section \ref{section:Statistical}) and has not been strongly variable over the
time interval from 2004 to 2012.

%======================================================================
\begin{table*}[]
\caption{Comparison of polarization measurements of Sgr~A* obtained in this paper with the ones reported in the literature.}% title of Table
\centering% used for centering table
\begin{tabular}{l l l l}% centered columns (4 columns)
\hline %inserts double horizontal lines
obs.date & degree~~~~~~~~~angle & degree~~~~~~~~~~~angle & degree~~~~~~~~~~~angle\\ [0.5ex]% inserts table
%heading
&Eckart+04&Zamaninasab+10&This work\\
\hline% inserts single horizontal line
13 June 2004&~$\sim$20$\%$~~~~$-20\,^{\circ}$-$+70\,^{\circ}$&~$\sim$20$\%$~~~~~~$-20\,^{\circ}$-$+70\,^{\circ}$&10$\%$-30$\%$~~$-60\,^{\circ}$-$+40\,^{\circ}$\\
30 July 2005&12$\%$-25$\%$~~~$40\,^{\circ}$-$+80\,^{\circ}$&12$\%$-25$\%$~~~~~~~~$\sim$$40\,^{\circ}$&~~5$\%$-20$\%$~~~~$10\,^{\circ}$-$+100\,^{\circ}$\\
15 May  2007&-&20$\%$$\pm$20$\%$~~$-50\,^{\circ}$-$+50\,^{\circ}$&~~5$\%$-20$\%$~~$-50\,^{\circ}$-$+20\,^{\circ}$\\
17 May  2007&-&20$\%$$\pm$10$\%$~~$-30\,^{\circ}$-$+20\,^{\circ}$&~~5$\%$-20$\%$~~$-50\,^{\circ}$-$+50\,^{\circ}$ \\ [1ex] % [1ex] adds vertical space

\hline%inserts single line
\end{tabular}
\label{table:nonlin3}
\end{table*}
%======================================================================

%======================================================================
%\begin{table*}[]% * makes wide table in 2 columns
%\caption{Best fit parameters for four different functions.}% title of Table
%\centering% used for centering table
%\begin{tabular}{c c c c c c c}% centered columns (4 columns)
%\hline %inserts double horizontal lines
%model & $\alpha$ & $f_{p,min}$ & $f_{p0}$ & $\mu$ & $\sigma$ & p-value\\ [0.5ex]% inserts table
%
%\hline% inserts single horizontal line
%plaw&4$\pm$ 0.145&0.5$\pm$0.266&0.02&-&-&$\leq$0.0001\\% inserting body of the table% 
%\hline%inserts single line
%\end{tabular}
%\label{table:nonlin4}
%\end{table*}
%======================================================================

%\begin{figure}
%  \centering
%  \includegraphics[width=0.45\textwidth]{Hist_ban_total.pdf}
%  \caption{\small
%  }
%  \label{histban4}    
%\end{figure}

%======================================

%=======================================================================
\begin{figure}
 \begin{center}
    \includegraphics[width=0.45\textwidth]{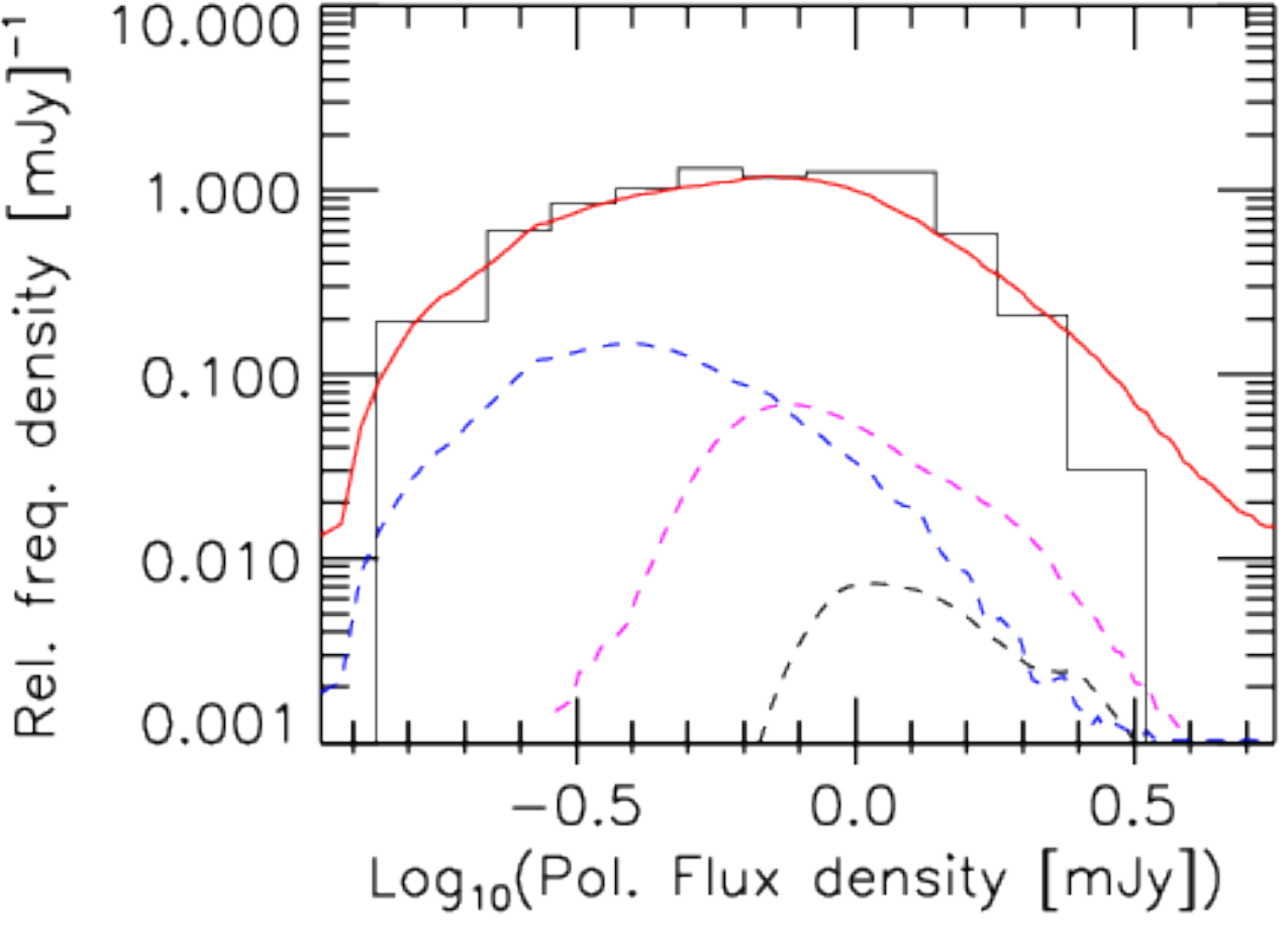}
 \end{center}
 \caption{Relative number frequency of polarized flux density as derived from the
light curves (Fig.\ref{histban3}) and from our heuristic modeling approach (straight red). The dashed curves
show the contributions from flux densities with polarization degrees of $\le$30\% (blue), $\ge$60\% (black) and
in between (magenta).}
 \label{fig:relait}
\end{figure}

\subsection{A heuristic analytic explanation of the polarized flux density distribution}
\label{section:heuristic}
\noindent
While the behavior of the entire sample of flares with significant polarized fluxes and 
polarization degrees is predicted by the simulations 
(see Figs. 5-7, and Tables 2 and 3.
in section \ref{section:Statistical}), the heuristic model we present here helps us to understand the
fact that the relative frequency density of measured polarized flux densities is much broader in
comparison to the relative frequency density of total flux density measurements 
as presented by \cite{Witzel2012} which was found to be consistent with a single-state emission process. 

In the following we call $D(F_{K, pol})$
the relative frequency density of measured polarized flux densities as shown in 
Fig.\ref{histban3}.
Keeping in mind the limited number of polarized flares that are available for our study this can be compared to the distribution shown for the total fluxes for a 
large sample of light curves by \cite{Witzel2012} (see Fig.3 therein). 
We now assume that 
for each randomly picked subset of $K_\mathrm{s}$-band flux densities, total flux density distribution has a similar shape.
We assume that for all flux densities all polarization degrees are possible.
This assumption is not fully justified and we comment on this later.
First, we pick values that belong to total flux densities $S_K$ that can be attributed to a polarization 
degree bin $p_{i}$ of width $\delta p$ from $p_{i}$ -$\delta p$/2 to $p_{i}$ +$\delta p$/2
the polarized flux densities would be $F_K \cdot p_{i}$.
We obtain from the distribution shown in Fig.\ref{fig:degrees}~(left) the weights $w_{i}(p_{i})$ for individual polarization states.
We can express the polarized flux density
distribution $D(F_{K,pol})$ as a product distribution that is a probability distribution 
constructed as the distribution of the product of (assumed to be) independent random variables 
$p_{i}$ and $D(F_K)$ that have known distributions. 

Using the weights $w_{i}(p_{i})$ for each polarization state the corresponding polarized flux density 
distribution $D(F_{K,pol_i})$ can be written as
\begin{equation}
D(F_{K,pol_i}) = w_i(p_{i}) \cdot D(F_K \cdot p_{i})~~~.
\end{equation}
The polarized flux density distribution $D(F_{K,pol})$ can then be written as a product distribution
by summing over all $N$ bins including all polarization degrees $p_{i}$:

\begin{equation}
D(F_{K,pol}) 
= \sum\limits_{i=1}^N D(F_{K,pol_i})
= \sum\limits_{i=1}^N w_i \cdot D(F_K \cdot p_{i})
\end{equation}
\noindent
In Fig.\ref{fig:relait} we show the result of this modeling approach.
In this figure we plot the measured (as in Fig.\ref{histban3}) and modeled relative number frequency of polarized flux density 
and show the combined contribution from different polarization states.
The steep drop towards higher polarized flux densities is due to the deficiency of bright flares with high polarization states. Of course this region is also affected by the fact that
the brightest flares tend to be statistically under-sampled.
In this region our initial assumption that for all flux densities all polarization degrees are possible
is not fulfilled. For fluxes above 5~mJy measured polarization degrees are below 30\%.
Therefore we overestimate the number of flare events with high polarized fluxes in this domain.
Except this deficit, however, the model closely describes the measured data which implies that 
the broader distribution as formally analyzed in section~\ref{section:polfluxdistrib}
can indeed be explained by combination of an intrinsic 
relative frequency total flux density histogram applied to the individual polarization states.

\subsection{Relation between total flux density and polarization degree}
\label{section:totdegr}

\begin{figure*}[]
     \begin{center}
        \subfloat{%
           \includegraphics[width=0.45\textwidth]{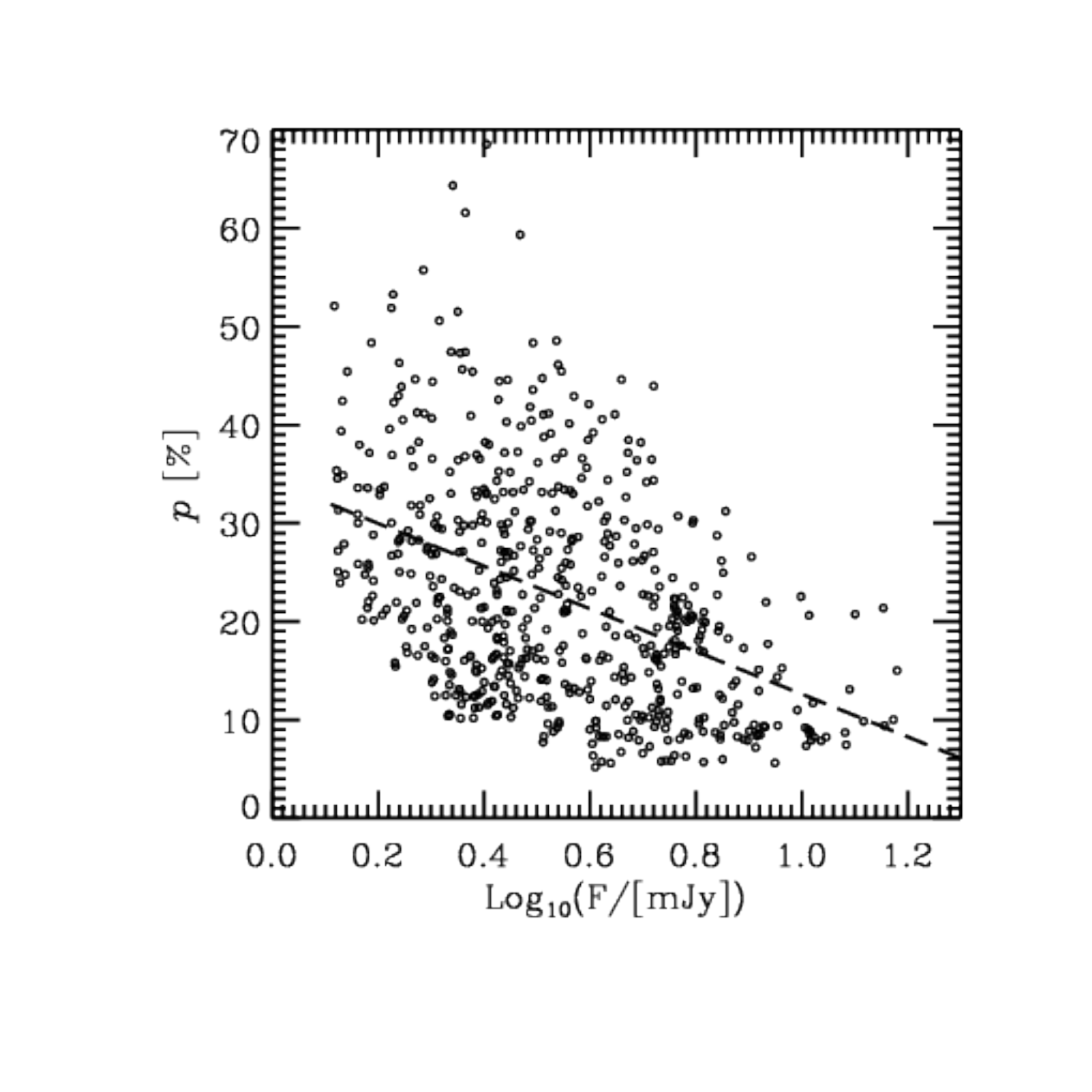}
        }
        \subfloat{%
            \includegraphics[width=0.45\textwidth]{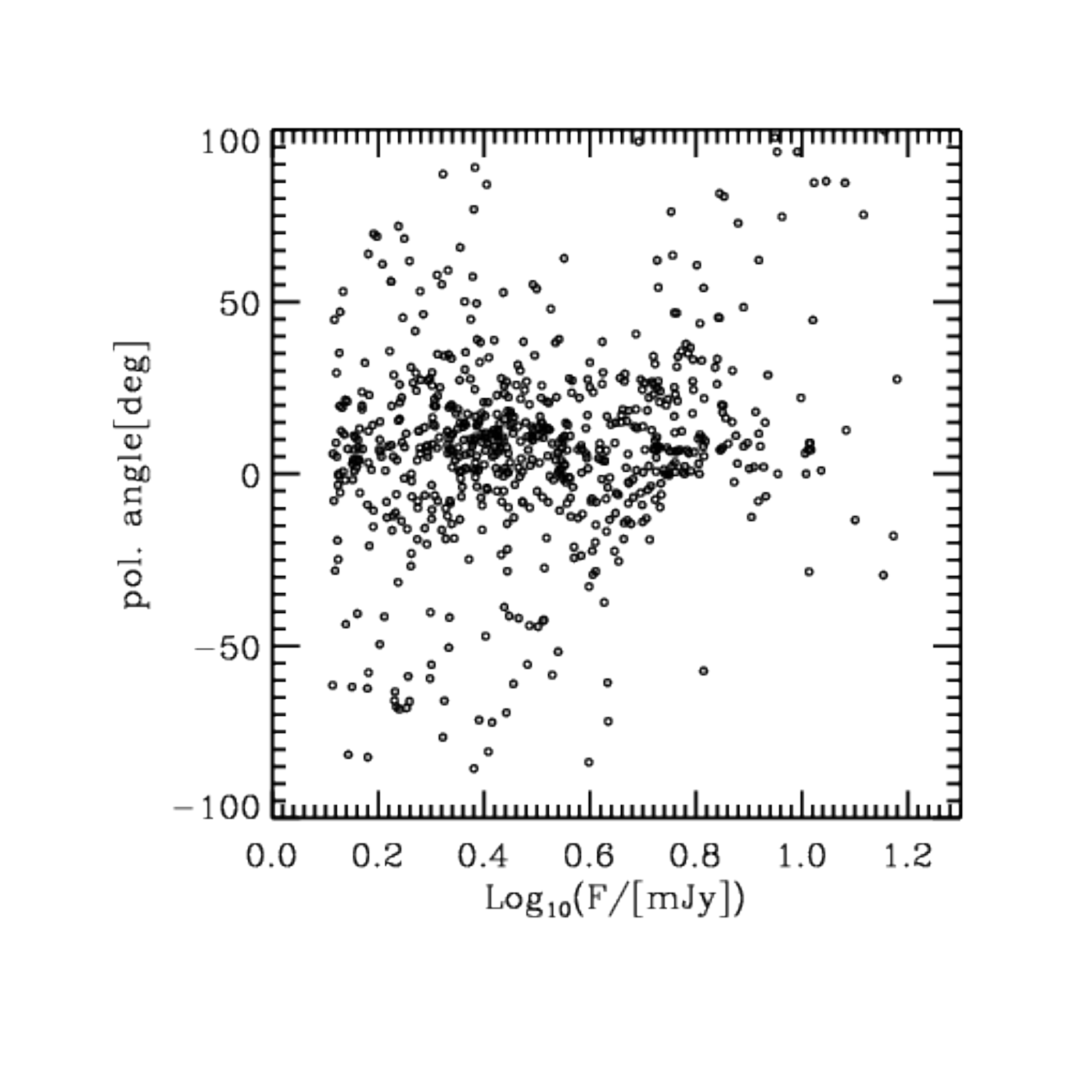}
        }\\ %  ------- End of the first row ----------------------%
 \end{center}
    \caption{Left: The relation between total flux density and polarization degree for significant data (based on Table \ref{table:nonlin}).
The dashed line shows the fitted line. 
  Right: The relation between total flux density and polarization angle for significant data.
     }
   \label{fig:relations}
\end{figure*}

Considering all the observed total flux densities for different epochs and their 
calculated polarization degrees, we plot the $K_\mathrm{s}$-band 
polarization degrees versus total flux densities in Fig.\ref{fig:relations}~(left). 
In this figure we consider the data that result in significant measurements 
of polarization degrees i.e. 
we excluded data that correspond to positions shown in non bold face in Table \ref{table:nonlin1}.
The same plot for the entire data set is shown in Fig.\ref{fig:App4}.
In both plots there is a clear trend that lower total flux densities have apparently higher polarization degrees, 
Polarization degrees are lower for higher flux densities for which the simulations show that the polarization data can be 
recovered very well.

In order to analyze if this trend is due to small-number statistics, 
we perform a bootstrap test. 
By minimizing the least square we fit a line to our significant data points 
(see bold face positions in Table \ref{table:nonlin})
and find a slope of
-21.7$\pm 1.25$\%($log_{10}$(F/[mJy]))$^{-1}$ 
and an intercept on the polarization degree axis of
34.3$\pm 0.74$\%.
The uncertainties are obtained by bootstrapping 
for $10^{4}$ times \citep{Wall&Jenkins2012}. 
We find the Pearson's correlation coefficient value, which measures 
how linearly dependent the parameters are, to be -0.43 for our anti-correlated data. 
The correlation coefficient for the original data could have been 
obtained by chance. Therefore, for measuring the strength of its value 
and its significance, we shuffle our data $10^{4}$ times 
and calculate the correlation coefficient each time for the resulted 
uncorrelated data sets. Then we calculate the mean and standard deviation 
($\sigma$) for the distribution of correlation coefficient values and 
check if the probability of the correlation coefficient of the original 
data was by chance or not. 
With a mean value of -10$^{-5}$ the correlation coefficient 
distribution is almost zero and the standard deviation is 0.042. 
Therefore, we find that the Pearson's correlation coefficient
for our data has a $\sim$10$\sigma$ offset from the mean value of 0.0 
that one finds for uncorrelated data sets.
The resulting correlation is represented by the dashed line in Fig.\ref{fig:relations} (left).
The simulations (see section \ref{section:Statistical}) show that 
this apparent correlation is most likely due to the  
asymmetrically distributed uncertainties that dominate the measurements at 
low flux density levels.

The apparent correlation is already implied by the results of the simulations
(see Figs.\ref{fig:simulation-1},\ref{fig:simulation-2} and Table \ref{table:nonlin}).
Table 2 exhibits a negative correlation between polarization degree and total flux density for intrinsic polarization degrees and total flux densities and it encloses the dashed line shown in Fig.\ref{fig:relations} (left).

%=======================================================================
%======================================

%#############################################################

%quark
\section{Summary and conclusions}
\label{section:summary}

We have analyzed the near-infrared polarization light curves obtained 
with NACO at the ESO VLT for Sgr A* at the center of the Milky Way.
Both the steep spectral index \citep[][and references there in]{bremer2011} and the strong variability in the NIR 
demonstrate that we are most likely dealing with optically thin synchrotron
radiation \citep{Eckart2012}. Therefore, all properties we can derive
based on the observation of variable polarized NIR radiation can directly 
be interpreted as source intrinsic properties.

For the entire data set the variation in polarization degree 
(mostly in the range of 10\%-50\% i.e. a factor of 5) 
is less pronounced than the variation in flare flux density 
(mostly in the range of $<$1~mJy to $>$15mJy i.e. a factor of $>$15).
However, simulations of the resulting observed poalrization parameters given the measured flux (and its error) in orthogonal polarized channels show that, at low flux densities a significant part of the variation in polarization degree
is due to measurement uncertainties.
The simulations presented in Section \ref{section:Statistical} also
show the uncertainty in recovering the intrinsic degree of polarization
for flux levels above 5~mJy is only of the order of 5\%. 
At this flux level, this value is considerably smaller than the measured
variation in polarization degree at high fluxes of up to 30\%. This implies an intrinsic variation of the polarization state of Sgr A* during flare events.

We show that there is a preferred polarization angle of 13$^o$$\pm$15$^o$
and that the polarized flux distribution is coupled to the total flux distribution 
analysed in \cite{Witzel2012}. 
The simulations presented in section \ref{section:Statistical} show that the 
uncertainty in the polarization angle even at high flux levels (e.g. $\ge$ 5~mJy) is dominated by the
statistical uncertainty (10$^o$ to 15$^o$) in recovering a value that is assumed to be constant.
Therefore, we can conclude that the intrinsic uncertainties on that quantity is smaller that 
the currently measured one and probably of the order of less than about 10$^o$.

The fact that for high fluxes the power-law slope of $\sim$4 of the number density of the 
polarized flux density is very close to the slope in number density distribution
of the total flux densities found by \cite{Witzel2012} also clearly indicates that there is
a preferred value or range of intrinsic degrees of polarization. 
The variation of this range must have been small over the past years (2004-2012)
else the slope in the number density of the polarized flux density would have been affected.
Combined with the result from comparing the distributions of polarization degrees and 
their relative uncertainties 
this is a clear indication of intrinsic variability in the degree of polarization.
Consistent with this we derived in section \ref{section:polfluxangle} for the entire set of significant polarization degrees its intrinsic 
relative variability must be of the order of 30\%.

Therefore, the entire distribution of polarized flux density as described formally in 
Section~\ref{section:polfluxdistrib}
can be thought of as being composed of the contributions of populations 
of flare events with varying flux densities but a rather constant polarization angle
and intrinsic polarization degrees 
that vary over a narrow range between about 10\% and 30\%.
A preferred range in polarization degree and a well defined preferred polarization angle
observed over a time span of 8 years (2004-2012) supports the assumption of a rather 
stable geometry for the Sgr A* system, i.e. a rather stable disk and/or jet/wind orientation.

In addition, a comparison of the observed data to the simulations presented in section \ref{section:Statistical} shows that the observed anti-correlation between polarization 
degree and total flux density is most likely dominated by an observational effect due to 
asymmetrically distributed uncertainties 
in the determination of the polarization degree for small flare fluxes (close to our acceptance flux and below). 
This means that, with the current instrumentation, it is impossible to know whether the polarization state of Sgr A* is intrinsically variable at flux densities below 2 mJy.
The observed apparent anti-correlation between polarization degree and total flux density means that the brighter flux 
density excursions are systematically less polarized compared to the lower flux densities. 
Such a behavior may be expected based on the findings by \citet{Zamaninasab2010} that the polarized 
flares in comparison with the randomly polarized red noise show a signature of radiating matter orbiting 
around the supermassive black hole.
The formation of partial Einstein rings and mild relativistic boosting during the approach of an orbiting 
source component will lead to a bright geometrically depolarized emission during a flare event. 
Polarization variations by increasing or decreasing of the flux densities have been reported 
in other publications \citep[see e.g.][]{Eckart2006b, Meyer2006a, trippe2007}.

%ffffffffffffffffffff
%%======================================

It is also instructive to ensure that the results presented in this work are consistent with those of
previous publications despite of different ways of extracting the polarization information.
Table \ref{table:nonlin3} presents a comparison of polarization angle and degree between this work
and former NIR polarization studies of Sgr~A*. Despite the strong source variability and different 
methods in deriving the observables \citep[see][for details]{Witzel2011} 
the agreement is satisfactory and as expected from 
the findings in Section~\ref{section:heuristic}
and shown in Fig.\ref{fig:relations}. The polarization angles of earlier analysis 
are also close to or approximately centered around a preferred value.

The preferred NIR polarization angle may also be linked to polarization properties in the 
radio cm- to mm-domain and to the orientation of the Sgr A* system.
In the radio cm-regime linear polarization of Sgr A* is very small, however, the source
shows a fractional circular polarization of around 0.4\% \citep{bower1999-523, bower1999-521}.
The circular polarization decreases towards short mm-wavelengths \citep{bower2003}, where \citet{macquart2006} report variable linear polarization from Sgr A* of a few percent in the mm-domain.
The polarization degree and angle in the sub-mm are likely linked to the magnetic field structure
or the general orientation of the source.
As in particular the NIR flare emission very likely originates from optically thin synchrotron
radiation (Eckart et al. 2012) one may expect a link between the preferred NIR polarization angle and the
NIR/radio structure of Sgr A*.
At millimeter wavelengths interstellar scattering is small and allows insight into the intrinsic source structure
of Sgr A*. \citet{bower2014} report an intrinsic major axis position angle of the structure of 
95$^o$ $\pm$10$^o$ (3 $\sigma$).
This angle of the radio structure is within the uncertainties orthogonal to the preferred infrared polarization angle.
It is currently unclear how the intrinsic angle of Sgr A* can with certainty be related to external structures.
This is particularly true if one takes into account the over 3 to 4 orders of magnitude in linear distance the
position angle changes.

In a range of position angles between 120$^o$ and 130$^o$
\cite{Eckart2006b, Eckart2006c}  
report an elongated NIR feature, an elongated X-ray feature \citep[see also][]{morris2004}
and a more extended elongated structure called LF, XF and EF in Figure.9
by \cite{Eckart2006b}.  
These features may be associated with a jet phenomenon. In this case the preferred NIR polarization angle
may be associated with the jet components close to or at the foot point of the jet as for jet components
the polarization may be along or perpendicular to the jet direction.

It is also possible that the NIR emission originates in hot spots on an accretion disk
in a sunspot like geometry in which the E-vector is mainly perpendicular to the equatorial plane.
Such a magneto-hydrodynamical model for the formation of episodic fast outflow is presented by \citet{yuan2009}.
Acceleration of coronal plasma due to reconnection may then drive a jet or wind perpendicular to the intrinsic radio
structure of the disk along the position angle of the NIR polarization.
The mini-cavity that may be due to the interaction of a nuclear wind from Sgr A* is located at a position
angle of about 193$^o$ (i.e. 13$^o$+180$^o$). The cometary tails of sources X3 and X7 reported by \citet{muzic2010}
also present additional observational support for the presence of a fast wind from Sgr A* under that position angle.

Further progress in investigating in particular the faint polarization states of Sgr A* in the NIR 
will require a higher angular resolution in order to better discriminate Sgr A* against background
contamination. 
It is also of interest to use the well defined NIR polarization properties of Sgr A* 
to better determine the apparent stability of the underlying geometrical 
structure of the system and potentially use variations in this stability to trace interactions 
of the super-massive black hole at the center of the Milky Way with its immediate environment.

%#############################################################
\begin{acknowledgements}

B.Sh. and N.S are members of the Bonn Cologne Graduate School (BCGS) for Physics and Astronomy supported by the Deutsche Forschungsgemeinschaft and acknowledge support from BCGS. B.Sh. gladly thanks E. Clausen-Brown, S. Trippe and M. Oshagh for fruitful discussions. 
We received funding from the
European Union Seventh Framework Programme (FP7/2007-2013)
under grant agreement No.312789.

This work was also supported in part by 
the Max Planck Society and the University of Cologne through
the International Max Planck Research School (IMPRS) for Astronomy and
Astrophysics.
Part of this work was supported by the German Deutsche Forschungsgemeinschaft, DFG, via grant SFB 956 and fruitful discussions
with members of the European Union funded COST Action MP0905: Black
Holes in a violent Universe and the
COST Action MP1104:
Polarization as a tool to study the Solar System and beyond.

\end{acknowledgements}
%\vspace*{0.5cm}
%#############################################################

%\vspace*{0.5cm}
\bibliographystyle{aa} % style aa.bst
\bibliography{shahzamanian} % your references Yourfile.bib

\newpage
\vfill

\begin{appendix}
\section{}

Here we provide additional plots similar to the ones shown throughout the main body of the text, showing the principal results from the analysis of Sgr A* observations taken in the polarization mode (NACO), but using the entire data set without limitation to the only significant values.

%======================================
\begin{figure*}[]
     \begin{center}
        \subfloat{%
            \includegraphics[width=0.45\textwidth]{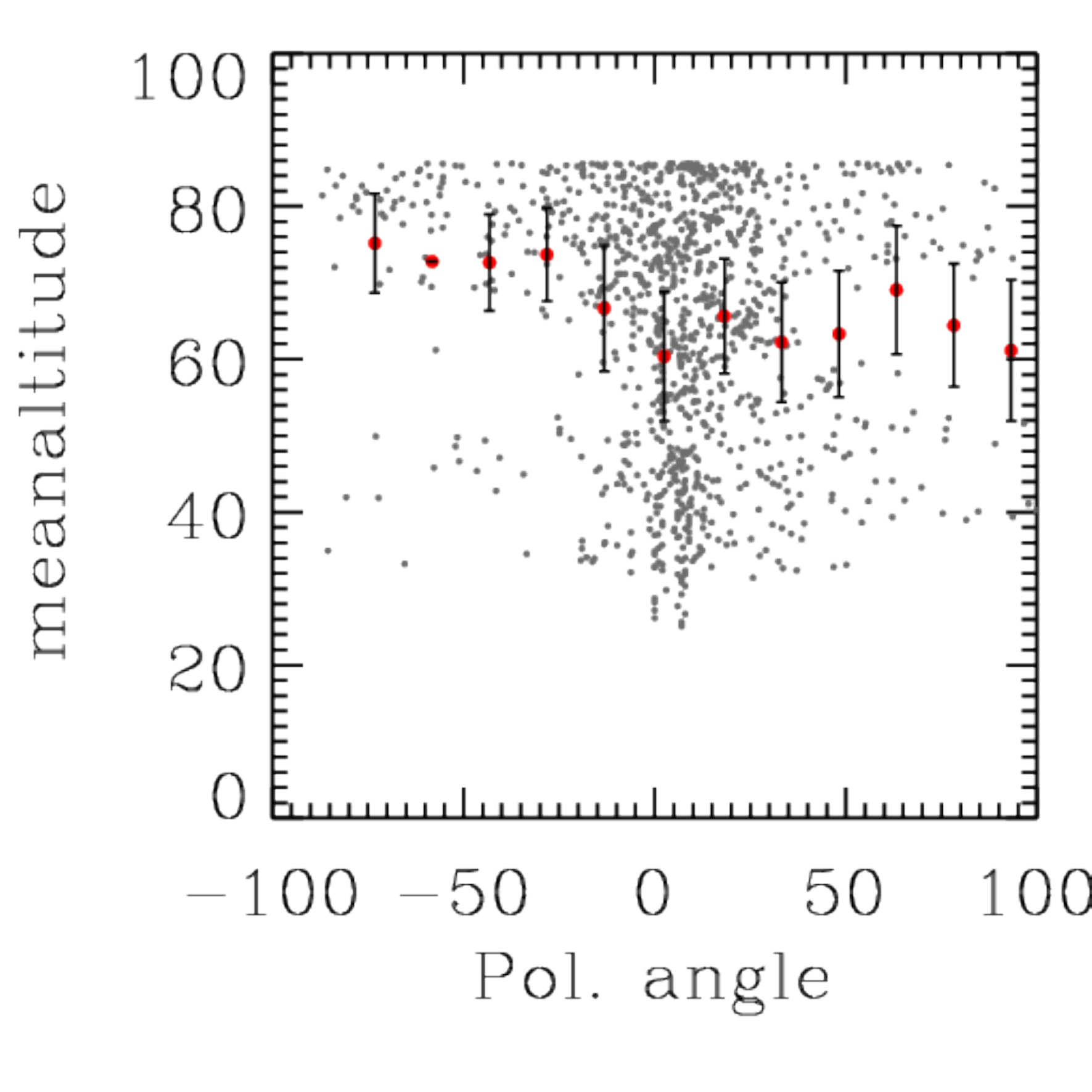}
        }
        \subfloat{%
           \includegraphics[width=0.45\textwidth]{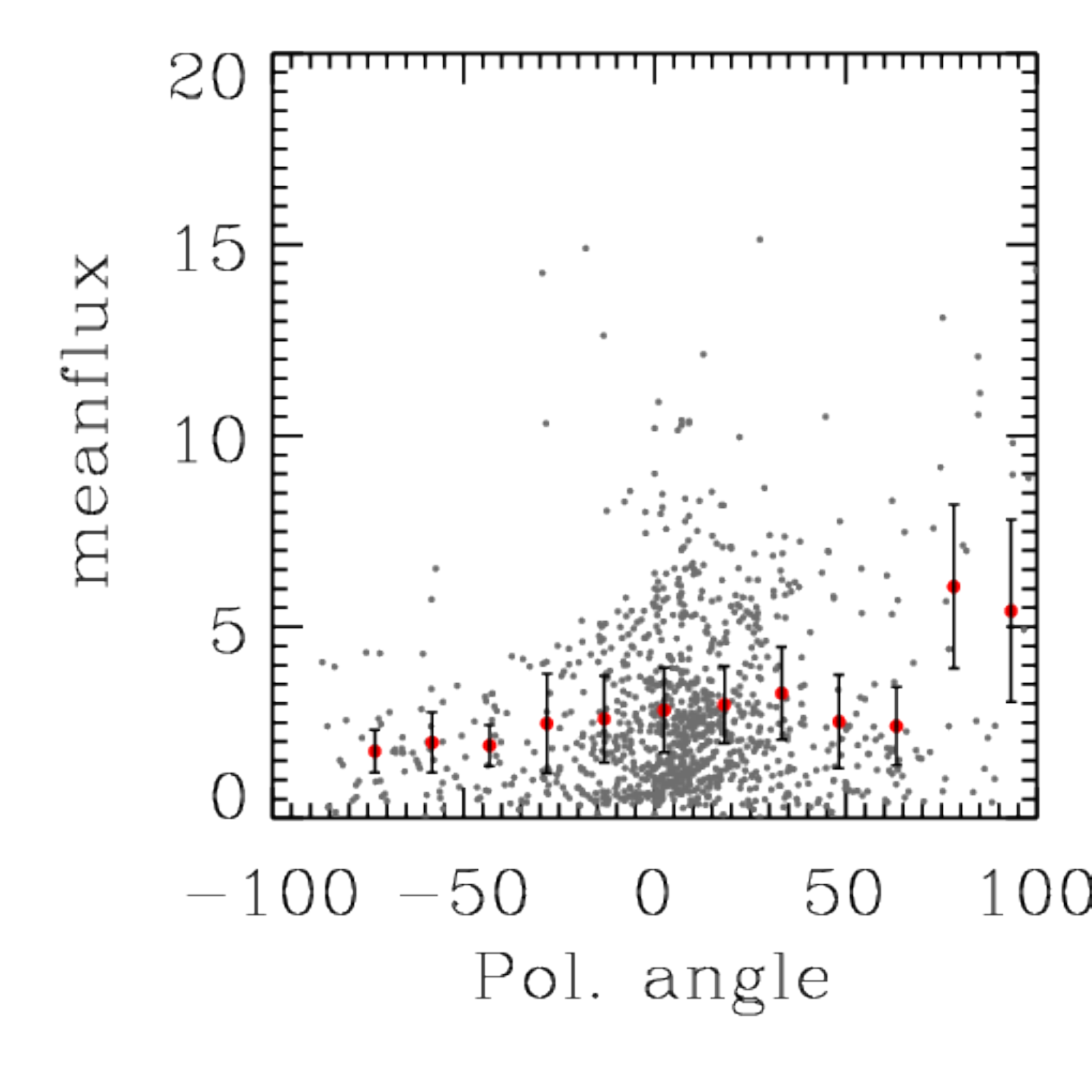}
        }\\ %  ------- End of the first row ----------------------%
 \end{center}
    \caption{Left: Altitude (elevation) of Sgr A* as a function of polarization angle
as derived from all data:
    Right: Total flux density as a function of polarization angle; 
    The bin width in polarization angle is 15$^o$.
    The values for individual measurements are shown as black dots. The mean values per bin
are shown as red dots with error bars indicating the standard deviation, if the number of data points
per bin is larger than 2.
     }
   \label{fig:App1}
\end{figure*}
%======================================
%======================================
\begin{figure*}[]
     \begin{center}
        \subfloat{%
            \includegraphics[width=0.45\textwidth]{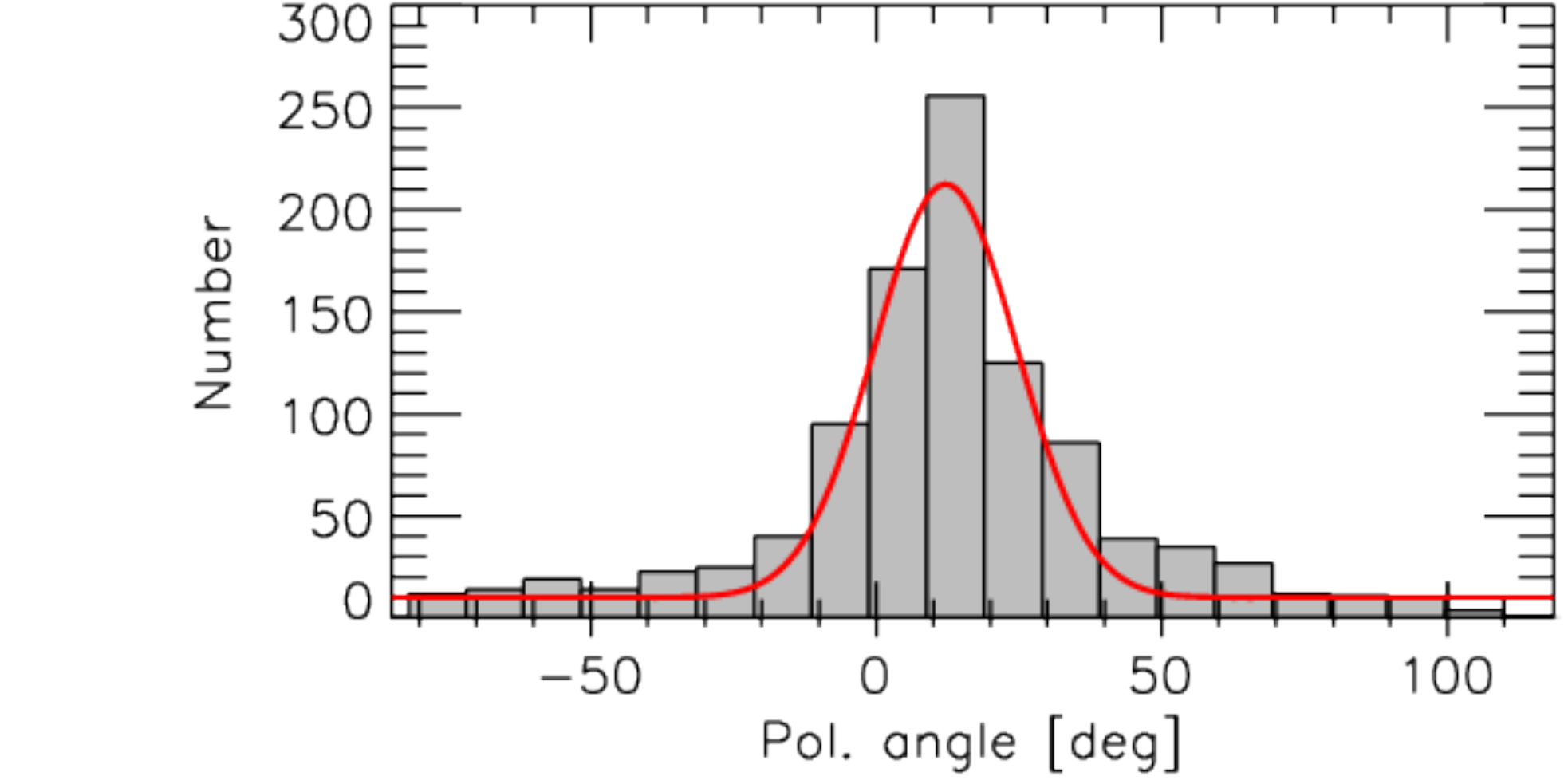}
        }
        \subfloat{%
            \includegraphics[width=0.45\textwidth]{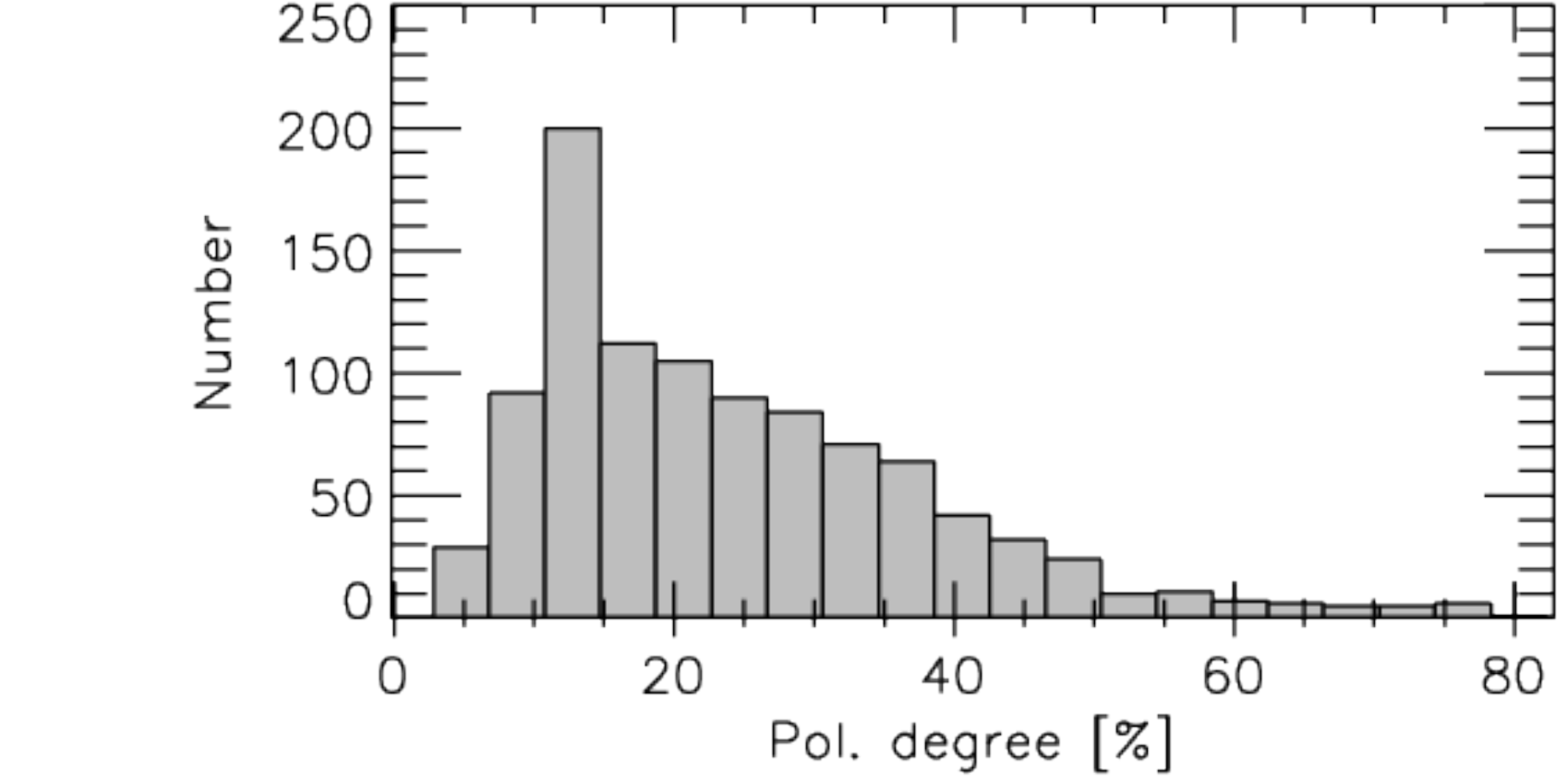}
        }\\ %  ------- End of the first row ----------------------%
 \end{center}
    \caption{Left: Distribution of $K_\mathrm{s}$-band polarization angles of Sgr~A* for the entire data set. 
The red line shows the fit with a Gaussian distribution.  
Right: Distribution of $K_\mathrm{s}$-band polarization degrees of Sgr~A* for the entire data set. 
     }
   \label{fig:App2}
\end{figure*}
%======================================
%======================================
\begin{figure*}[]
     \begin{center}
        \subfloat{%
  \includegraphics[width=0.45\textwidth]{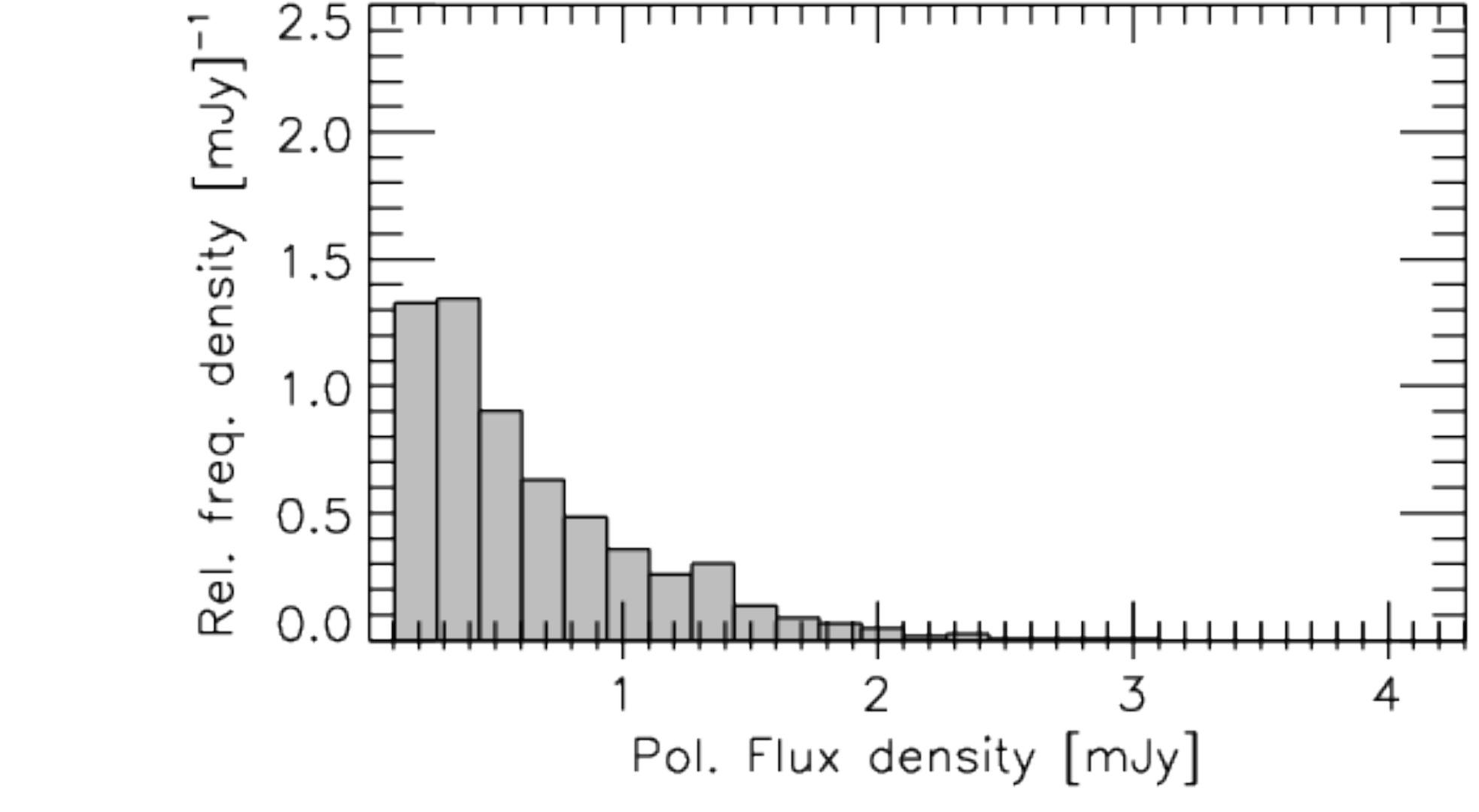}
        }
        \subfloat{%
  \includegraphics[width=0.45\textwidth]{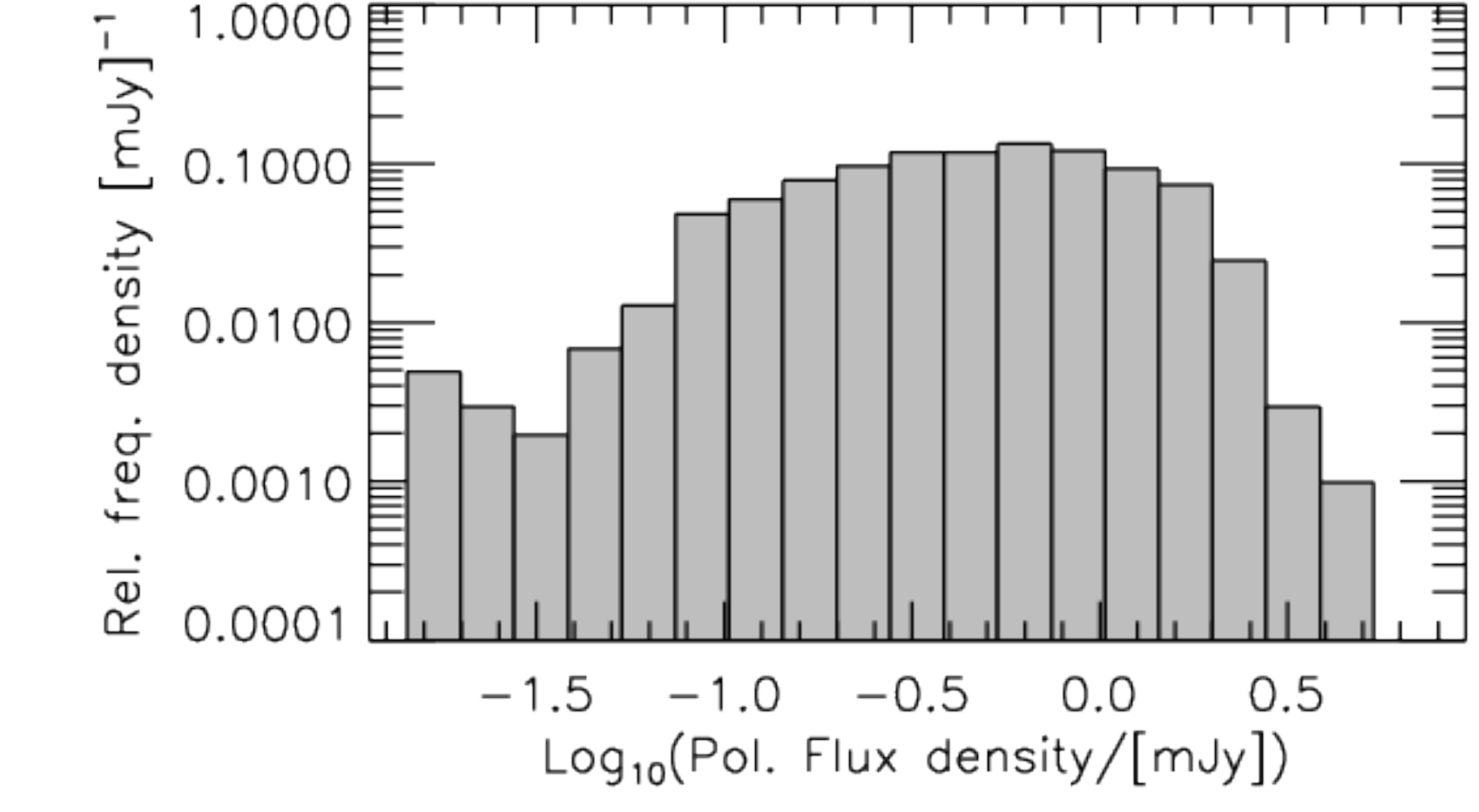}
        }\\ %  ------- End of the first row ----------------------%
 \end{center}
  \caption{\small Left: Histogram of polarized flux density (total flux density times polarization degree) for the whole data set after correction for stellar contamination.
  Right: the same plot in logarithmic scale.
  }
   \label{fig:App3}
\end{figure*}
%======================================
%======================================
\begin{figure*}[]
     \begin{center}
        \subfloat{%
            \includegraphics[width=0.45\textwidth]{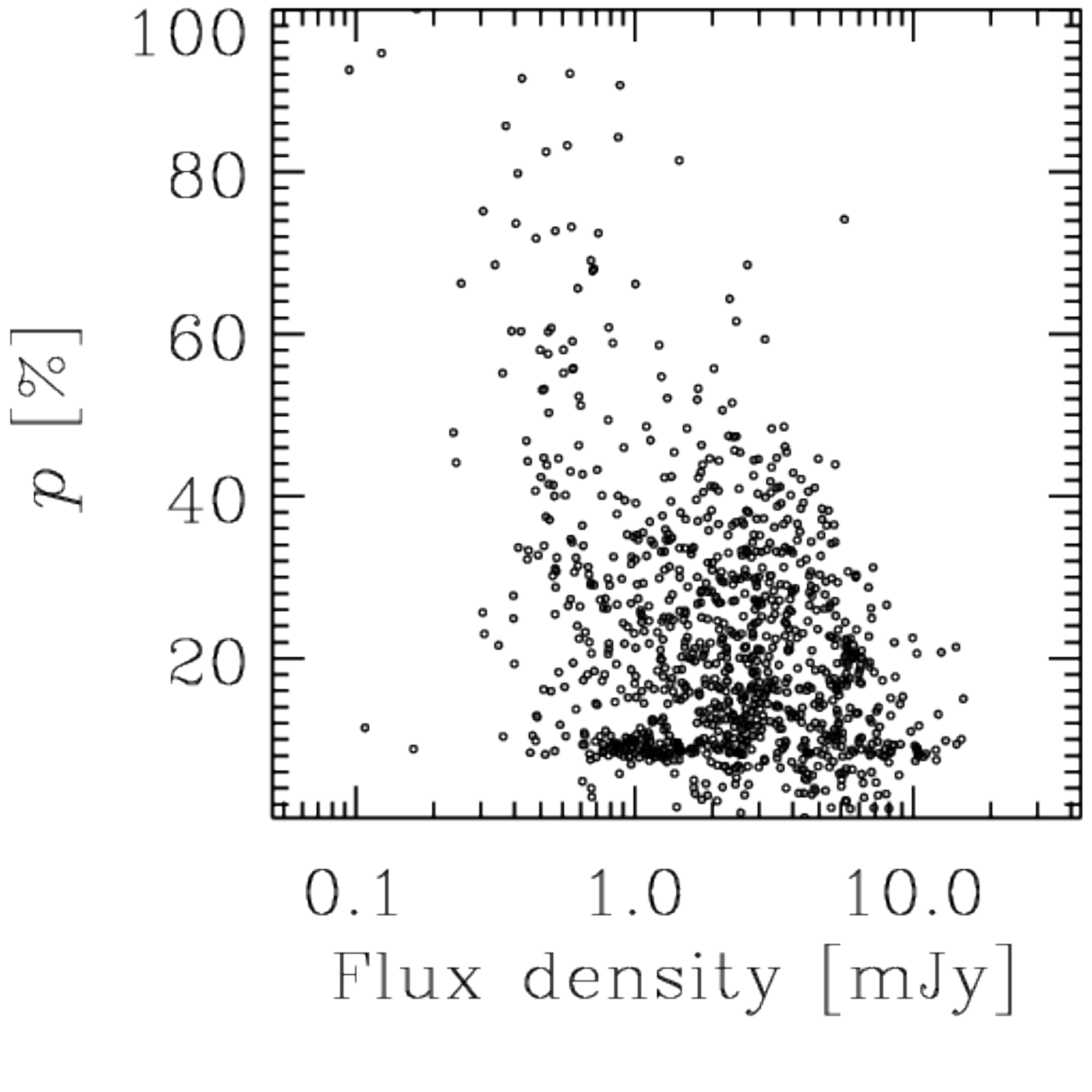}
        }
        \subfloat{%
            \includegraphics[width=0.45\textwidth]{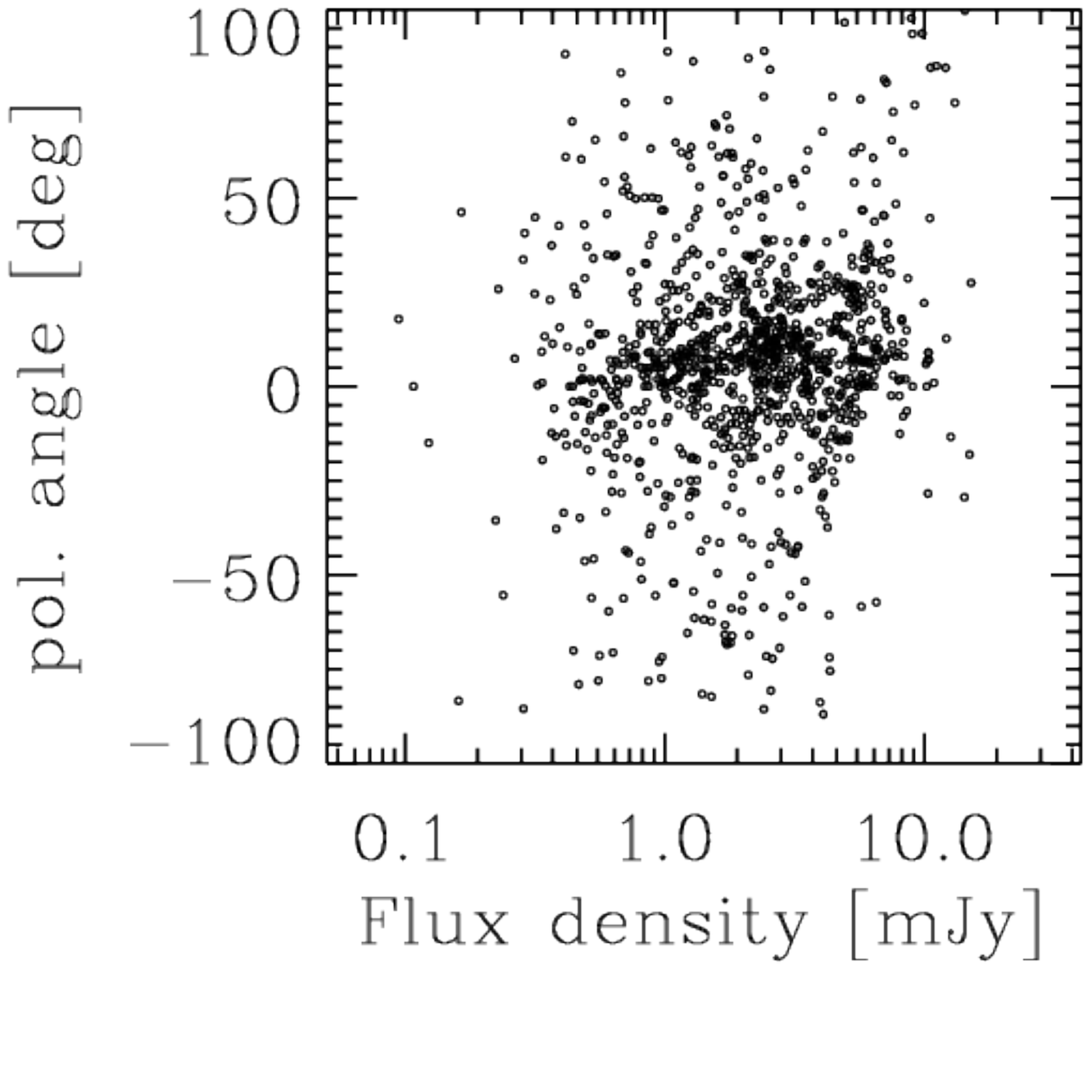}
        }\\ %  ------- End of the first row ----------------------%
 \end{center}
    \caption{Left: Total flux density and degree of polarization relation for the entire data after correction for the offset. 
  Right: Total flux density and angle of polarization relation for the entire data  after correction for the offset.
     }
   \label{fig:App4}
\end{figure*}
%======================================
\begin{figure*}[]
     \begin{center}
%        \subfloat{%
%            \includegraphics[width=0.45\textwidth]{angdeguncetrain.pdf}
%        }
        \subfloat{%
            \includegraphics[width=0.45\textwidth]{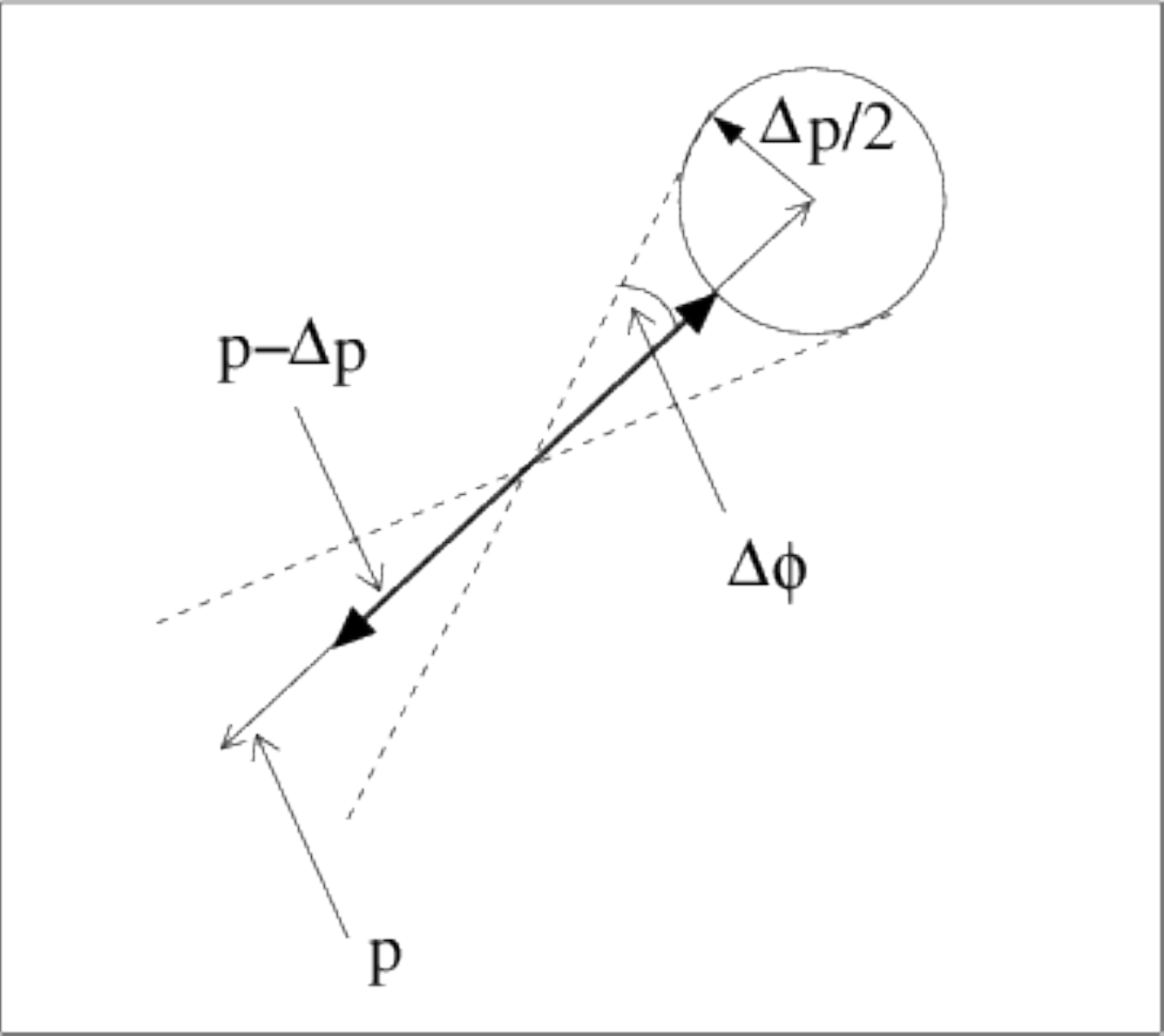}
        }\\ %  ------- End of the first row ----------------------%
 \end{center}
    \caption{Approximate relation $\Delta$$p$$\sim$$p$~tan($\Delta$$\phi$) 
between the mean uncertainty of the polarization angle $\Delta$$\phi$ and
the polarization degree $\Delta$$p$. Here polarization degree $p$ and polarization angle $\phi$ are projected on the sky.
     }
   \label{fig:App5}
\end{figure*}

\end{appendix}

\end{document}